\documentclass[10pt,preprint]{emulateapj}

\shorttitle{Photospheric Properties of T Tauri Stars}
\shortauthors{Herczeg}

\begin{document}

\title{An Optical Spectroscopic Study of T Tauri Stars.  I.  Photospheric Properties}

\author{Gregory J. Herczeg\altaffilmark{1,2,3,4} \& Lynne A. Hillenbrand\altaffilmark{2}}
\altaffiltext{1}{Kavli Institute for Astronomy and Astrophysics, Peking University, Yi He Yuan Lu 5, Haidian Qu, Beijing 100871, People's Republic of China}
\altaffiltext{2}{Caltech, MC105-24, 1200 E. California Blvd., Pasadena, CA 91125, USA}
\altaffiltext{3}{Max-Planck-Institut f\"ur extraterrestrische Physik, Postfach 1312, 85741 Garching, Germany}
\altaffiltext{4}{Visiting Astronomer, LERMA, Observatoire de Paris, ENS, UPMC, UCP, CNRS, 61 avenue de l'Observatoire, 75014 Paris, France}

\begin{abstract}
Measurements of the mass and age of young stars from their location in the HR diagram are limited by not only the typical observational
uncertainties that apply to field stars, but also by large systematic uncertainties related to circumstellar phenomena.  In this paper, we analyze flux calibrated optical spectra to measure accurate spectral types and extinctions of 283 nearby T Tauri stars.  The primary advances in this paper are (1) the incorportation of a simplistic accretion continuum in optical spectral type and extinction measurements calculated over the full optical wavelength range and (2) the uniform analysis of a large sample of stars, many of which are well known and can serve as benchmarks.  Comparisons between the non-accreting TTS photospheric templates and stellar photosphere models are used to derive conversions from spectral type to temperature.  Differences between spectral types can be subtle and difficult to discern, especially when accounting for accretion and extinction. 
The spectral types measured here are mostly consistent with spectral types measured over the past decade.  However, our new spectral types are 1-2 subclasses later than literature spectral types for the original members of the TW Hya Association and are discrepant with literature values for some well known members of the Taurus Molecular Cloud.  
Our extinction measurements are consistent with other optical extinction measurements but are typically 1 mag.~lower than near-IR measurements, likely the result of methodological differences and the presence of near-IR excesses in most CTTSs. 
As an illustration of the impact of accretion, spectral type, and extinction uncertainties on the HR diagrams of young clusters, we find that the resulting luminosity spread of stars in the TW Hya Association is 15-30\%.  The luminosity spread in the TWA and previously measured for binary stars in Taurus suggests that for a majority of stars, protostellar accretion rates are not large enough to significantly alter the subsequent evolution.
\end{abstract}
 

%
%


\section{INTRODUCTION}

Classical T Tauri stars are the adolescents of stellar evolution.  The star is near the end of its growth and almost fully formed, with a remnant disk and ongoing accretion.  The accretion/disk phase typically lasts $\sim 2-5$ Myr, though some stars take as long as 10 Myr before losing their disks and emerging towards maturity.    Strong magnetic activity leads to pimply spots on the stellar surfaces.  Some T Tauri stars are still hidden inside their disks, not yet ready to emerge.   Manic mood swings change the appearance of the star and are often explained with stochastic accretion.  Depression has been seen in lightcurves on timescales of days to years.  
Sometimes every classical T Tauri star seems as uniquely precious as a snowflake.

T Tauri star properties were systematically characterized in seminal papers by e.g.  \citet{Cohen1979,HBC1988,Basri1990,Valenti1993,Hartigan1995,Kenyon1995,Gullbring1998}.  In the last decade, dedicated optical and IR searches revealed thousands of young stars, typically confirmed with spectral typing \citep[e.g.][]{Hillenbrand1997,Briceno2002,Luhman2004,Rebull2010}.  However, significant differences in extinction and accretion properties between different papers and methods has lead to confusion in the properties of even the closest and best studied samples of young stars. 

Some of this confusion is exacerbated by stochastic and rotation variability of T Tauri stars.  While manic and depressive periods provide fascinating diagnostics of the stellar environment and star-disk interactions, they also pose significant problems for assessing the stellar properties and evolution of the star/disk system.  How disk mass, structure and accretion rate change with age and mass requires accurate spectral typing and luminosity measurements \citep[e.g.][]{Furlan2006,Sicilia2010,Oliveira2013,Andrews2013}.  While median cluster ages provide an accurate relative age scale between regions \citep[e.g.][]{Naylor2009}, age spreads within clusters may be real or could result from observational uncertainties \citep[e.g.][]{Hartmann1998,Hillenbrand2008,Preibisch2012}.

\begin{table*}[!t]
\caption{Observation Setup and Log}
\label{tab:obslog.tab}
\begin{tabular}{cccccccccccc}
&&&& \multicolumn{3}{c}{Blue Setup} &&& \multicolumn{3}{c}{Red Setup}\\
Telescope & Dates& Instrument & Slit & Grating & Wavelength & Res.  &&&Grating & Wavelength & Res.  \\
\hline
Palomar & 18-21 Jan. 2008 & DoubleSpec & 1--4$^{\prime\prime}$ & B600 & 3000--5700 & 700 &&& R316& 6200--8700 & 500 \\
Palomar & 28-30 Dec. 2008  & DoubleSpec & 4$^{\prime\prime}$ & B600 & 3000--5700 & 700 &&& R316& 6200--8700 & 500 \\
Keck I & 23 Nov.~2006 & LRIS & $0\farcs7-1^{\prime\prime}$ &B400 & 3000--5700 & 900 &&& R400 & 5700--9400 & 1000 \\
Keck I & 28 May~2008   & LRIS & $1^{\prime\prime}$ &B400 & 3000--5700 & 900 &&& R400 & 5700--9400 & 1000 \\
\hline
\end{tabular}
\end{table*}

The uncertainties in stellar parameters affect our interpretation of stellar evolution.  For example, \citet{Gullbring1998} found accretion rates an order of magnitude lower than those of \citet{Hartigan1995} and attributed much of this difference to lower values of extinction. 
The \citet{Gullbring1998} accretion rates of $10^{-8}$ M$_\odot$~yr$^{-1}$ means that steady accretion in the CTTS phase accounts for a negligible amount of the final mass of a star. 
 However, subsequent near-IR analyses have revised extinctions upward \citep[e.g.][]{White2001,Fischer2011,Furlan2011}.  These higher extinctions would yield accretion rates of $10^{-7}$ M$_\odot$~yr$^{-1}$, fast enough that steady accretion over the $\sim 2-3$ Myr CTTS phase would account for $\sim 20-50$\% of the final stellar mass, or more with the older ages measured by \citet{Bell2013}.  The uncertainties in stellar properties introduce skepticism in our ability to use young stellar populations to test theories of star formation and pre-main sequence evolution.

For classical T Tauri stars, minimizing the uncertainties in spectral type, extinction, and accretion (often referred to as veiling of the photosphere by accretion) requires fitting all three parameters simultaneously \citep[e.g.][]{Bertout1988,Basri1989,Hartigan2003}.  In recent years, such fits have received increasing attention and have been applied  {\it HST} photometry of the Orion Nebula Cluster \citep{daRio2010,Manara2012}, broadband optical/near-IR spectra of two Orion Nebular Cluster stars \citep{Manara2013b}, and to near-IR spectroscopy \citep{Fischer2011,McClure2013}.

In this project, we  analyze low resolution optical blue-red spectra to determine the stellar and accretion properties of 283 of the nearest young stars in Taurus, Lupus, Ophiucus, the TW Hya Association, and the MBM 12 Association.  This first paper focuses on spectral types and extinctions of our sample.  The primary advances are the inclusion of blue spectra to complement commonly used red optical spectra and accretion estimates to calculate the effective temperatures and luminosities with a single, consistent approach for a large sample of stars.  Discrepancies are found between our results and near-IR based extinction measurements.  We then discuss how these uncertainties affect the reliability of age measurements.  This work was initially motivated to calculate accretion rates from the excess Balmer continuum emission, which will be described in a second paper.  A third paper in this series will discuss spectrophotometric variability within our sample.

\begin{table}
\caption{Flux Calibration}
\label{fluxcal.tab}
\begin{tabular}{ccc}
Wavelength & G191B2B & LTT 3864 \\
\hline
\AA & \multicolumn{2}{c}{Absolute Scatter in Fluxes} \\
\hline
3500  &  0.067  & 0.087\\
4300  &   0.056 & 0.046\\
5400  &   0.041 & 0.047\\
6300  &   0.063 & 0.091 \\
8400  &   0.061 & 0.089  \\
\hline
Flux ratio & \multicolumn{2}{c}{Scatter in Flux Ratios} \\
\hline
$F_{7020}$/$F_{7140}$  &   0.007 & 0.005\\
$F_{8400}$/$F_{6300}$  &   0.016 & 0.014\\
$F_{6300}$/$F_{5400}$  &   0.057 & 0.101 \\
$F_{4300}$/$F_{5400}$    &   0.034 & 0.051\\
$F_{3500}$/$F_{5400}$    &   0.048 &  0.087\\
\hline
\end{tabular}
\end{table}

\section{OBSERVATIONS}

We obtained low resolution optical spectra with the Double Spectrograph \citep[DBSP][]{Oke1982} on the Hale 200 inch telescope at Palomar Observatory on 18-21 Jan.~2008 and 28-30 Dec.~2008, and with the Low Resolution Imaging Spectrograph \citep[LRIS][]{Oke1995,McCarthy1998} on Keck I on 23 Nov. 2006 and 28 May 2008.  The entire sample of the 2006 Keck observations was published in \citet{Herczeg2008}.  The latest spectral types of the May 2008 run were published in \citet{Herczeg2009}.  The Atmospheric Dispersion Corrector \citep{Phillips2006} was used for the May 2008 run but was not yet available in November 2006.  Both DBSP and LRIS use a dichroic to split the light into red and blue beams at $\sim 5600$ \AA.  
Details of the gratings and spectral coverage are listed in Table~\ref{tab:obslog.tab}.

On DBSP, the blue light was recorded by the CCD 23 detector, with $15\mu$m ($0\farcs389$) pixels in a $2048\times4096$ format.  The red light was recorded by the Tektronix detector, with 24 $\mu$m ($0\farcs468$) pixels in a $1024\times1024$ format.  The red detector has since been replaced.  On LRIS, the blue E2V and the red LBNL detectors both have $2048\times4096$ pixels with a plate scale of $0\farcs135$.

Our typical observing strategy consisted of 3--10 short (1-60s) red exposures and 1--2 long (60--900s) blue exposures obtained simultaneously.  Most DBSP observations in Jan. 2008 were obtained with the $2\farcs$-width slit, though a few sources were observed with the $1\farcs$ or $4\farcs$-width slits, adjusted for seeing.  All Dec.~2008 observations were obtained with the $4\farcs$-width slit.  Our LRIS observations were obtained with the $0\farcs7$ and $1\farcs0$ slits.  Seeing during both Palomar runs typically varied from 2--4$^{\prime\prime}$, though for a few hours the seeing reached $\sim 1^{\prime\prime}$.  The seeing was $\sim 0\farcs8$ and $\sim 0\farcs 7$ during our Nov.~2006 and May~2008 Keck runs, respectively.  Seeing was often worse than these measurements for objects at high airmass.  The position angle of the slit was set to the parallactic angle for all observations of single stars to minimize slit loss.  For binaries, the position angle may be misaligned with the parallactic angle.
These observations were timed to occur at low airmass or when the parallactic angle matched the binary position angle.

The images were overscan-subtracted and flat-fielded.  Most DBSP spectra were extracted using a 21-pixel ($10^{\prime\prime}$ window centered on the source, followed by subtracting the sky as measured nearby on the detector.  Binaries with separations $<5^{\prime\prime}$ were extracted simultaneously by assuming a wavelength-dependent point spread function determined from an observation of a single star observed close in time.  The counts from one source are subtracted from the image, yielding a clean extraction of counts from the other source.
 In several cases the counts are extracted on only half of the line spread function to further minimize contamination from the nearby component.  
Each spectrum is then corrected for light loss outside the slit and outside our extraction window based on the measured seeing as a function of wavelength and under the assumption that the point spread function is Gaussian.   The light loss is typically 3--10\% and increases to short wavelengths.

\subsection{Flux calibration}

To calibrate fluxes, spectrophotometric standards (G191B2B,  LLT 3864, Hz 44, Feige 110, and Feige 34, see Oke~1990) were observed $\sim8-13$ times on most nights.  On 21 Jan.~2008, G191B2B was observed twice and the night ended early because of snow.  The 2006 Keck run included only two spectrophotometric standards and has a large uncertainty in the flux calibration.  These spectra were also used to correct telluric features in the red, particularly H$_2$O bands at 7200 and 8200 \AA.  Windows between 7580--7680 and 6860-6890 \AA\ are severely contaminated by deep telluric absorption and not used.  A different atmospheric transmission curve was calculated for every night and was applied to each spectrum.  The correction at 3500 \AA\ ranged from 0.5-0.65 mag/airmass at Palomar and 0.4 mag/airmass at Keck.

The standard deviation in count rates and flux ratios for our 47 DBSP spectra of G191B2B and 9 DBSP spectra of LTT 3864 are listed in Table~\ref{fluxcal.tab}.  The flux calibration is based on multiple G191B2B spectra each night, so the standard deviation in flux are not completely independent.  The LTT 3864 spectra were observed at airmass $\sim 3$ and are all independent data points.  The flux calibration within the red channel is $<2\%$.  The absolute flux uncertainty, cross-calibration between the red and blue spectra, and the relative flux calibration withinin the blue channel are accurate to $\sim 5$\%.  The quality of the calibration degrades to $\sim 10$\% at high airmass.
When extracting close binaries the absolute accuracy in flux is $\sim 30\%$, particularly for secondaries that are much fainter than the primary or for observations where the seeing was larger than the binary separation.

Fringing is often apparent in DBSP spectra at $<3700$ \AA\ for observations obtained at high airmass and is likely a result of telescope vignetting.  However, accurate continuum fluxes in this region are still measurable in large wavelength bins.

\begin{table*}
\caption{Spectral indices}
\label{tab:indices.tab}
\begin{tabular}{cccccccccc}
Name      &  Continuum Range (C) &  Band Range (B) & Feature & x & SpT & Range & Zero-pt & rms$^a$\\
\hline
G-band      & 4550--4650        & 4150--4250 & G-band & C/B &$-25.6+29.96x$ & G & G0 & $\sim 1$\\
 R5150      & 4600--4700        & 5050--5150   &  MgH & $\frac{F(5100)}{F(4650)} \frac{F_{\rm line}(4650)}{F_{\rm line}(5100)}^b$ & $-29.7+28.3x$ & K0-M0 & K0 & 1.0\\
TiO 6250  & 6430--6465        &  6240--6270 & TiO   & $\log$ ($\frac{C}{B}-1$) & $3.20-5.43x+1.73x^2 $ & (M0--M4) & M0 &--\\
TiO 6800  &  6600--6660,6990--7050    & 6750--6900  & TiO  & C/B & $-15.37 + 19.77x$ & K5--M0.5 & K0 &--\\
TiO 7140  & 7005--7035        &  7130--7155 & TiO   &   $\log$ ($\frac{C}{B}-1$)& $4.36 + 6.33x + 1.57 x^2$ & M0--M4.5 & M0 & 0.42$^c$ \\
TiO 7700  & 8120--8160$^d$        &  7750--7800 & TiO   &  C/B &$0.11 + 2.27 x$  & M3--M8 &M0 &  0.21 \\
TiO 8465  & 8345--8385        &  8455--8475 & TiO   & C/B & $-0.74 + 4.21 x$  & M4--M8 & M0 & 0.18\\
\hline
\multicolumn{9}{l}{All SpT indices are calculated from the median flux in the given spectral range.}\\
\multicolumn{9}{l}{$^a$SpT rms calculated from literature SpT (R5150, TiO 6800), Kirkpatrick SpT (TiO 7140), and Luhman SpT (TiO 7700, 8465)}\\
\multicolumn{9}{l}{$^b$The observed flux ratio is divided by the same ratio obtained from a linear fit to the 4650--5300 \AA\ region, see text.}\\
\multicolumn{9}{l}{$^c$0.3 between M1 and M4, 0.8 earlier than M1}\\
\multicolumn{9}{l}{$^d$The 8120--8160 \AA\ continuum range should be used only for spectra that are corrected for telluric H$_2$O absorption.}\\
\end{tabular}
\end{table*}

\subsection{Sample Selection}

At Palomar, we tried to observe all visually bright targets in Taurus with spectral type (SpT) between K0--M4 and that were known as of 2008 
\citep[see review by][]{Kenyon2008}.
A few Taurus objects with spectral types earlier than M4 were missed due to clerical errors.  Many new Taurus members were identified after 2008 and are not included here.  For later spectral types, our sample is far from complete and is biased to the targets that were optically brightest because they had the best chance of having U-band detections.  We also obtained a complete sample of the known objects in MBM 12 \citep{Luhmanmbm12} and some of the TW Hya Association.   In some cases, the membership of the star in the parent cloud is uncertain.  The stars from the HBC with numbers between 352-357 that were observed here are consistent with low gravity but are likely not members of Taurus \citep{Kraus2009}.

During our Keck runs, we observed many brown dwarfs to measure accretion rates at the lowest mass end of the initial mass function.  Our 2006 Keck run was focused on Taurus, while the 2008 run included objects in the Ophiucus, Lupus, and Corona Australus molecular clouds and the Upper Sco OB Association. 

The source list and final properties for the young stars in our sample are listed in Appendix C (see Table~\ref{tab:props.tab}).  
Multiple spectra were obtained for 62 targets, including $>3$ observations on 31 bright and famous targets.
 We also obtained spectra of 40 main sequence K and M-dwarfs with known spectral type \citep{Kirk93}.  The spectra from these stars are used when describing field star spectra in \S 3 but are otherwise not discussed. 

Our sample includes some brown dwarfs.  For simplicity, all objects are referred to as stars regardless of their estimated mass.

\section{ESTABLISHING SPECTRAL TEMPLATES FOR T TAURI STARS}

\begin{figure}
\epsscale{1.}
\plotone{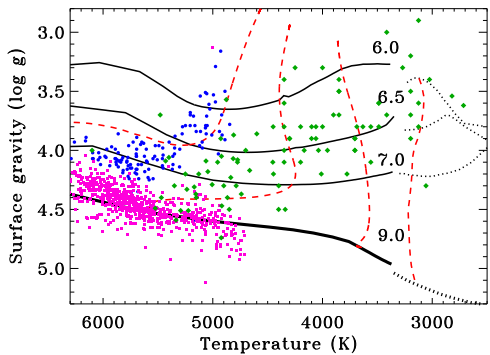}
\caption{Isochrones of gravity versus temperature from \citet{Tognelli2011} for M$_\ast>0.3$ M$_\odot$ (solid horizontal lines) and  \citep{Baraffe2003} for M$_\ast<0.3$ M$_\odot$ (dotted horizontal lines), plotted analgous to an HR diagram.  Pre-main sequence tracks of 2.0, 1.0, 0.5, and 0.2 M$_\odot$ stars are shown from left to right (red dashed lines).  The gravity increases as the star contracts during the pre-main sequence evolution ($\log$ age as labeled).
The gravities of pre-main sequence stars \citep[green diamonds from][]{Stassun2007,Santos2008,Dorazi2009,Dorazi2011,Biazzo2011,Biazzo2012} are similar to the gravities of luminosity class IV subgiants (blue circles, $\log g=3.5-4.5$, Valenti et al.~2005) and less than the main sequence (thick horizontal line at the bottom, main sequence data points as purple squares from Valenti et al.~2005).}
\label{fig:grav1}
\end{figure}

The necessary ingredients for age and accretion calculations are the stellar mass, radius, and accretion luminosity.  These parameters require measurements of the stellar effective temperature, the photospheric flux, and the extinction.  While analysis of optical spectra of main sequence stars from photospheric features is usually straightforward, the lower gravity and presence of accretion complicates the measurement of stellar properties of young stars.  Pre-main sequence stars have similar surface gravity to cool subgiants of luminosity class IV and are offset from the luminosity class V field dwarfs (Fig.~\ref{fig:grav1}).  However, gravity measurements for pre-main sequence stars are challenging because, unlike subgiants, they are fast rotators.

A typical weak lined T Tauri star spectrum is covered with photospheric absorption in molecular bands and atomic lines, along with chromospheric emission in H Balmer and \ion{Ca}{2} H \& K lines.  Accretors usually show strong emission in those lines, along with weak emission in the \ion{Ca}{2} infrared triplet, in \ion{He}{1} lines, in an accretion continuum, and often in forbidden lines.  Some accretors show many additional lines, mostly of Fe \citep[e.g.][]{Hamann1992,Beristain1998}.  The accretion continuum reduces the depth of photospheric absorption lines, a process that is called ``veiling''.  The veiling is defined as $r_\lambda=F_{veil}/F_{phot}$ at a given wavelength $\lambda$.  The veiling at 5700 \AA, $r_{5700}$, is typically between 0.1--1, though in rare cases the veil may cover the photospheric emission \citep{Hartigan1995,Fischer2011}.
The flux in the photospheric and emission lines are often reduced by extinction.

 In this section, we describe our initial approach for measuring the properties of the stars in our sample, with an emphasis on quantifying the approach for measuring SpT, $A_V$, and the accretion continuum flux.  The analysis in this section results in a grid of extinction-corrected spectral templates and an approach for including the accretion in spectral type and extinction measurements, which are then applied to the full dataset in \S 4.

\begin{figure}
\epsscale{1.}
\plotone{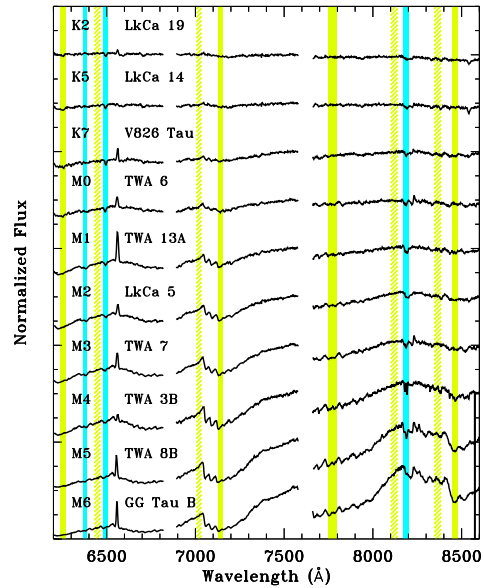}
\caption{Red spectral sequence from K3--M5.5.  Regions used for TiO band indices are highlighted in yellow.  Selected gravity-sensitive absorption in CaH $\lambda6382$, \ion{Fe}{1} $\lambda6497$, and the \ion{Na}{1} $\lambda8189$ doublet are highlighted in blue.}
\label{fig:tiobands}
\end{figure}

\subsection{Quantification of Spectral Indices}

In this subsection, atlases of low resolution optical spectra are used to establish a set of quantified spectral indices for young stars.  The following descriptions are divided by spectral type, each of which is sensitive to a different spectral index.  Spectral typing of young stars has typically relied on eyeball comparisons to a sequence of spectral standards.  While that approach can be very accurate, a quantified approach allows for greater consistency between different sets of eyes.  A quantified approach also readily accounts for accretion and extinction by calculating over a grid of values to find a best fit solution.

The full set of spectral indices discussed in this paper is listed in Table~\ref{tab:indices.tab}.  By design, our focus is on K and M stars.  The M-dwarf spectral types rely on  the depth of TiO and VO absorption bands (hereafter referred to as TiO), which start to become detectable at $\sim$ K5.  For K-dwarfs, a spectral type index is developed based on the 5200 \AA\ absorption feature, which is a combination of MgH, Mg b, and Fe I \citep[e.g.][]{Rich1988}.  The spectral typing of BAFG stars relies on a visual comparison of the G-band and absorption in H and Ca lines.

\begin{figure*}
\epsscale{1.}
\plottwo{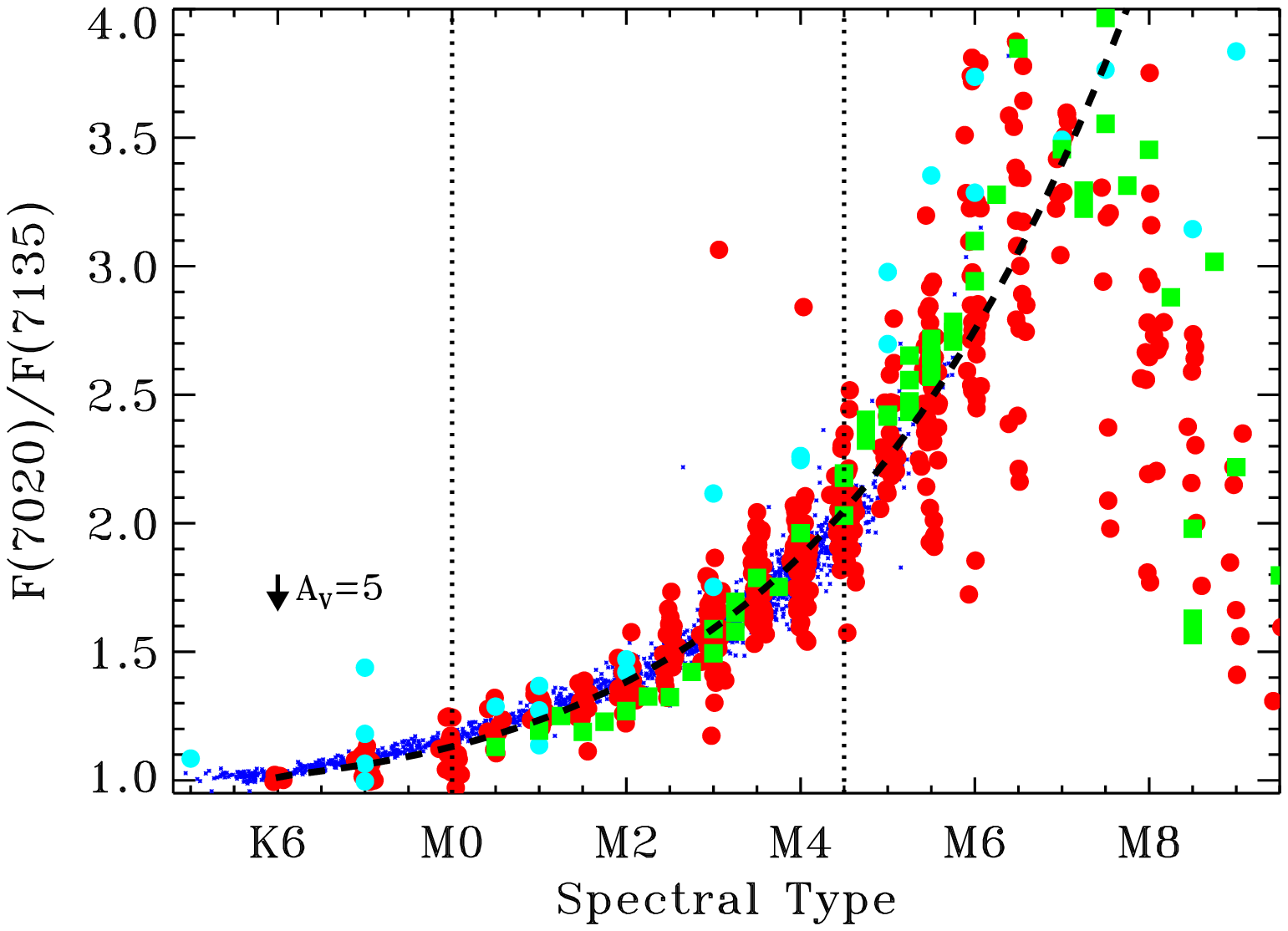}{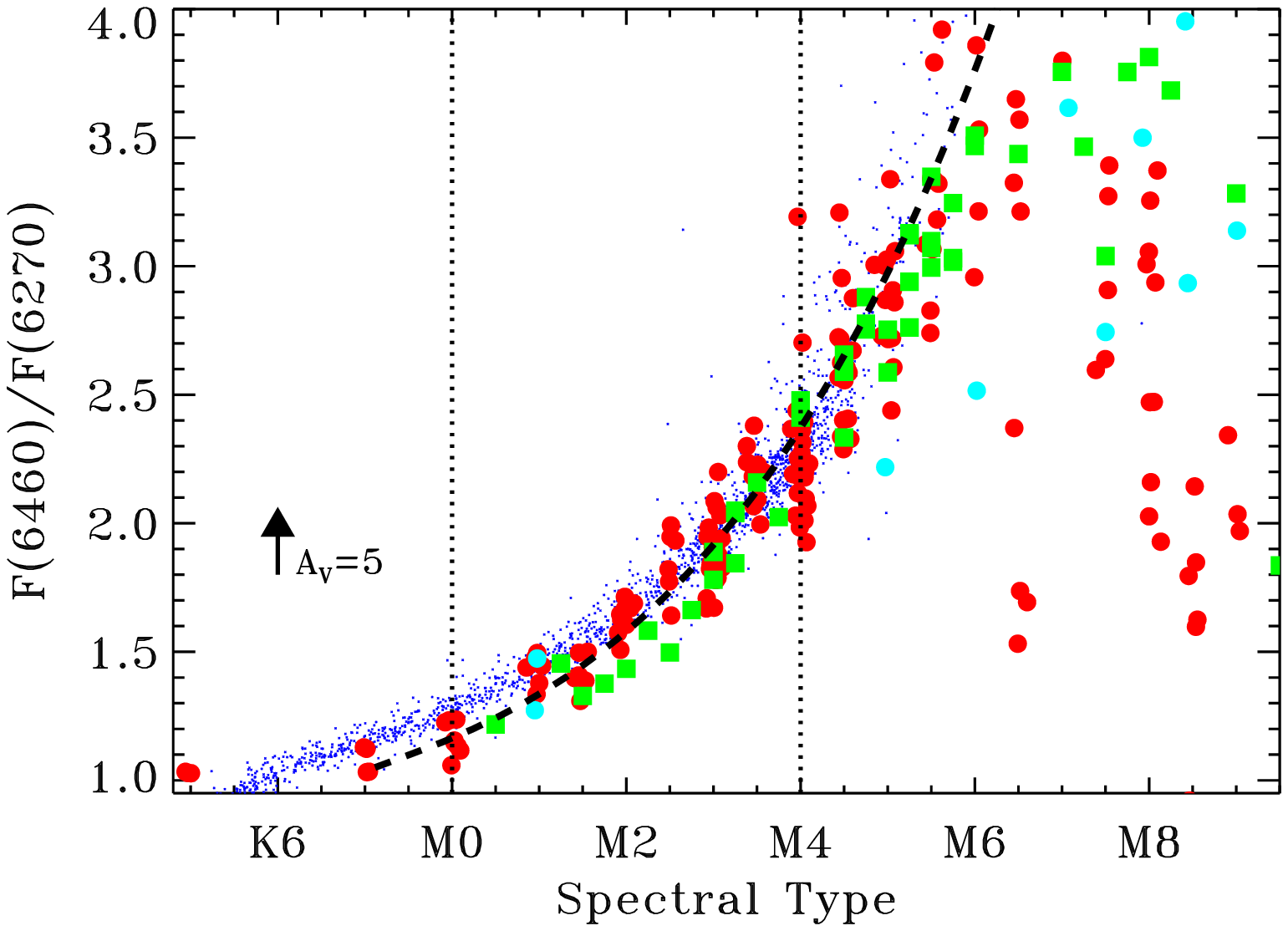}
\plottwo{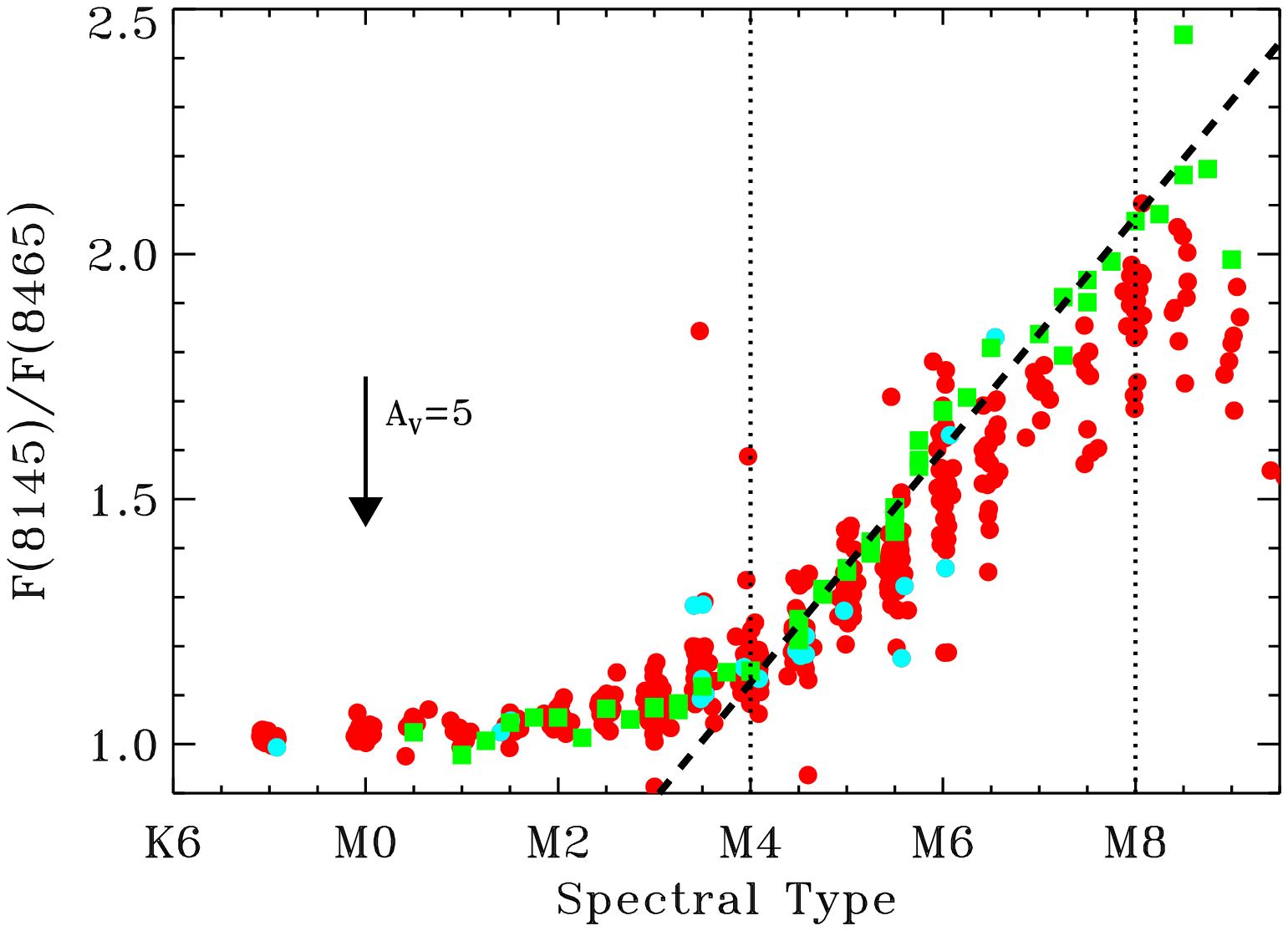}{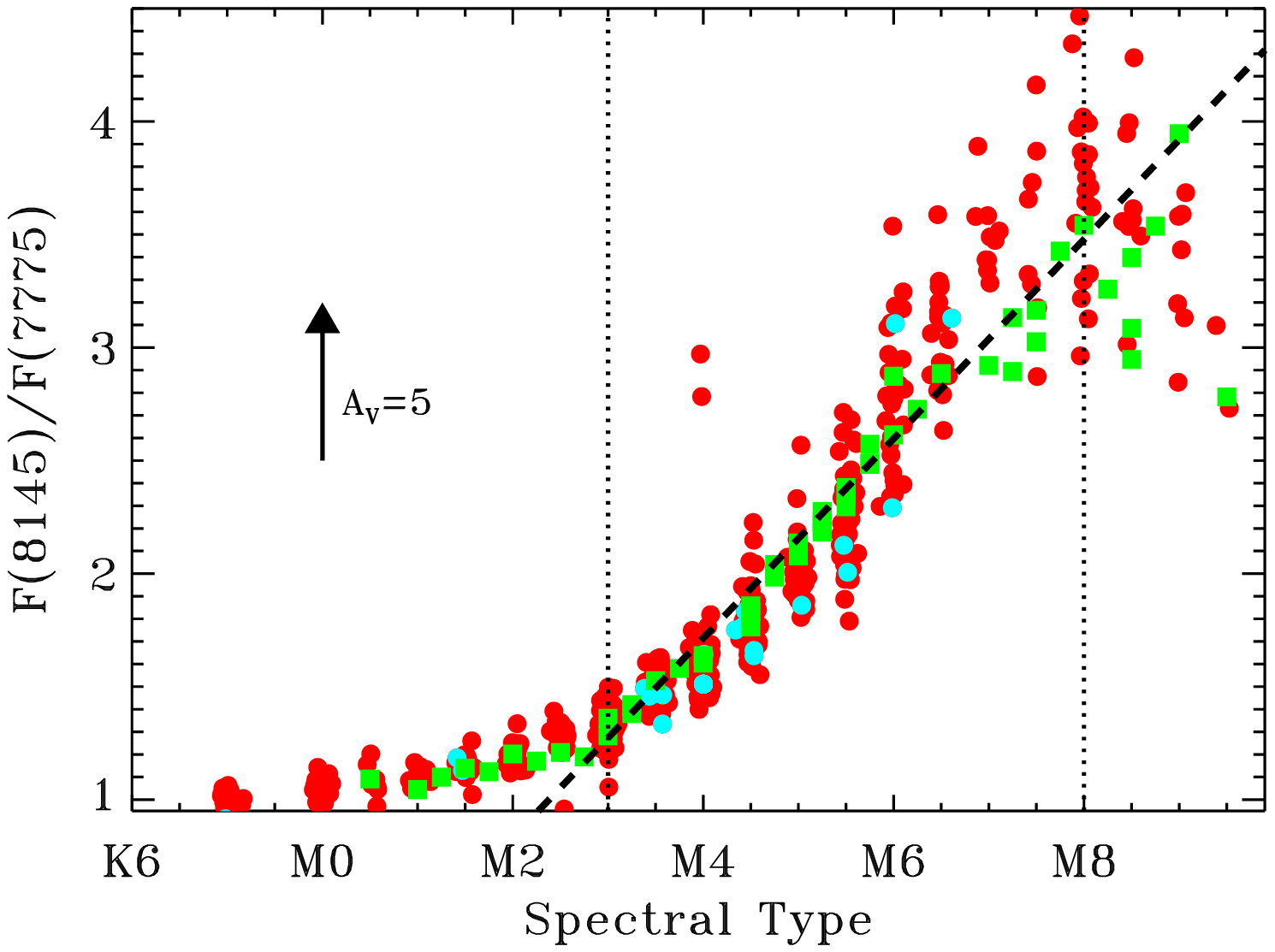}
\caption{Relationships between spectral index and spectral type for four TiO bands.  Large circles are calculated from Kirkpatrick (red are main sequence and cyan are giant stars), small blue dots are from PMSU, and green points are from spectra provided by Luhman.  Best fit conversions between spectral index and spectral type are shown as the dashed line and quantified in Table~\ref{tab:indices.tab}.  The vertical dotted lines show the spectral type range where these relationships are used.  The arrows show how the listed extinction $A_V=5$ mag.~would shift the index.}
\label{fig:sptanal}
\end{figure*}

The quantification of spectral typing provides an objective and repeatable method to measure spectral types with precision.
The quantified prescriptions of M-dwarfs are similar to those of \citet{Slesnick2006} and \citet{Riddick2007}, while the prescriptions for earlier spectral types are similar to those developed by, e.g., \citet{Worthey1994} and \citet{Covey2007}.  The spectral indices described here are tailored to low spectral resolution.  These spectral indices are then combined with an accurate flux calibration and blue spectra to measure accretion and extinction simultaneously (see \S 4).  In several cases, the spectral index is changed to a $\log$ scale to provide a better fit between spectral type and spectral index.
The TiO-7700 spectral index defined here uses a continuum region that overlaps with telluric H$_2$O absorption and should only be used when telluric calibrators are  obtained contemporaneously.  
Use of indices can also be problematic if the spectrum is either not flux calibrated or not corrected for extinction.  Converting spectral indices to accurate spectral types requires high S/N in the $\sim 30$ \AA\ integration bins and an accruate relative flux calibration (for example, see Table 2 for our flux calibration relevant to the TiO-7140 index).  A 2\% error in the TiO indices typically leads to an error of 0.1-0.2 subclasses in spectral type.

Scatter in these quantified relationships are caused by metallicity and gravity differences between stars.  The metallicity of nearby young associations is uniform \citep[e.g.][]{Padgett1996,Santos2008,Dorazi2011}.  Gravity differences between 1--10 Myr may be significant and are discussed but are not fully investigated.

In the following subsections, we describe how these spectral indices are used to measure spectral types.  Each spectral index is sensitive to different spectral types and is discussed separately, beginning with the coolest stars in our sample.

\begin{figure*}
\plottwo{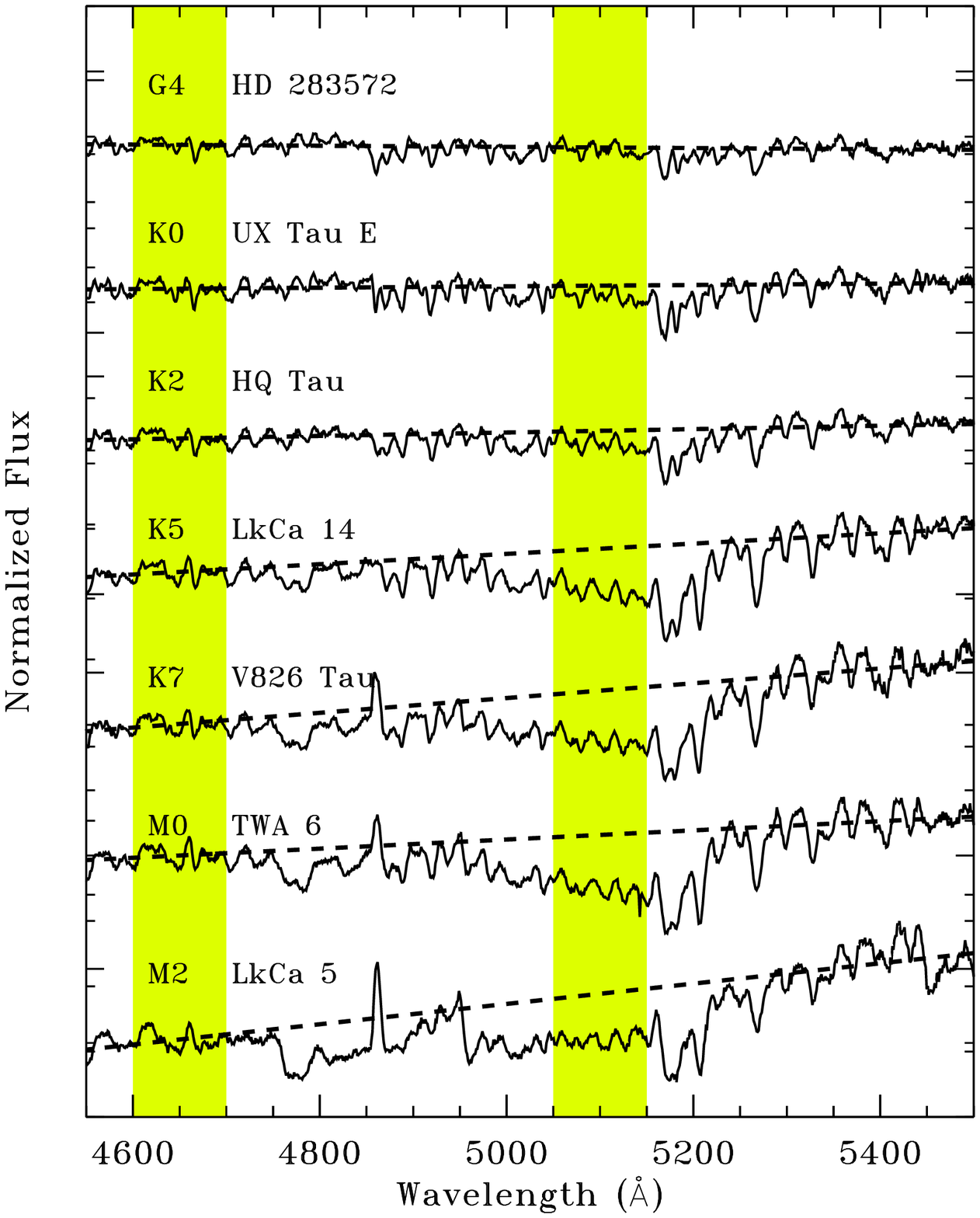}{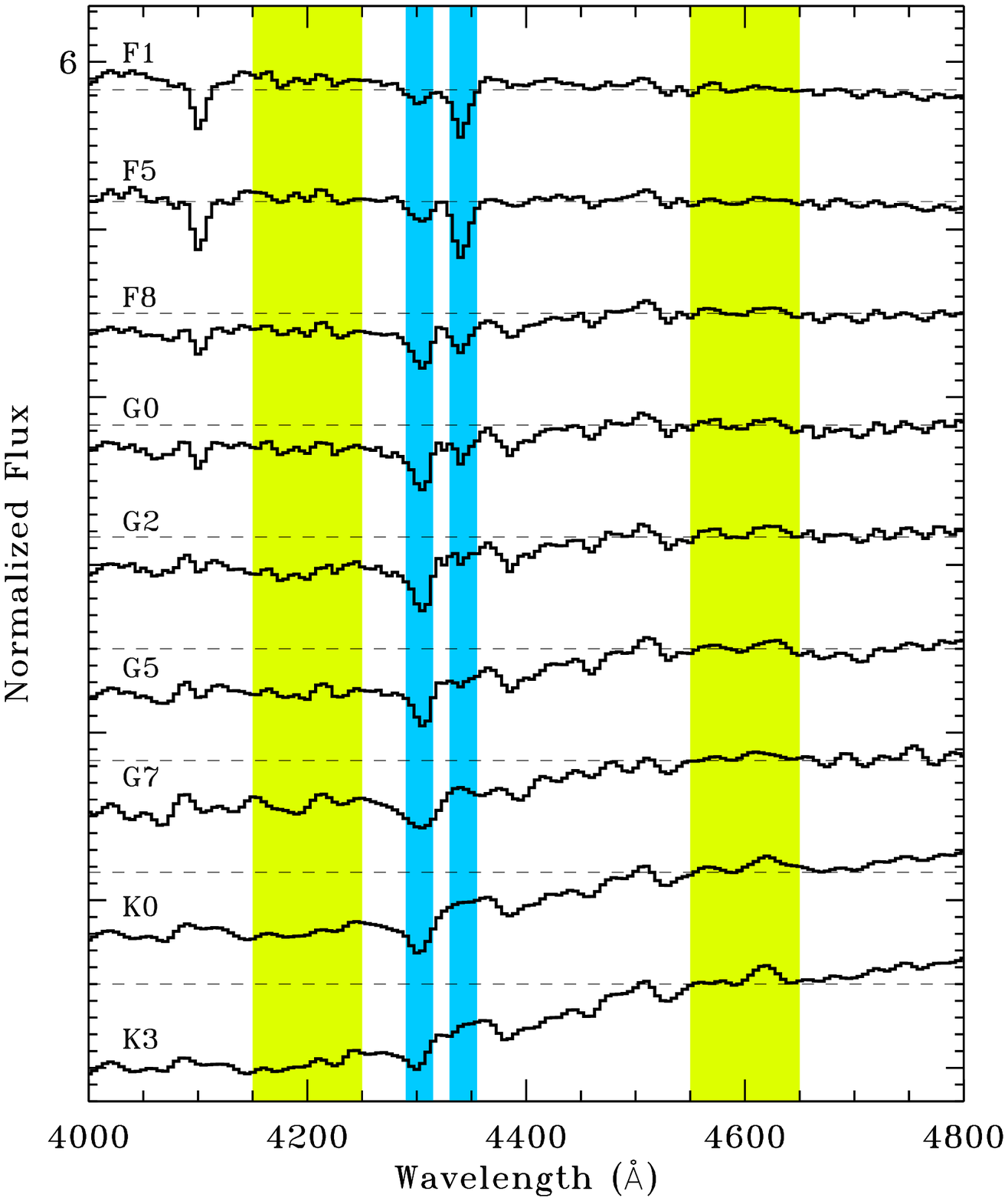}
\caption{The blue spectral sequence used to measure SpT for K stars (left, from this work) and FG stars (right, luminosity class IV stars from Pickles).  K-dwarf spectra show a dip in flux just shortward of the \ion{Mg}{1} $\lambda5150$ line, which gets stronger to later K and to higher gravity.  The right panel shows how the H$\gamma$ and the \ion{Ca}{1} $\lambda4227$ lines (shaded blue regions) vary with SpT from F1 to K3.  The shaded yellow regions show the ranges used to calculate spectral indices for G and K dwarfs.}
\label{fig:kdwarf}
\end{figure*}

\subsubsection{Spectral Types of M stars}
The majority of stars in our sample are M stars.  At optical wavelengths, M stars are easily identified from the presence of strong TiO absorption bands.  \citet{Kirk91} and \citet{Kirk93}, hereafter Kirkpatrick, established a grid of M-dwarf spectral type standards from field stars.  \citet{Reid95}, hereafter PMSU, quantified relationships between spectral type and the depth of TiO  bands at 7100 \AA\ from moderate resolution optical spectra based on the \citet{Kirk91} sequence.   

\citet{Luhman1999} and \citet{Luhman2003}, hereafter Luhman (also includes, e.g., Luhman et al. 2004, 2006), recognized that for pre-main sequence stars, the depth of TiO features deviates from dwarf stars because of lower gravity (see also Gullbring et al.~1998).  Luhman developed a spectral type sequence for young M-dwarfs later than M5 based on a hybrid of field dwarf and giant stars, since TTSs are typically luminosity class IV.  For stars earlier than M5, Luhman relied on the Kirkpatrick grid along with the \citet{Allen1995} red spectroscopic survey of standards. 
Although the Luhman spectral sequence is well accepted and widely used, it has no standards or quantified conversions between spectral index and spectral type.  As a consequence, spectral types based on the Luhman method are likely less precise when applied by authors other than Luhman himself.

Our quantified spectral type sequence is derived from the methods established in those seminal works.
In the following analysis, the objects in the PMSU catalog are all assigned a spectral type based on their TiO5-SpT conversion, which is accurate to $\sim 0.5$ subclasses between K7--M6.  The TiO5 spectral index, the flux ratio of 7130--7135 to 7115-7120 \AA, requires flux measurements in narrow regions and is not possible to calculate from our low resolution spectra.  The Luhman sequence discussed here is from a set of 54 young stars spanning M0.5--M9.5 provided by Luhman (private communication).

Four prominent TiO bands are present in our red spectra (see Fig.~\ref{fig:tiobands} and Table~\ref{tab:indices.tab}).  Figure \ref{fig:sptanal} compares the spectral types and four spectral indices for the PMSU, Kirkpatrick, and Luhman samples.  
For stars earlier than M5, the Luhman relationship between SpT and TiO depth for young stars was intended to follow the \citet{Kirk91} results.  
However, for objects from M0 to M3, the Luhman spectral types are $\sim 0.5$ subclasses later than the median Kirkpatrick spectral type (TiO-7140 and TiO-6200 spectral indices).  For the spectral types later than M5, gravity differences between field M-dwarfs dwarfs and pre-main sequence M-dwarfs lead to the Luhman spectral types being slightly earlier than the median Kirpatrick object (as discussed by Luhman).  
For M-dwarfs earlier than M4, we adopt the spectral type sequence of Kirkpatrick, which may introduce a small offset between our spectral types and Luhman spectral types.  For M-dwarfs later than M4, we adopt the spectral type sequence of Luhman.  Several additional TiO/VO bands are detected at blue wavelengths and are not well studied (see Fig.~\ref{fig:kdwarf}).  While our initial approach does not consider these bands, the final spectral types are calculated from a best fit to a spectral sequence using the full optical spectrum.
\footnotetext[1]{The TiO 7140 index was developed by \citet{Slesnick2006}.  Our definition uses a slightly different continuum region.}

{\it M4--M8:}  Objects later than M4 have spectral types assessed from the TiO 7700 and 8500 \AA\ bands, with a conversion from spectral index to spectral type calculated from the sequence of objects provided by Luhman.
 An uncertainty of 0.2 subclasses is assigned based on the change in feature strength verus subclass and on the standard deviation in the fits to the Luhman objects.  This uncertainty is consistent with that assigned by Luhman.

\begin{figure}
\plotone{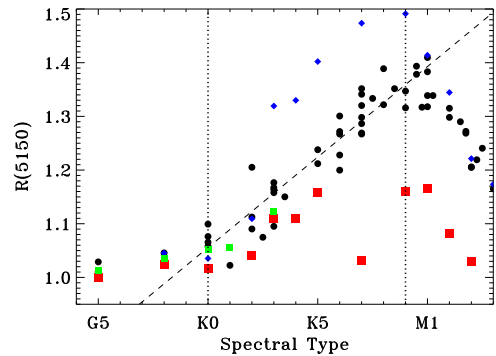}
\caption{Spectral type versus the spectral index $R5150$ for young K-dwarfs (see Fig.~\ref{fig:kdwarf}) for young stars in our sample (black circles), Pickles templates of luminosity class IV (green squares), V (blue diamonds), and the average of Pickles templates of luminosity class III and V (red squares).   The vertical dotted lines show the spectral type range where these relationships are used.  The SpT used in this analysis are all from the literature and may differ from the SpT calculated in this work.  The turnover around M1 occurs when TiO absorption becomes prominent enough to affect the flux ratio.}
\label{fig:kdwarf_rats}
\end{figure}

{\it M0--M4:}  The TiO band at 7140 \AA\ is most reliable for early-to-mid M-dwarfs. 
 Within the Kirkpatrick sample between M0--M4.5, the standard deviation of the index-determined SpT and adopted SpT is 0.4 subclasses.  The relative accuracy of spectral typing within a single star-forming region is likely better than 0.40 subclasses because the Kirkpatrick lists SpT at only 0.5 subclass intervals and because
the metallicity should be uniform in samples of nearby star forming regions but not in field dwarfs.
The TiO 6200 \AA\ band is also sensitive to early-to-mid M-dwarfs, 
with a standard deviation of  0.22 subclasses for spectral types  M0-M4 within the Kirkpatrick sample.  However, few Kirkpatrick objects were observed at 6200 \AA, and the PMSU sample is systematically offset from the Kirkpatrick sample in this TiO feature.  As a consequence, we do not use this relationship here to derive spectral types.

\begin{figure*}[!th]
\epsscale{1.}
\plottwo{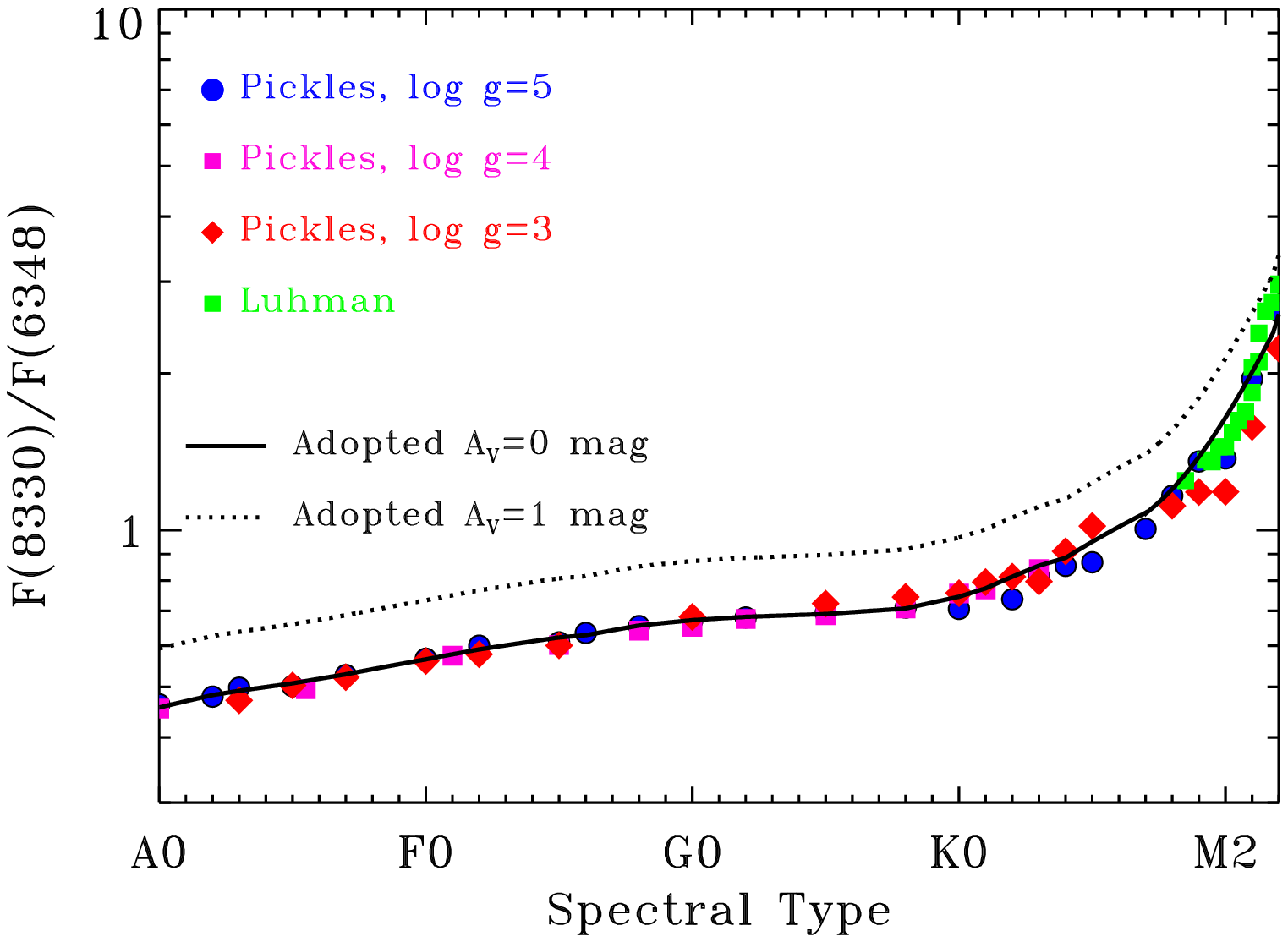}{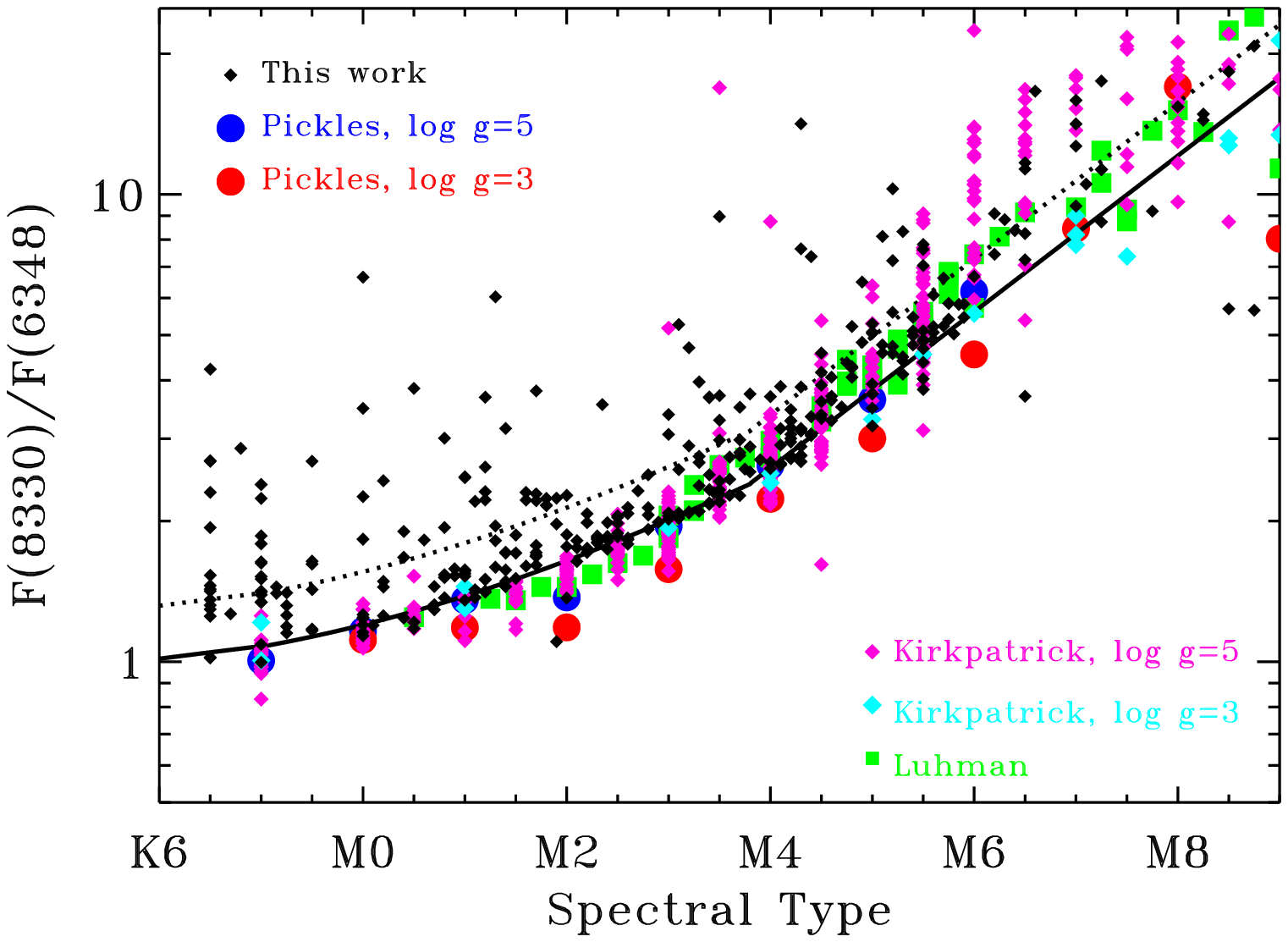}
\caption{The flux ratio $F_{red}=F(8330)/F(6448)$ versus SpT for A-M stars (left) and focused in on K-M stars (right), for SpT calculated from the spectral indices in described in \S 3.  Extinction can then be measured by comparing $F_{red}$ to the expected ratio for a given spectral type.}
\label{fig:extinct1}
\end{figure*}

\subsubsection{K0-M0.5 Spectral Types}

Figure~\ref{fig:kdwarf} (left panel) shows that K-dwarfs are characterized by MgH and Mg b absorption at $\sim 5150$ \AA.
This dip is not present in G-type stars.  We define a spectral index, $R(5150)$,
\begin{equation}
R(5150)=\frac{F(5100)}{F(4650)}\times\frac{F_{\rm line}(4650)}{F_{\rm line}(5100)},
\end{equation}
where F($\lambda$) is the flux in a 100 \AA-wide band around $\lambda$, and $F_{rm line}(4650)/F_{\rm line}(5100)$ is the flux ratio expected at those same wavelengths based on the spectral slope obtained in a linear fit to the $\lambda=4650$ and $\lambda=5450$ \AA\ spectral regions.  Dividing by the F(5100)/F(4650) ratio calculated from the linear fit accounts for extinction.

Figure~\ref{fig:kdwarf_rats} shows the relationship between R$(5150)$ versus literature spectral type for stars with little or no accretion.  The spectral types earlier than M0 are obtained from the literature, usually from high-resolution spectra \citep{Basri1990,White2004,White2007}, and are supplemented by some low resolution spectral types from Luhman.  Spectral types later than M0 are calculated from the TiO spectral indices. Figure~\ref{fig:kdwarf_rats} also shows $R(5150)$ versus SpT from the compilation of low resolution spectral atlases by \citet{Pickles98}.  The R$(5150)$ index is similar to that of luminosity class IV subgiants and to the average index obtain by adding spectra of dwarfs and giants (luminosity class III + luminosity class V).
 The relationship is gravity-sensitive and should be applied only to pre-main sequence K-stars.

The standard deviation between calculated and literature spectral types between K0 and M0 is 1.0 subclasses.  Some of this scatter is attributable to uncertainty in literature spectral type, which typically claim an accuracy of 1--2 subclasses, and to studies listing integer steps in subclass.
We assign an uncertainty of 1 subclass between K0--M0 for this relationship.

The TiO 6800 \AA\ absorption band is detectable for spectral types K5 and later.  
From K5-M0, the Kirkpatrick objects are about 1 subclass later than the Pickles libraries.   The PMSU data are also shown, though the PMSU TiO-5 index is not reliable at spectral types earlier than K7.  We include a spectral type K8 as an intermediate between K7 and M0.  
Spectral types between between K6--M0.5 are assigned an uncertainty of $\sim 0.5$ subclasses.  Most accreting stars in this spectral type range have a spectral type uncertainty of 1 subclass.   Within this range there may be an additional systematic uncertainty of $\sim 0.5$ subclasses between our spectral types and those of Luhman.

\subsubsection{B, A, F, and G Spectral Types}

By design, only a few objects in our sample have a spectral type earlier than K.  Spectral types for these objects are measured by visual comparison to Pickles templates.  The shape of the G-band helps to determine G spectral types, while the absence of the G-band requires that the star be F or later \citep[e.g.][]{Fraunhofer1814,Cannon1912,Covey2007}.  Both G and F spectral types are also measured from from the relative strengths of the 4300 \AA\ line and the nearby H$\gamma$ line.  Hotter stars have spectral types measured from the strength of Balmer lines and the \ion{Ca}{2} H \& K lines.   The strength of the \ion{Ca}{2} K line is particularly important for discriminating between B and early A spectral types \citep[e.g.][]{Mooley2013}, although the absorption may be filled in with emission.
More rigorous approaches to spectral typing large samples of BAF stars are described by \citet{Hernandez2004} and \citet{Alecian2013}.  Some uncertainty in our classification is introduced by emission and possible wind absorption in H and Ca lines.  

\subsection{Photospheric Extinction Measurements}

Extinction measurements require a comparison of observed flux ratios or spectral slopes to the same flux ratios or slopes from a star with the same underlying spectrum and a known extinction.  For non-accreting stars, this flux ratio can be compared to a photospheric template with similar gravity and negligible extinction.  The effect of accretion on photospheric extinction measurements is discussed in \S 3.4.

The extinction curve used in this paper is from \citep{Cardelli1989} with the average interstellar value for total-to-selective extinction, $R_V=3.1$.  The value for $R_V$ increases to 5.5 for larger dust grains found deep in molecular clouds when $A_V\sim20$, far larger than any extinction measured in this optical sample \citep{Indebetouw2005,Chapman2009}.  To keep the amount of analysis reasonable and for consistency, $R_V$ is assumed to be constant throughout our sample when possible.  A few stars could only be fit with higher $R_V$ (see Appendix C).

Initial extinctions calculated in this paper and applied to a spectral template grid are based on the flux ratio $F_{red}=\frac{F(8330)}{F(6448)}$ (flux at 8330 \AA\ to that at 6448 \AA), although our final extinctions use the full blue-red spectra (see \S 4).   The ratio $F_{red}$ is affected by the photospheric temperature, accretion spectrum, and extinction.  These wavelengths are selected to avoid telluric and TiO absorption bands and to maximize the wavelength difference of the two bands while requiring both to be in the red detector.

Figure~\ref{fig:extinct1} shows $F_{red}$ versus spectral type for the full range of SpT (left) and for late-K and M-dwarfs (right).  The curve of $F_{red}$ versus SpT for $A_V=0$ for stars earlier than M0 is based on the Pickles spectral atlas, with giants and dwarf having similar values.  
 Objects provided by Luhman are also included to help fill the grid for stars with SpT later than M5.  The value of $F_{red}$ diverges between the young star and the field dwarf sample at SpT later than M4, which confirms the approach of Luhman to calculate a new SpT-effective temperature conversion for young stars.
Within this range, a 0.25 uncertainty in SpT subclass leads to a 0.15 mag uncertainty in $A_V$.

At spectral types earlier than K5, we lack the necessary coverage in spectral types of unreddened stars to establish a reliable baseline in $F_{red}$ versus spectral type to calulate extinctions.  Instead, we interpolate $F_{red}$ over the spectral type grid from the flux-calibrated Pickles compilation of stars with luminosity class V.  Most objects in the Pickles compilation have fluxes accurate to $\sim$1\%.   For K-dwarfs, $F_{red}$ is about 5\% larger for objects of luminosity class III and IV relative to V.  We therefore multiply the interpolated curve by 3\%, intermediate between luminosity classes III and V and assess a 3\% uncertainty in the flux baseline.  This uncertainty introduces a 0.12 mag uncertainty in $A_V$ measurements.    For F and G-dwarfs, $F_{red}$ is not very sensitive to changes in SpT, with an average change of 2\% per subclass so that a 1-subclass SpT uncertainty leads to a 0.07 mag.~uncertainty in $A_V$.

\begin{figure}
\plotone{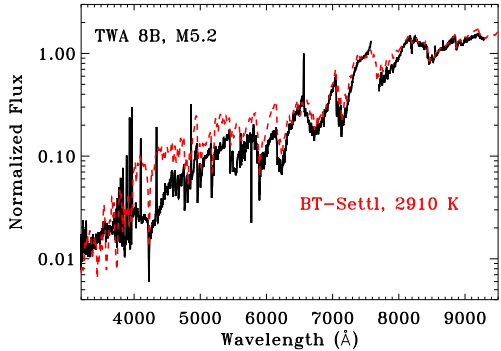}
\caption{The spectrum of the M5.2 WTTS TWA 8B, compared with BT Settl models of the best-fit temperature.  The spectra are scaled to unity at 7325 \AA.  The synthetic spectrum significantly overproduces emission between 4000--6000 \AA.  The TiO bands are deeper in the observed spectrum than in the synthetic spectrum.}
\label{fig:twa8b.ps}
\end{figure}

\subsection{A Grid of Pre-Main Sequence Spectral Types}
Based on the previous descriptions, a grid of photospheric spectral templates are established and listed in \S~\ref{tab:gridstars}.  Templates at
spectral types earlier than K0 are obtained from the Pickles library because of very sparse coverage in our own data.  At K0 and later, weak lined T Tauri stars with low extinctions are selected from our spectra for use as photospheric templates.  This criterion leads to the selection of many TWA objects for our grid.  The conversion from the spectra to temperature and luminosity are described in the following two subsubsections.  

This set of stars is then combined into a grid.  Two separate spectral sequences are calculated from stars at $\sim 1$ subclass intervals.  Between K5--M6, the grids comprise of every second star in Table~\ref{tab:gridstars} and are therefore independent.  The two grids are then linearly interpolated at 0.1 suclasses (earlier than M0) and 0.05 subclasses (later than M0) and are averaged to create a final 
spectral grid.  
The photospheric template at all classes between K6--M5.5 therefore includes the combination of 3--4 stars.  This method minimizes the problems introduced by any single incorrect spectral type or extinction within this spectral sequence.

\begin{figure}
\plotone{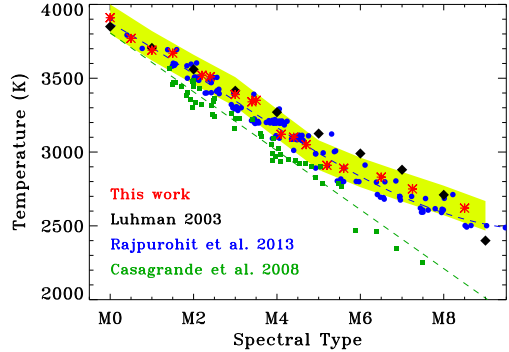}
\caption{The conversion from spectral type to effective temperautre from Luhman et al.~2003 (black diamonds), Rajpurohit et al.~2013 (blue circles, with best fit polynomial shown as dashed blue line), Casagrande et al.~2008 (green squares, with best fit line shown as dashed green line), and this work (red asterisks).  Small spectral type and temperature changes are randomly applied to the Rajpurohit et al.~data so that each point is visually displayed.  The shaded yellow region shows the approximate temperature range at constant spectral type derived from atmosphere models, accounting for uncertainties in the comparison between model and observed spectra.}
\label{fig:sptteffconv.ps}
\end{figure}

Unresolved binarity affects photospheric measurements of both our spectral grid and our target stars.  
Among the known multiple systems in our grid, V826 Tau is a near-equal mass spectroscopic binary, so the combined optical spectrum would have a very similar spectrum as both components.  LkCa 5 has a very low-mass companion \citep{Kraus2011} that contributes a negligible amount of flux at optical wavelengths.  Although LkCa 3 is a quadruple system consisting of two spectroscopic binaries \citep{Torres2013}, the global spectral type and extinction is reasonable compared to other stars of similar SpT.  In the spectral fits described in \S 4, the combined use of multiple templates for any given star should minimize the problems introduced by known and unknown binarity in the templates.

\begin{table}
\caption{Derived Parameters for grid of weak-lined T Tauri Stars}
\label{tab:gridstars}
\begin{tabular}{lccc}
 
Star         & SpT$^a$ & $A_V$ (mag)$^a$ & Teff (K)\\
\hline
HBC 407 & K0 & 0.80 & 5110 \\
HBC 372 & K2 & 0.63 & 4710 \\
LkCa 14 & K5  & 0.00 & 4220\\
MBM12 1 & K5.5 & 0.00 & 4190\\
TWA 9A   & K6.5 & 0.00 & 4160 \\
V826 Tau & K7 & 0.38 & 4020\\
V830 Tau & K7.5 & 0.40 & 3930\\
TWA 6   & M0 & 0.00 & 3950\\
TWA 25 & M0.5 & 0.00 & 3770\\
TWA 13A & M1.0 & 0.00 & 3690\\
LkCa 4 & M1.5 & 0.00  & 3670\\
LkCa 5 & M2.2 & 0.27 & 3520\\
LkCa 3 & M2.4 & 0.00 & 3510\\
TWA 8A & M3.0 & 0.00 & 3390\\ 
TWA 9B & M3.4 & 0.00 & 3340 \\
J1207-3247 & M3.5 & 0.00 & 3350\\
TWA 3B & M4.1 & 0.00 & 3120 \\
 XEST 16-045  & M4.4 & 0.00 & 3100\\
J2 157 & M4.7 & 0.41 & 3050\\
TWA 8B & M5.2 & 0.00 & 2910 \\
MBM12 7 & M5.6 & 0.00 &  2890\\
V410 X-ray 3 & M6.5 & 0.25 & 2830\\
Oph 1622-2405A & M7.25 & 0.00 & 2750\\
2M 1102-3431 & M8.5 & 0.00 & 2590\\
\hline
\multicolumn{4}{l}{$^a$From red spectrum, may differ from final SpT, $A_V$.}
\end{tabular}
\end{table}

\subsubsection{Conversion from Spectral Type to Effective Temperature}

The standard conversion from SpT to effective temperature for young stars is based on the work of \citet{Schmidt1982} and \citet{Straizys1992}, as compiled by \citet{Kenyon1995}.  Luhman updated this conversion for M-dwarf T Tauri stars, based on a scale intermediate between giants and dwarfs.   Synthetic M dwarf spectra from model atmospheres have advanced considerably since \citet{Luhman2003} established this conversion.  \citet{Rajpurohit2013} recently obtained a new scaling between spectral type and temperature for M dwarfs by comparing BT-Settl synthetic spectra calculated from the Phoenix code \citep[e.g.][]{Allard1995,Allard2012} to observed low-resolution spectra.
A similar approach by \citet{Casagrande2008} with the Cond-GAIA synthetic spectra yielded much lower temperatures than \citet{Rajpurohit2013} for the same spectral type.

An initial comparison between our standard grid and Phoenix/BT-Settl synthetic spectra with CFITSIO opacities and gravity $\log g=4.0$ \citep{Allard2012,Rajpurohit2013} reveals good agreement between the observed and synthetic spectra for temperatures higher than $\sim 3200$ K.  Discrepancies in between the observed and modeled depths of TiO absorption bands are problematic at cooler temperatures (Fig.~\ref{fig:twa8b.ps}).  
We speculate that some of these differences may be explained with uncertainties in the strengths of TiO transitions and 
 in the strength of continuous optical emission produced by warm dust grains in the stellar atmosphere.  
Details of these comparisons and fits of the synthetic spectra to observed spectra are described in Appendix B.

An effective temperature scale for pre-main sequence stars is derived by fitting Phoenix/BT-Settl synthetic spectra to 
 our spectral type grid (K5-M8.5) and Pickles luminosity class IV stars (F-K3).  Figure~\ref{fig:sptteffconv.ps} and Table~\ref{tab:newtemps} compares our new K and M-dwarf temperature scale to other pre-main sequence and dwarf temperature scales$^2$.  
Our scale matches the Luhman scale between M0-M4 and deviates at later spectral types.  The differences between our scale and the \citet{Rajpurohit2013} scale are likely attributed to gravity differences between pre-main sequence and dwarf stars.   The K-dwarf temperature scale is shifted to lower temperatures relative to the scale used by \citet{Kenyon1995}.  
\footnotetext[2]{The scales for \citet{Rajpurohit2013}, \citet{Casagrande2008}, and our work were calculated by using best fit polynomials to the data points of spectral type versus effective temperature.  For \citet{Casagrande2008}, the data were obtained from tables of \citet{Rajpurohit2013}.}

\begin{table}
\caption{Spectral Type to Temperature Conversions}
\label{tab:newtemps}
\begin{tabular}{lccccccc}
\hline
SpT & CK79$^a$  & B$^a$  & KH95$^a$  & C08$^a$  & R13$^a$  & L03& Here\\
\hline
F5 & -- &  -- & 6440 &--&--& --& 6600\\
F8 & -- & -- & 6200 &--&--& --& 6130\\
G0 & 5902 & 6000 & 6030 &--&--& --& 5930 \\
G2 & 5768 & -- & 5860 &--&--& --& 5690\\
G5 & -- &  5580 &  5770 &--&--& --& 5430\\
G8 & 5445 & -- & 5520 &--&--& --& 5180\\
K0 & 5236 & --  & 5250 &--&--& --&  4870\\
K2 & 4954 & 5000 & 4900 & -- & -- & -- & 4710\\
K5 & 4395 & 4334 & 4350 &--&--& --& 4210\\
K7 &3999 &  4000 & 4060 &--&--& --& 4020\\
M0 & 3917 & 3800 & 3850 &  -- & 3975 &  --   & 3900\\
M1 &3681 & 3650 & 3720 & 3608 & 3707  & 3705 & 3720\\
M2 & 3499 & 3500 & 3580 & 3408 & 3529  & 3560 & 3560\\
M3  & 3357 & 3350 & 3470 & 3208 & 3346 & 3415 & 3410\\
M4  & 3228 &3150 &  3370 & 3009 & 3166 & 3270 & 3190\\
M5  & 3119 & 3000 & 3240  & 2809  &2993 & 3125 & 2980\\
M6  & -- & -- & 3050 & 2609 & 2834  & 2990 & 2860\\
M7  & -- & -- & -- & 2410 & 2697  & 2880 & 2770\\
M8  & -- & -- & -- & 2210 & 2588  & 2710 & 2670\\
M9  & -- & -- & -- & --    & 2511  & 2400 & 2570 \\
\hline
\multicolumn{8}{l}{$^a$Conversions developed for field dwarfs}\\
\multicolumn{8}{l}{CK:  \citet{Cohen1979}}\\
\multicolumn{8}{l}{B:\citet{Bessell1979} and \citet{Bessell1991}}\\
\multicolumn{8}{l}{KH: Adopted by \citet{Kenyon1995}}\\
\multicolumn{8}{l}{~~~from Schmidt-Kaler (1982) and Straizys (1992)}\\
\multicolumn{8}{l}{C08:  \citet{Casagrande2008}}\\
\multicolumn{8}{l}{R13:  \citet{Rajpurohit2013}}\\
\multicolumn{8}{l}{L03:  \citet{Luhman2003}}\\
\end{tabular}
\end{table}

\subsubsection{Photospheric Luminosities}

\begin{figure}
\plotone{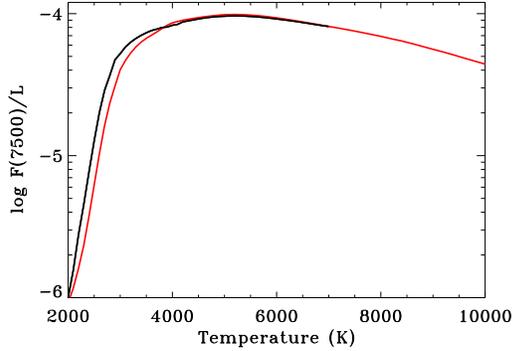}
\caption{The bolometric correction applied to our dataset, calculated by comparing the flux at 7510 \AA\ to the total luminosity in the BT-Settl stellar model (black, adopted for $T<7000$ K) and NextGen stellar model (red, adopted for $T>7000$ K).  The bolometric correction is modified from the models to account for the weakness of the VO 7510 \AA\ absorption band, relative to model predictions.}
\label{fig:bolcor.ps}
\end{figure}

Stellar photospheric luminosities, $L_{phot}$, are calculated using the Phoenix/BT-Settl models with CFITSIO opacities \citep{Allard2012} for effective temperatures $<7000$ K.  Similar bolometric corrections for hotter stars are calculated from the NextGen model spectra \citep{Hauschildt1999}.  The flux ratio $F_{7510}/L_{phot}$ is used as the bolometric correction and applied to the measured photospheric fluxes (Figure~\ref{fig:bolcor.ps} and Table~\ref{tab:bolcor}).  These conversions all assume a gravity of $\log g=4$.  Differences due to subtracting the a flat continuum in late M-dwarfs (\S 3.2.1) amount to $\sim 1$\% of the total stellar luminosity and are ignored here.

At temperatures $<3500$, the opacity in a VO absorption band at 7500 \AA\ is much larger in synthetic spectra than in the observed spectra.  At these temperatures, the bolometric correction for 7510 \AA\ flux is calculated by fitting a line between the flux at 7300 and 7580 \AA, omitting the VO opacity from the fit.

\begin{table}[!t]
\caption{Bolometric Corrections from $F_{7510}$}
\label{tab:bolcor}
\begin{tabular}{cc|cc|cc}
\hline
T$_{phot}$ & $F_{7510}/F_{phot}$ &T$_{phot}$ & $F_{7510}/F_{phot}$ & T$_{phot}$ & $F_{7510}/F_{phot}$\\
\hline
 2400 &   8.98e-06 &  4300 &   8.90e-05 &  6400 &   8.78e-05 \\
 2500 &   1.49e-05 &  4400 &   9.04e-05 &  6600 &   8.55e-05 \\
 2600 &   2.35e-05 &  4500 &   9.18e-05 &  6800 &   8.32e-05 \\
 2700 &   3.37e-05 &  4600 &   9.31e-05 &  7000 &   8.12e-05 \\
 2800 &   4.26e-05 &  4700 &   9.41e-05 &  7200 &   7.90e-05 \\
 2900 &   5.30e-05 &  4800 &   9.50e-05 &  7400 &   7.66e-05 \\
 3000 &   5.80e-05 &  4900 &   9.56e-05 &  7600 &   7.41e-05 \\
 3100 &   6.41e-05 &  5000 &   9.59e-05 &  7800 &   7.17e-05 \\
 3200 &   6.98e-05 &  5100 &   9.64e-05 &  8000 &   6.92e-05 \\
 3300 &   7.52e-05 &  5200 &   9.66e-05 &  8200 &   6.66e-05 \\
 3400 &   7.93e-05 &  5300 &   9.64e-05 &  8400 &   6.42e-05 \\
 3500 &   8.20e-05 &  5400 &   9.61e-05 &  8600 &   6.15e-05 \\
 3600 &   8.43e-05 &  5500 &   9.56e-05 &  8800 &   5.88e-05 \\
 3700 &   8.58e-05 &  5600 &   9.51e-05 &  9000 &   5.61e-05 \\
 3800 &   8.73e-05 &  5700 &   9.44e-05 &  9200 &   5.37e-05 \\
 3900 &   8.80e-05 &  5800 &   9.36e-05 &  9400 &   5.12e-05 \\
 4000 &   8.89e-05 &  5900 &   9.28e-05 &  9600 &   4.86e-05 \\
 4100 &   8.83e-05 &  6000 &   9.18e-05 &  9800 &   4.63e-05 \\
 4200 &   8.75e-05 &  6200 &   8.99e-05 & 10000 &   4.41e-05 \\
\hline
\multicolumn{6}{l}{$T<7000$ K from BT-Settl models, corrected for scaling}\\
\multicolumn{6}{l}{~~~~~factor listed in Table~\ref{tab:sptteff1.tab}
}\\
\multicolumn{6}{l}{$T>7000$ K from Phoenix models.}\\
\end{tabular}
\end{table}

\pagebreak

\subsection{Including the Accretion Continuum in Spectral Fits}

Many young stars have strong enough accretion to partially or, in rare cases, even fully mask the photosphere.  Fully masked stars have no detectable photosphere and have no measured spectral type, but an extinction is still measureable with an estimate for the shape of the accretion continuum.  Even for lightly veiled stars, the extinction should be measured by comparing the observed colors to a combination of the photospheric spectrum and the accretion continuum spectrum.  This subsection describes how to estimate the shape and strength of the accretion continuum so that it can be included in extinction measurements.

\subsubsection{The shape of the accretion continuum spectrum}

Including the accretion spectrum into the spectral fit requires an estimate for veiling versus wavelength.
The measureable accretion continuum is produced by H recombination to the $n=2$ level (Balmer continuum, $\lambda<3700$ \AA) and to the $n=3$ level (Paschen continuum $\lambda<8200$ \AA), plus an H$^-$ continuum \citep[for detailed descriptions, see][]{Calvet1992,Calvet1998}.  
The ratio of these different components depends on the temperature, density, and optical depth of the accreting gas and heated chromosphere.   The size of the Balmer jump between stars is different \citep[e.g.][]{Herczeg2009}, which forces this analysis to be restricted to the shape of the continuum either at $<3700$ \AA\ or between 3700--8000 \AA.   Here we concentrate on the emission at $>4000$ \AA.  The spectral slope at $<3700$ is uncertain in the observed spectra due to the large wavelength dependence in the telluric correction near the atmospheric cutoff.

\begin{figure}
\epsscale{1.}
\plotone{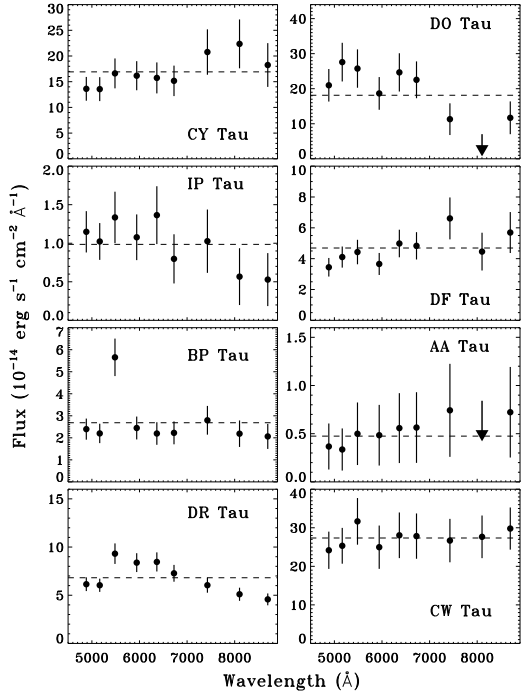}
\caption{The accretion continuum flux of eight stars.  For the top six stars, the accretion continuum is calculated from veiling measurements in \citet{Fischer2011} and the stellar flux measured here and then corrected for extinction (method b in the text).  For the heavily veiled stars DR Tau and CW Tau in the bottom panel, the accretion continuum is calculated by multiplying the measured veiling to photospheric templates (method a).  Uncertainties are assessed by including a 0.05 uncertainty in the veiling measurement and a 10\% uncertainty in the flux.
Differences in the slope of the accretion continuum between sources (e.g., slightly rising with wavelength for AA Tau but slightly falling for IP Tau)  may be real or may be introduced by observational uncertainties.  Our assumption that the optical accretion continuum is flat (shown here as the dashed horizontal line) is roughly consistent with these calculations and with the optical veiling measurements of \citet{Basri1990}.}
\label{fig:veil1.ps}
\end{figure}

The exact shape of the accretion continuum likely depends on the properties of the accretion flows.  
 Models of the accretion continuum typically assume a single shock structure.  Fitting the Balmer continuum emission leads to model spectra where at $>4000$ \AA, the flux decreases with increasing wavelength (see Fig.~3 from Ingleby et al.~2013 for spectra from isothermal models at different densities).  These synthetic spectra underestimate the observed veiling at red wavelengths (see, e.g., models of Calvet \& Gullbring 1998 and measurements by Basri \& Batalha~1990 and Fischer et al.~2011).  \citet{Ingleby2013} explains this problem by invoking the presence of accretion columns with a range of densities, some of which are lower density than has been typically assumed and produces cooler accretion shocks.  This physical situation may be expected if accretion occurs in several different flows or if a single flow has a range of densities, perhaps because the magnetic field connects with the disk at a range of radii.
The weaker shocks yield cooler temperatures and produce redder emission, thereby recovering the measured veilings around $1$ $\mu m$.  

Empirical measurements of the accretion continuum flux are shown in Fig.~\ref{fig:veil1.ps}.  The fraction of emission attributed to the accretion continuum is calculated from the optical veiling measurements of \citet{Fischer2011} and 
the relationship $(r_\lambda=\frac{F_{acc}}{F_{phot}})$.  This fraction is then converted to the accretion continuum flux by two different methods:  (a) multiplying the fraction by our flux-calibrated spectra, corrected for extinction, and (b) multiplying the veiling by the flux from the template spectrum for the relevant spectral type from Table~\ref{tab:gridstars}.  Uncertainties in the accretion continuum are estimated to be $0.05$ times the flux from the calibrated spectrum and $0.05$ times the total standard+accretion flux, respectively.$^3$  These two methods are somewhat independent from each other but yield similar results.
\footnotetext[3]{The uncertainty in veiling measurements is typically 0.05-0.1 for moderately veiled stars and much larger for heavily veiled stars because the definition is the accretion flux divided by the photospheric flux.  In either case, the uncertainty in the flux is 5--10\% of the total observed flux and not 5--10\% of the flux attributed to the accretion continuum.}

Table~\ref{tab:fischer} compares the $\chi^2$ from linear fits ($\chi^2_{\rm line}$) and flat fits ($\chi^2_{\rm flat}$) to the accretion continuum fluxes for methods (a:)  veil*flux and (b:) veil*template described above.  Most cases are consistent with a flat accretion continuum.  The veiling measurements tend to be smaller at longer wavelengths because the photospheric flux is brighter in the red than in the blue.  This exercise presents somewhat circular logic because the spectral type and extinction are both calculated assuming that the accretion continuum is flat.   Method (b) depends less on the assumption of a flat continuum but is sensitive to gravity and spectral type uncertainties in the optical colors.
However, the results from both approaches demonstrate that the assumption of a flat accretion continuum is reasonable and self-consistent.

\begin{table}
\caption{Slopes of the Accretion Continuum}
\label{tab:fischer}
\begin{tabular}{c|ccc|ccc}
\hline
 & \multicolumn{3}{c}{a:  veil * flux} & \multicolumn{3}{c}{b:  veil * template}\\ 
Star & $F4/F8^a$ & $\chi^2_{\rm line}$ & $\chi^2_{\rm flat}$ & $F4/F8^a$  &
 $\chi^2_{\rm line}$ & $\chi^2_{\rm flat}$ \\
\hline
AA Tau  & 0.79 & 0.4 & 0.4 & 0.64 & 0.5 & 0.4\\
BP Tau   & 1.09 & 0.4 & 0.5 & 0.70 & 3.7 & 1.4\\
CW Tau$^b$  & 1.03 & 1.4 & 1.6 & 1.03 & 0.8 & 0.9\\
CW Tau$^b$ & 1.08 & 1.1 & 1.2 &  1.08 & 0.5 & 0.5\\
CY Tau  & 1.39 & 4.3 & 5.7 & 1.00 & 8.7 & 10\\
DF Tau & 0.53 & 17 & 6.8 & 0.3 & 61 & 7.4\\
DO Tau & 1.29 & 1.7 & 0.8 & 0.76 & 8.9 & 5.2\\
IP Tau & 1.59 & 1.7 & 1.9 & 1.46 & 1.3 & 2.0\\
DG Tau & 1.04 & 2.0 & 2.3 & 0.80 & 3.7 & 2.1\\
DK Tau & 0.87 & 2.2 & 2.1 & 0.72 & 3.2 & 1.4\\
DL Tau & 0.98 & 4.2 & 4.7 & 0.98 & 1.4 & 1.6\\
DR Tau & 0.83 & 12.6 & 10 & 1.42 & 3.9 & 0.9\\
HN Tau A & 1.42 & 3.9 & 0.9 & 1.46 & 5.5 & 1.8 \\
\hline
\multicolumn{7}{l}{Spectral slopes of the accretion continuum for two methods}\\
\multicolumn{7}{l}{~~~of converting the veiling to flux, see also Fig.~\ref{fig:veil1.ps}}\\
\multicolumn{7}{l}{ Flat $\chi^2$ has 8 degrees of freedom, linear fit has 7.}\\
\multicolumn{7}{l}{$^a$$F4/F8$:  flux ratio of accretion continuum at 4000 \AA\ to 8000 \AA.}\\
\multicolumn{7}{l}{$^b$\citet{Fischer2011} observed CW Tau twice.}\\
\end{tabular}
\end{table}

Based on these calculations and the results of \citet{Ingleby2013}, we make the simplifying approximations that the shape of the accretion continuum is (a) the same for all accretors and (b) that the accretion continuum is constant, in erg cm$^{-2}$ s$^{-1}$ \AA$^{-1}$, at optical wavelengths.
In contrast, \citet{Hartigan2003} assumes that the accretion continuum is a line with a slope that differs from star to star.   Real differences between spectra are surely missed in our approach.  However, our approach is simpler and reproduces the observed spectra with fewer free parameters.
A more rigorous asssessment of the accretion continuum spectrum is possible from broad band high resolution spectra, which has been applied to small samples  \citep[e.g.][]{Fischer2011,McClure2013} but is time consuming, has not yet been implemented for large datasets, and suffers from the same degeneracies and systematic trades between surface gravity, reddening, and veiling by emission from the accretion shock and the warm inner disk.

\begin{figure}
\plotone{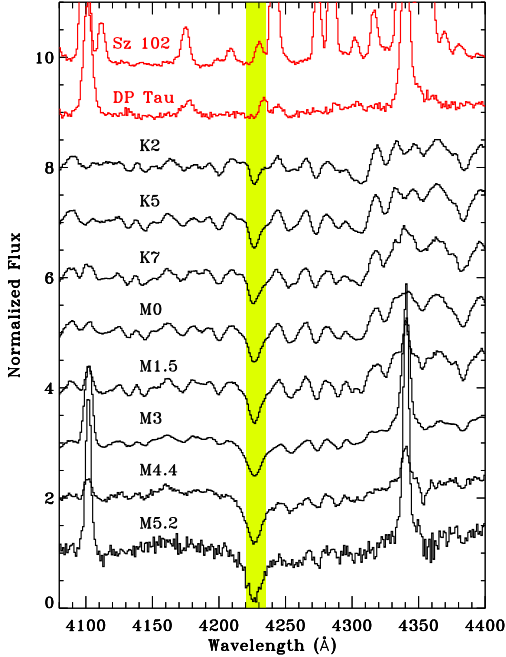}
\caption{The \ion{Ca}{1} $\lambda4227$ photospheric absorption line (shaded yellow region)  versus spectral type for WTTSs from early K through mid-M for stars listed in Table~\ref{tab:gridstars}.  In low-resolution spectra, emission lines produced by accretion processes can fill in the photospheric absorption, as shown for heavily veiled spectra of Sz 102 and DP Tau (top red spectra).}
\label{fig:ca4200plots.ps}
\end{figure}

\begin{figure}
\plotone{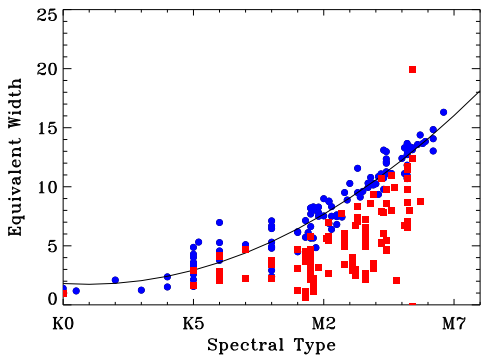}
\caption{The \ion{Ca}{1} $\lambda4227$ line equivalent width versus SpT, for WTTSs (blue circles) and CTTSs (red squares) in our sample.  The accretion continuum at 4227 \AA\ is calculated by comparing the equivalent width in the line to that expected at a given SpT.  For CTTSs, this line is often shallower than expected because of accretion, a fact which we explot to measure the strength of the accretion continuum from low-resolution spectra.}
\label{fig:ca4200.ps}
\end{figure}

\subsubsection{Veiling Estimates at Ca I 4227}

Veiling can be accurately measured from high resolution spectroscopy \citep[e.g.][]{Basri1990,Hartigan1991} or estimated from low resolution spectrophotometric fitting \citep[e.g.][]{Fischer2011,Ingleby2013}.  Here we develop an intermediate approach to measure the veiling by comparing the depth of a single, strong absorption line to its depth in a template star.  While accurate veiling measurements require high resolution spectroscopy, veiling may be estimated by measuring the depth of strong photospheric lines in low resolution spectra.  This section concentrates on the strong \ion{Ca}{1} $\lambda4227$ line.

Figure ~\ref{fig:ca4200plots.ps} shows spectra of the \ion{Ca}{1} region versus spectral type.  The \ion{Ca}{1} equivalent width depends on spectral type as
\begin{equation}
{\rm EW(Ca I)} = -189.218 + 7.36 x -0.072 x^2
\end{equation}
where $x=50$, 58, 63 for K0, M0, and M5, respectively (Fig.~\ref{fig:ca4200.ps}).  Values lower than this equivalent width indicate that the depth of the photospheric absorption is reduced because of extra emission from the accretion continuum.  This difference yields the strength of the accretion continuum at 4227 \AA.

\begin{figure*}
\vspace{-90mm}
\hspace{-20mm}
\epsscale{0.9}
\plotone{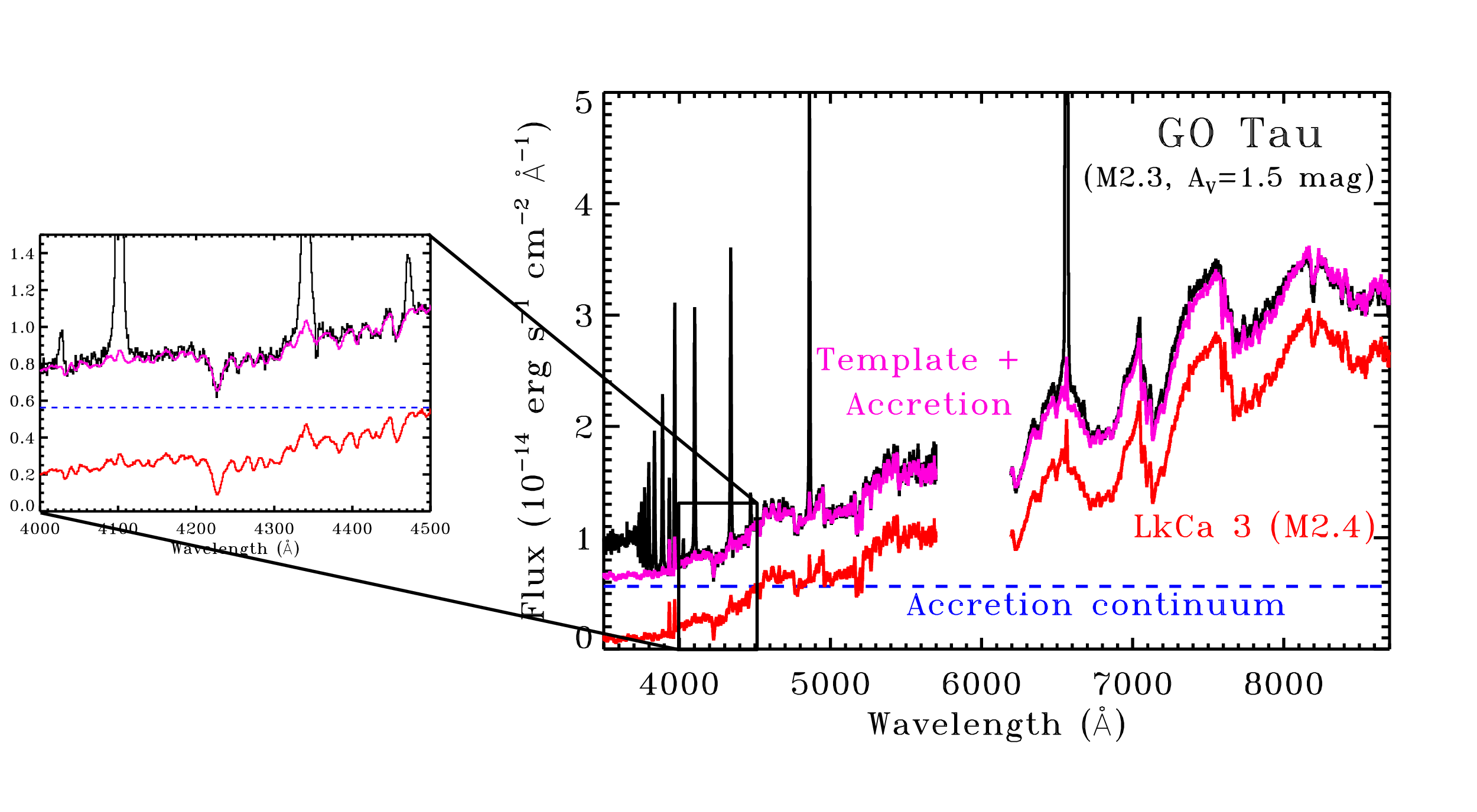}
\caption{Our method of fitting a photosphere, an accretion continuum and an extinction, demonstrated for the CTTS GO Tau, as described in \S 4.  The photospheric template (red spectrum) is determined from the molecular band indices (see \S 3.1), with a strength that is reduced by a flat accretion continuum spectrum (blue dashed line).  The accretion continuum flux is estimated from the depth of the \ion{Ca}{1} $\lambda4227$ line (see \S 3.4).  The photospheric temperature, stellar luminosity, accretion continuum strength, and extinction are all free parameters that vary until a best fit (purple spectrum) is found and visually confirmed.}
\label{fig:gotau_veil.ps}
\end{figure*}

Fig.~\ref{fig:gotau_veil.ps} shows an example of how the veiling at 4227 \AA\ is estimated for each source.  A flat accretion spectrum is added to the template photospheric spectrum so that the combination matches the observed line depth.  The \ion{Ca}{1} absorption line is sometimes filled in by emission from nearby Fe lines, thereby affecting the veiling estimate \citep{Gahm2008,Petrov2011,Dodin2012}.  Although this particular line is also sensitive to surface gravity, the use of temperature matched WTTS as templates should mitigate  gravity dependent line depth systematics.   For cooler stars, calculating the accretion continuum flux at 4227 \AA\ maximizes the sensitivity to accretion for cool stars because the ratio of accretion flux to photospheric flux is higher at short wavelengths.

\section{Final Assessment of Stellar Parameters}

The previous section provides a grid of spectral templates (Table~\ref{tab:gridstars}), a method to estimate the strength of the accretion continuum emission, and a description of extinction.  In this section, we apply these analysis tools to simultaneously measure the spectral type, extinction, and accretion for our sample.  Our procedures for K and M spectra with with zero to moderate veiling are discussed in \S 4.1.   Heavily veiled stars require a different approach and are discussed separately in \S 4.2.  We then describe in \S 4.3 how these methods are implemented for several selected stars.  In \S 4.4-4.5, spectral types and extinctions are compared to selected measurements in the literature.  

Our final spectral types, extinctions, veilings, and photospheric parameters are presented in Appendix C.  
Some stars have extinction values that are measured to be negative.  These extinctions are retained for statistical comparisons to other studies but are unphysical and treated as zero extinction when calculating luminosities.

\subsection{Stars with zero to moderate veiling}

A best-fit SpT, $L_{phot}$, $A_V$, and accretion continuum flux (veiling)  is calculated for each star by fitting 15 different wavelength regions from 4400--8600 \AA.  The wavelength regions are selected by concentrating on obtaining photospheric flux measurements both within and outside of absorption bands.  For stars with spectra covered by emission lines (more than the H Balmer, \ion{He}{1}, and \ion{Ca}{2} lines), the bluest regions are excised from the fit and the remaining wavelength regions are altered to focus on continuum regions.  The wavelength regions incorporate the spectral type indices described previously.  The accretion continuum flux is initially estimated from the equivalent width of the \ion{Ca}{1} line and is manually adjusted.  All fits are confirmed by eye.  This approach is similar to that taken by \citet{Hartigan2003} to analyze spectra of close binaries in Taurus, although our spectral coverage is broader and our grid of WTTSs is more complete.

Spectral types and extinctions are calculated from the spectral grid established in \S 3.3.  The spectral types are listed to 1, 0.5, and 0.1 subclasses for spectral types earlier than K5, K5--M0, and later than M0.  Extinction is calculated at intervals of $A_V=0.02$ mag. and listed to the closest 0.05 mag.  For M-dwarfs, these values approximately Nyquist sample the uncertainties of $\sim 0.2-0.3$ subclasses in SpT and $\sim 0.2-0.3$ mag.~in $A_V$.
The accretion continuum is fixed to 0 for stars with no obvious signs of accretion.  For accreting stars, the accretion continuum is also initially a free parameter.  Comparing the results of this grid yield an initial best fit to the spectral type, accretion continuum strength, and extinction.  This initial spectral type measurement is then used to constrain the accretion continuum from fitting to the \ion{Ca}{1} 4227 \AA\ line (\S 3.4).  With this new accretion continuum, a new best-fit spectral type and extinction are calculated.  For stars used as templates, our best fit SpT and $A_V$ are calculated here from the full photospheric grid and may therefore differ slightly from the values listed Table~\ref{tab:gridstars}.

As a check for self-consistency, most of our final SpT and $A_V$ agree with earlier measurements in this project to 0.1 subclass in SpT and 0.1 mag., respectively, for M-dwarfs.  The previous measurements were based on a slightly different spectral grid.  This comparison defines our internal precision for extinction and spectral type.

Fig.~\ref{fig:gotau_veil.ps} demonstrates how this method is implemented for the CTTS GO Tau.  The accretion continuum and photospheric spectrum together match the \ion{Ca}{1} 4227 \AA\ line, many other bumps in the blue spectrum, and the TiO absorption in the red spectrum.  In some cases the best fit accretion continuum is found to differ from the \ion{Ca}{1} absorption depth calculated based on Fig.~\ref{fig:ca4200.ps}, likely because weak line emission fills in the photospheric absorption line.  Indeed, emission in several lines around 4227 \AA\ is seen from many heavily veiled stars.
Accounting for veiling is particularly important for stars with moderate or heavy veiling ($>0.1$ at 7000 \AA), as described for several individual sources in \S 4.3.  Even for lightly veiled stars, accretion can affect the SpT and extinction.

\begin{figure}
\epsscale{1.}
\plotone{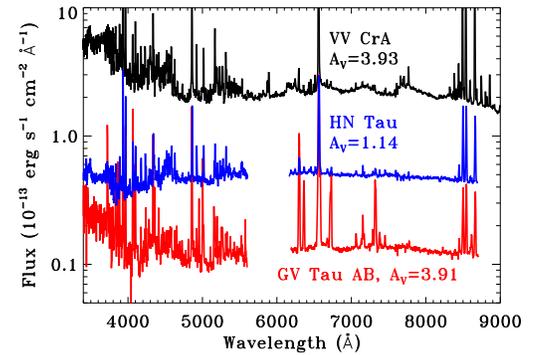}
\caption{The extinction-corrected spectra of heavily veiled stars.  These stars cannot be spectral typed from low-resolution spectra and are classified as ``continuum'' objects.  Extinctions for heavily veiled sources are measured by assuming that the observed continuum is flat.  In our low-resolution spectra, this continuum is measurable at $> 4600$ \AA.  The blue spectra of heavily veiled objects are typically covered in a forest of emission lines.}
\label{fig:bband.ps}
\end{figure}
 
Multiple spectra were obtained for 62 targets in our sample.  Repeated observations yields more accurate photospheric measurements because the best fit spectral type should be consistent despite changes in the veiling.  The spectral type, extinction, and accretion continuum flux were initially left as free parameters for each spectrum.   No significant change in SpT was detected.  The final SpT was set to the average SpT of all spectra of the object.  The spectra were then fit again with this SpT, leaving extinction and the accretion continuum flux as the free parameters.   The average extinction value was then applied to all observations of a given object, when possible, to calculate the accretion and photospheric luminosity.  In 3 cases, definitive changes were detected and a single extinction could not be applied.  This approach of trying to find a single extinction to explain repeated spectra should miss some real changes in extinction.  Changes in the strength of the accretion continuum are frequently detected among the different spectra, usually with amplitude changes of less than a factor of $\sim 3$.
  No star in our sample changes between lightly and heavily veiled.  Spectral variability will be discussed in detail in a subsequent paper.

\subsection{Heavily Veiled Stars}

Heavily veiled stars have spectra dominated by emission produced by accretion, with little or no detectable photospheric component and a forest of emission lines at blue wavelengths (Fig.~\ref{fig:bband.ps}).  In some cases, high resolution spectra can yield enough photospheric absorption lines to reveal a spectral type \citep{White2004}.  In our work, the 5200 \AA\ band and the TiO bands can be detected from some objects despite high veiling.  RNO 91 has few photospheric features but many fewer emission lines than the other stars.  A weak 5200 \AA\ bump of RNO 91 suggests a SpT of K0--K2, while a stronger bump for HN Tau A suggests K2--K5.  Both RNO 91 and HN Tau A lack detectable TiO absorption.  DL Tau has weak TiO absorption and has a SpT between K5--K8.   Two cases, CW Tau and DG Tau, are assigned spectral types of K3 and K6.5 and are discussed in detail in subsections 4.2.5--4.2.6.

The stars VV CrA, GV Tau AB, AS 205A, and Sz 102 have no obvious photospheric features in our spectra and are listed here as continuum sources.  These sources likely have spectral types between late G and early M, since earlier and later spectral types are unlikely based on indirect arguments.  At spectral types earlier than late G, the photosphere is bright enough that it dominates the spectrum for reasonable accretion rates.  For stars cooler than early M, the photosphere is so red that the TiO bands are always easily detected at red wavelengths \citep[e.g.][]{Herczeg2008}.  An M5 star with veiling of 30 at 8400 \AA\ would still have a TiO band depth of 3\%, which is detectable in most of our spectra because of sufficient signal to noise and flux calibration accuracy.  Such a high veiling is not expected for M5 stars because the accretion rate correlates with $M_\ast^2$.
Indeed, several mid-late M-dwarfs (e.g., GM Tau, CIDA 1, 2MASS J04141188+2811535, J04414825+2534304)  have blue spectra with high veiling and are covered by a forest of chromospheric emission lines, similar to the cases of CW Tau and DG Tau, but have red spectra with easily identified TiO absorption. 
 Only between K0--M2 are photospheric features are weak enough and the photosphere faint enough that it could be fully masked by a strong accretion continuum.$^4$
\footnotetext[4]{These arguments do not apply at the earliest stages of protostellar evolution or for outbursts, when accretion rates are much higher than those typically measured in the T Tauri phase.  In these cases, the accretion luminosity may be much brighter than any photospheric luminosity, regardless of the underlying spectral type on an unheated photosphere, if present.}

For heavily veiled stars, the extinction is calculated by assuming that the accretion continuum is flat (see CW Tau and DR Tau in Fig.~\ref{fig:veil1.ps}) and dominates the optical emission.  Extinction corrected spectra for three stars are presented in Fig.~\ref{fig:bband.ps}.  Fits to the continuum were made to avoid emission lines and TiO emission (see \S 5.4.2).
The extinction is likely underestimated to stars such as HL Tau, Sz 102, GV Tau, and 2MASS J04381486+2611399 because of edge-on disks and/or remnant envelopes.  The optical flux from these sources is very faint but appears to have no or little extinction.  These three objects also have forbidden emission lines with large equivalent widths, characteristic of sources where the edge-on disk occults the star but not the outflow.

\begin{figure}
\plotone{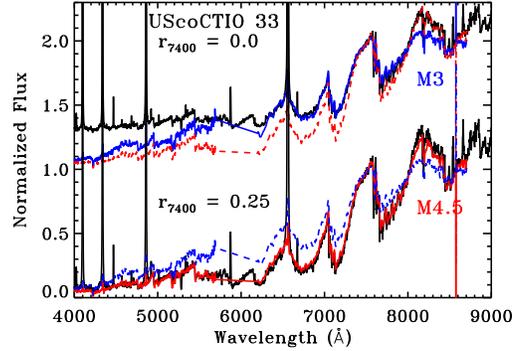}
\caption{The spectrum of UScoCTIO 33 compared to M3 and M4.5 spectra with veiling $r_{7510}=0.0$ (top) and $0.25$ (bottom).  If veiling by accretion is not accounted for, the 6000-9000 \AA\ spectrum appears to be an M3 (top).  However, the M3 template badly underestimates the blue emission.  If we estimate and subtract off an accretion continuum, the spectrum is consistent with an M4.5 star (bottom).}
\label{fig:ctio33.ps}
\end{figure}

\subsection{Examples of specific stars}

These subsections illustrate how the logic described above is implemented for several example stars, which cover a range of spectral type and accretion rate.  The pre-main sequence tracks applied here to calculate masses and ages are combined from \citet{Tognelli2011} and \citet{Baraffe2003}, as described in Appendix C.

\begin{figure*}
\epsscale{1.2}
\plotone{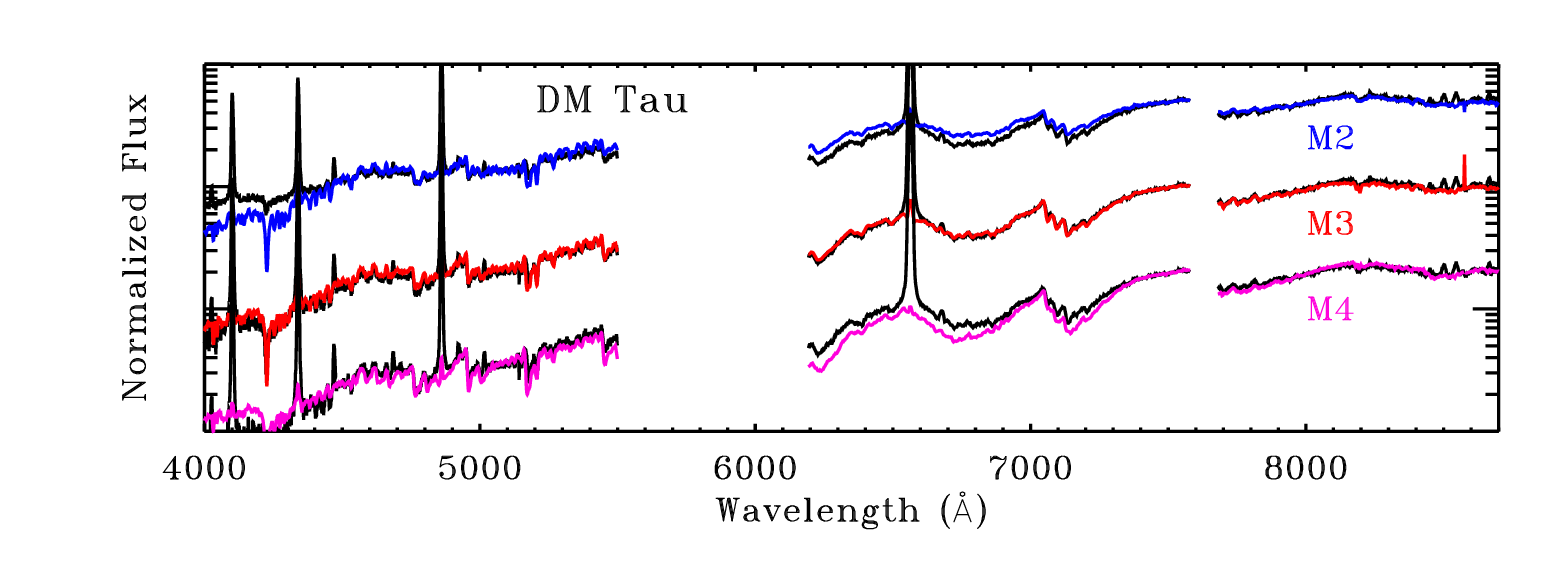}
\plotone{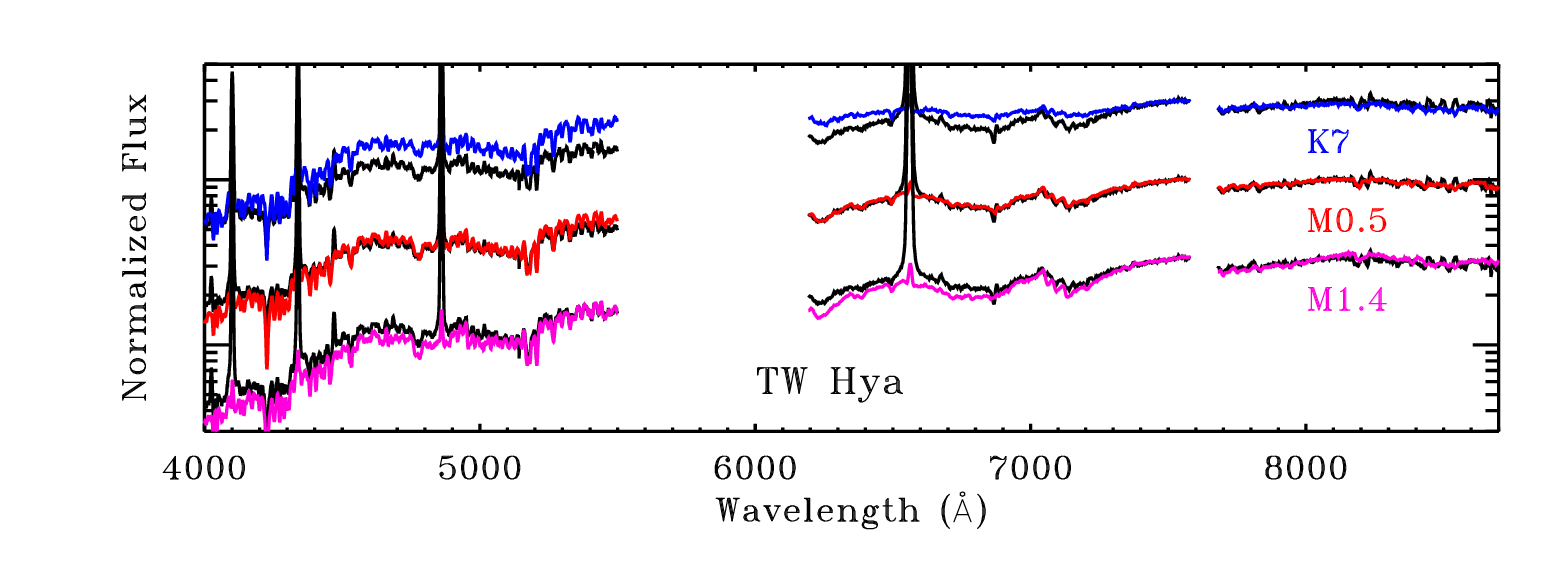}
\caption{Otical spectra of DM Tau (top, M3.0) and TW Hya (bottom, M0.5), after subtracting a flat accretion continuum and compared with templates of different spectral type (colored spectra).  
TW Hya is consistent with a spectral type of M0.5, intermediate between previous measurements of M2 and K7.  The M3 spectral type of DM Tau is later than literature measurements of M1.}
\label{fig:twa1.ps}
\end{figure*}

\subsubsection{UScoCTIO 33}

UScoCTIO 33 was originally identified as a possible member of the Upper Scorpius OB Association in a photometric survey by \citet{Ardila00}.  A spectroscopic survey by \citet{Preibisch02} confirmed membership, classified the star as M3, and found strong H$\alpha$ emission indicative of accretion.  

Figure~\ref{fig:ctio33.ps} shows our Keck spectrum of UScoCTIO 33 compared with M3 and M4.5 stars with a veiling $r_{7525}$=0.0 and 0.25.  If the veiling is 0, the red spectrum is best classified as an M3 spectral type, with only small inconsistencies between the template and the spectrum.  However, the M3 template spectrum is much weaker than the observed blue emission.
The veiling $r_{7525}$=0.25 is calculated from the depth of the \ion{Ca}{1} $\lambda4227$ line.  Subtracting this accretion continuum off of the observed spectrum yields photospheric lines that are deeper than the uncorrected observation.  The consequent M4.5 spectral type with veiling improves the fit to the red and blue spectra.

The M4.5 SpT leads to a mass of 0.11 $M_\odot$ and an age of 5 Myr.  The M3 SpT and no veiling yields a mass of 0.32 $M_\odot$ and an age of 35 Myr, assuming no change in $A_V$.

\subsubsection{DM Tau}
The literature spectral type of M1 for DM Tau traces back to \citet{Cohen1979}.  Despite significant interest as the host of a transition disk \citep[e.g.][]{Calvet2005}, its spectral type has not been reassessed using modern techniques.  

Fig.~\ref{fig:twa1.ps} shows the veiling-corrected DM Tau spectrum 29 Dec.~2008, compared with M2, M3, and M4 spectra.  The veiling is calculated from the depth of the Ca I 4227 \AA\ line.  The veiling $r_{7510}=0.17$ leads to SpT of M3 and $A_V=0.08$.  If the composite photospheric+accretion spectrum is not constrained by a good fit to the \ion{Ca}{1} $\lambda4227$ line, then $r_{7510}$ could range from 0.09, with SpT M2.7 and $A_V=-0.20$, to 0.31, with SpT M3.4 and $A_V=0.50$.  In this case, the extinction increases with later spectral type because the veiling has increased (see also the case of DP Tau).  If the blue side is ignored entirely, then a veiling of $r_{7510}=0$ would yield M2.5 and $A_V=0.06$ while an upper limit on veiling of $r_{7510}=0.39$ would yield M4.1 and $A_V=-0.06$.  In these latter cases, the resulting red spectrum looks reasonable.  The uncertainties in SpT and veiling are about half the size of these ranges when using the blue and red spectra together.  Even with the blue+red spectrum, differences between M2 and M4 are subtle and are likely undetectable with a cruder method, such as photometry.

The change from M1 to M3 for DM Tau leads to a younger age (17 versus 4.9 Myr) and a lower mass (0.62 versus 0.35  M$_\odot$), assuming no change in $A_V$.  The luminosity does not change significantly because the bolometric correction from the red photospheric flux is similar for an M1 and an M3 star.

\subsubsection{TW Hya}

Despite being the closest and possibly the most studied CTTS, the spectral type of TW Hya has been the subject of some controversy.  The original spectral type of K7 was obtained from low resolution spectroscopy by \citet{delaReza1989}.   \citet{Yang2005} used high-resolution optical spectra to measure an effective temperature of $4126\pm24$ K, equivalent to K6.5.  They caution that the uncertainty in effective temperature likely underestimates systematic uncertainties.  This spectral type is consistent with the K7 SpT derived by \citet{Alencar2002}, also from high-resolution spectra.  In contrast, \citet{Vacca2011} relied on low resolution spectra from 1--2.4 $\mu$m to obtain a new spectral type of M2.5.  \citet{McClure2013} found that TW Hya is consistent with roughly M0 spectral type at $1.1$ $\mu$m.
\citet{Debes2013} argued that the 5500--10200 \AA\ spectrum is a composite K7+M2 in the near-IR, with the warmer component related to accretion.  \citet{Debes2013} did not consider veiling by the accretion continuum, which would preferentially cause the measured SpT at short wavelengths to be earlier than the actual spectral type.

Fig.~\ref{fig:twa1.ps} shows that the optical spectrum is consistent with an M0.5 spectral type, which corresponds to $\sim 3810$ K, with $A_V=0.0$ mag.~and veilings that range $r=0.09-0.21$.  This spectral type is consistent with all of our TW Hya spectra, obtained on 7 different nights in Jan., May, and Dec.~2008.  The effective temperature is significantly lower than that from \citet{Alencar2002} and \citet{Yang2005}.  While spectral fits to high resolution spectra may suffer from emission lines filling in photospheric absorption for strong accretors \citep[e.g.][]{Gahm2008,Dodin2012}, this problem is not expected to be significant for a weakly accreting star like TW Hya.  However, the high veiling during the \citet{Yang2005} observation, three times higher than the median veiling measured here and by \citet{Alencar2002}, may have complicated their temperature measurements.

\begin{figure*}[!t]
\epsscale{1}
\plottwo{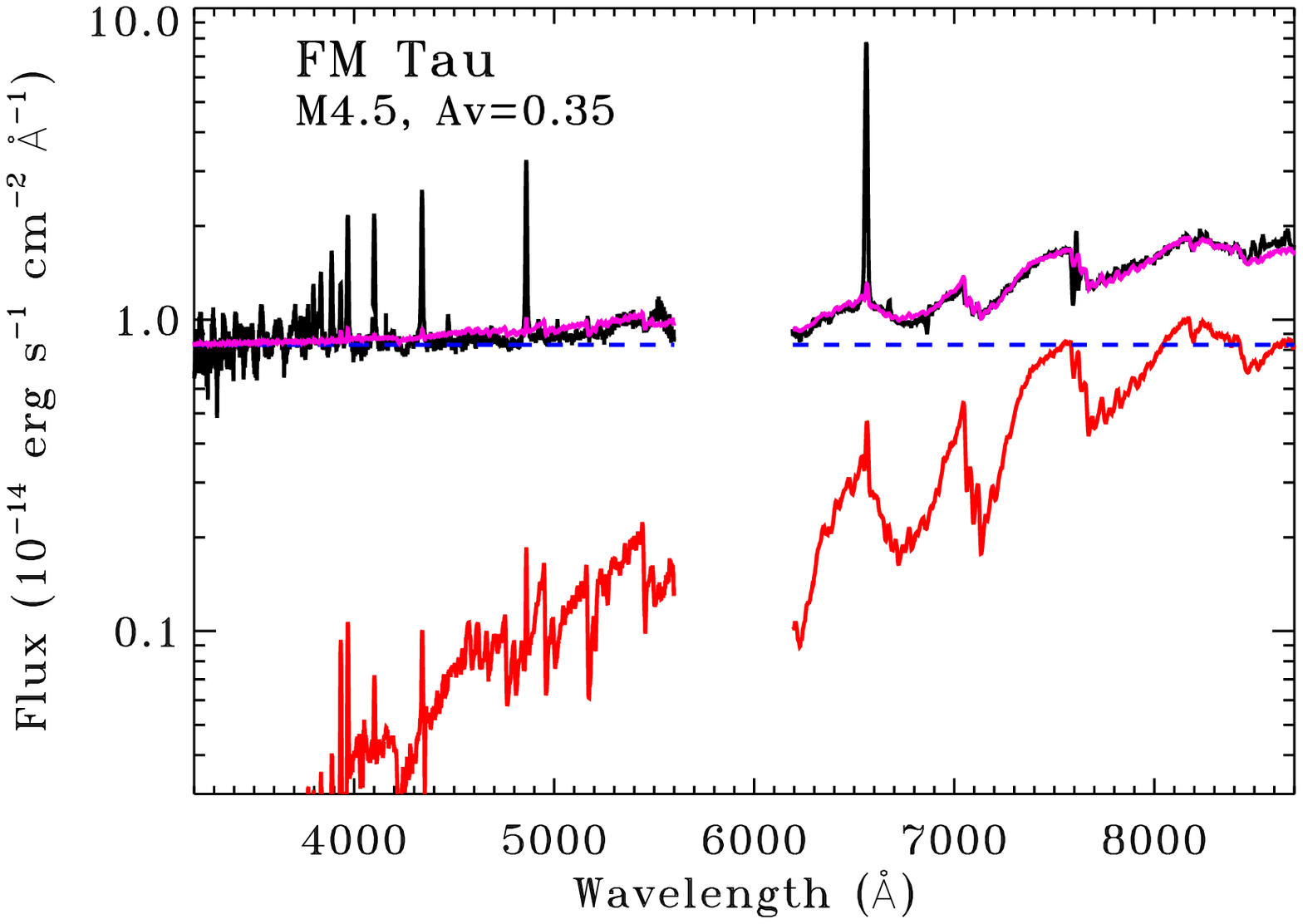}{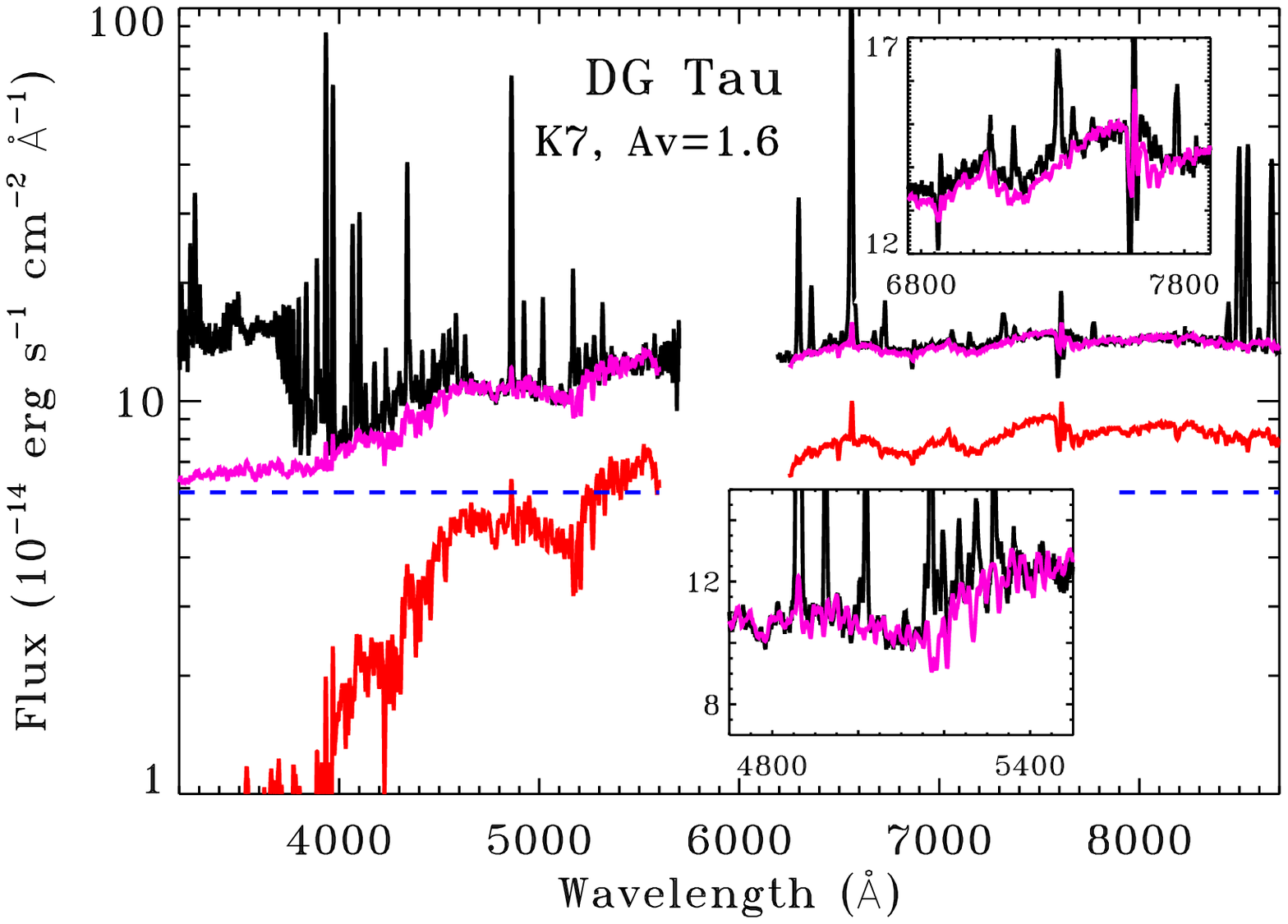}
\caption{Fits to extinction-corrected optical spectra of two heavily veiled stars, FM Tau and DG Tau.  The black line is the observed spectrum, the red line is the photospheric template, the dashed blue line shows the flat accretion continuum, and the purple line shows the best fit template plus accretion continuum.}
\label{fig:fmtau.ps}
\end{figure*}

Our spectral type is inconsistent with the late spectral type of \citet{Vacca2011}.  An M2 spectral type could only be recovered for our TW Hya spectra if the accretion continuum were three times larger in the red than that measured in the blue, which is inconsistent with both models and previous measurements of the accretion continuum.  The spectral templates of \citet{Vacca2011} were dwarf stars, which may differ in certain near-IR features from TTSs.  
Visual inspection of these templates do not reveal significant differences between TW Hya and an M0.5 dwarf star, except in the H$_2$O band at 1.35 $\mu$m.  \citet{McClure2013} also found that all K7--M0 CTTSs appear as M2 dwarf stars in one of their most prominent line ratios, which demonstrates the need to use WTTSs as templates.
  A composite spectrum of photospheres with different temperatures is not needed to explain the optical spectrum at $<10000$ \AA, although magnetic spots are expected to affect effective temperature measurements.

\subsubsection{FM Tau}
FM Tau is our most extreme example of a change in spectral type.  The most commonly used spectral type of M0 can be traced back to \citet{Cohen1979}.  However, the \citet{Cohen1979} spectra cover 4500--6600 \AA, where FM Tau looks like an M0 star because of high veiling.  \citet{Hartigan1994} twice obtained FM Tau spectra from 5700--7000 \AA\ and classified FM Tau as M0 and M2.   The prominent TiO absorption bands are readily detected at $>7000$ \AA, where the red photosphere is stronger than the accretion continuum. 

FM Tau is classified here as an M4.5$\pm0.4$.  The large uncertainty in spectral type is caused by the high level of veiling.  The measured extinction of $A_V=0.35\pm0.2$ is mostly constrained by fitting to the accretion continuum rather than the photosphere.  The systematic uncertainty in $A_V$ is caused by the uncertain shape of the accretion continuum.  An M0 star is a reasonable approximation for the FM Tau colors, so our extinction is similar to literature values \citep[e.g.][]{Kenyon1995,White2001}.

\subsubsection{DG Tau}

DG Tau is the source of a famous and well-studied jet \citep[e.g.][]{Eisloffel1998,Bacciotti2000}.  Literature spectral types range from K3--M0 \citep{Basri1990,Kenyon1995,White2004}, including what should be a reliable spectral type of K3 from high-resolution optical spectra \citep{White2004}.  The discrepancies are caused by the high veiling.  \citet{Gullbring2000} called DG Tau a continuum star, implying that no photosphere is detectable.

At low resolution, the spectrum shows the 5200 feature, which is typical of K stars, and TiO bands, which are seen only in stars with SpT K5 and later (Fig.~\ref{fig:fmtau.ps}).  The SpT is K7$^{+1}_{-1}$, with $A_V=1.6\pm0.15$ mag.  The extinction depends mostly on the shape of the accretion continuum and does not significantly change with a small change in SpT.  The SpT is limited to earlier than or equal to M0 by the shallow depth of the TiO bands.

\subsubsection{CW Tau}

CW Tau is a heavily veiled T Tauri star with a jet \citep[e.g.][]{Coffey2008}.  \citet{Cohen1979} classify CW Tau as a K3 star, which is consistent with our measurement.  \citet{Horne2012} and \citet{Brown2013} found absorption in CO v=1-0 transitions, indicating that our line-of-sight passes through the surface layers of the circumstellar disk.  The spectral analysis presented here is based on the 18 Jan.~2008 spectrum.  The three spectra obtained in Dec.~2008 are $\sim 8$ times fainter because of a change in extinction.  This variability will be discussed further in a future paper.

Unlike DG Tau and DL Tau, the CW Tau spectrum does not show any TiO absorption, which restricts the SpT to earlier than K5.  The presence of the 5200 \AA\ bump suggests that CW Tau is later than K0.  Our best fit is a K3 star with $A_V=1.74$.  The acceptable SpT range is from K1--K4 with corresponding $A_V$ of 1.9 and 1.6, respectively.  As with other heavily veiled stars, the methodological uncertainty in $A_V$ is smaller than would be expected for this range in SpT because of high veiling.  However, the systematic uncertainty in extinction may be higher because the extinction relies on the assumption that the accretion continuum is flat.

The observed spectrum is weaker than the model spectrum at $>8000$ \AA, a difference which is also detected in some other accreting stars.  The weaker flux indicates a smaller veiling, which could be attributed to the Paschen jump.

\subsubsection{DP Tau}

DP Tau is a 15 AU binary system \citep{Kraus2011} that appears as a heavily veiled star.  The spectral type assigned here of M0.8 is based on the depth of the TiO bands for accretion continuum veilings of $r_{7510}=$0.36, 0.38, and 0.40 for our three spectra.  The uncertainty in spectral type is $\sim 0.5$ subclasses and is dominated by the uncertainty in the accretion continuum.  The extinction of $A_V=0.78$ mag.~is  calculated by comparing the observed spectrum to a combined accretion plus photospheric spectrum.

DP Tau is highlighted here as an example of a counterintuitive parameter space, like DM Tau but with higher veiling, where a later spectral type leads to a higher extinction.  This behavior for heavily veiled stars is the opposite of expectations for stars without accretion.    For DP Tau, if the accretion continuum is increased so that $r_{7510}=0.55$, then the increased depth of the TiO features leads to a SpT of M1.9.  However, the combined spectrum of accretion plus photospheric template is bluer than the M0.8+accretion spectrum, so the $A_V=0.98$ mag.  Similarly, a low veiling of $r_{7510}=0.28$ leads to M0.6 and $A_V=0.56$ mag.  These fits are the limiting cases for reasonable fits to the observed spectrum.  The extinction measurements are similar because they are based largely on the shape of the blue accretion continuum, which is assumed to be flat.

\subsection{Comparison of Spectral Types to Previous Measurements}

In this subsection, we compare our spectral types to selected literature measurements.  Our internal precision in SpT is $\sim 0.2$ subclasses for M-dwarfs and 0.5--1 subclass for earlier spectral types, based on the repeatability of SpT from independent multiple observations of the same stars.  In general the spectral types agree with literature values to $\sim 0.5$ subclasses, as demonstrated in our comparison of spectral types of stars the MBM 12 Association with \citet{Luhmanmbm12}.  However, significant discrepancies exist for members of the TW Hya Association and for some well known members of Taurus. The K5--M0.5 range in SpT may also have systematic offsets of 0.5--1 subclass in SpT relative to other studies.

\subsubsection{Comparison to Luhman Spectral Types}

The spectral type sequence described here is based largely on that established by Luhman.  Table~\ref{tab:luhmanspt} compares 29 stars with spectral types and extinctions measured here, in a survey of the MBM 12 Association by \citet{Luhmanmbm12}, and in a survey of Spitzer IRAC/X-ray excess sources by \citet{Luhman2009}.   

The median absolute difference between our and Luhman M-dwarf spectral types is 0.25 subclasses.  The standard deviation is 0.37 subclasses.  Six objects (20\% of the sample) differ by more than 0.5 subclasses.  Three of those six objects have spectral types in the K7--M1 range, as might be expected given the possible differences in our SpT scales (see \S 3.1). 

\begin{table}
\caption{Comparison to Luhman Spectral Types}
\label{tab:luhmanspt}
\begin{tabular}{lcccc}
\hline
Star & \multicolumn{2}{c}{This Work} & \multicolumn{2}{c}{Luhman} \\
 & SpT & $A_V$ & SpT & $A_V$\\
\hline
MBM 1 &  K5.5 & 0.08 & K6 & 0.39\\
MBM 2 & M0.3 & 1.64 &  M0 & 1.17\\
MBM 3 & M2.8 &0.54 & M3 & 0.0 \\
MBM 4 & K5.5 & (-0.24) & K5 & 0.85\\
MBM 5 & K2  & 0.88 & K3.5 & 1.95\\
MBM 6 & M3.8 & 0.50 & M4.5 & 0.0 \\
MBM 7 & M5.6 &(-0.08) & M5.75 & 0.0\\
MBM 8 & M5.9 &0.28 &  M5.5 & 0.0\\
MBM 9 & M5.6 & 0.10 & M5.75 & 0.0\\
MBM 10 & M3.4 & 0.60 &M3.25 & 0.18\\
MBM 11 & M5.8 & (-0.08) & M5.5 & 0.0\\
MBM 12 & M2.6 & 0.24 & M3 & 1.77\\
\hline
FU Tau & M6.5 & 1.20 &M7.25 & 1.99\\  
V409 Tau & M0.6 & 1.02 & M1.5 & 4.6 \\
XEST 17-059 & M5.2 & 1.02 & M5.75 & 0.0\\
XEST 20-066 & M5.2& (-0.14) &M5.25 & 0.0\\
XEST 16-045 & M4.5 & (-0.06) & M4.5 & 0.0\\
XEST 11-078 & M0.7 & 1.54 &M1 & 0.99\\
XEST 26-062 & M4.0 & 0.84 & M4 & 1.88\\
XEST 09-042 & K7 & 1.04 & M0 & 0.39\\
XEST 20-071& M3.1 & 3.02 & M3.25 & 2.77\\
2M 0441+2302 & M4.3 & (-0.15) & M4.5 & 0.39\\
2M 0415+2818 &M4.0  & 1.80& M3.75 & 1.99 \\
2M 0415+2746 &M5.2 & 0.58 & M5.5 & 0.0 \\
2M 0415+2909 & M0.6 & 2.78 & M1.25 & 1.99\\  
2M 0455+3019 & M4.7 & 0.70 & M4.75 & 0.0\\
2M 0455+3028 & M5.0 & 0.18 & M4.75 & 0.0\\  
2M 0436+2351 & M5.1 & -0.18 & M5.25 & 0.34\\  
2M 0439+2601 & M4.9 & 2.66 & M4.75 & 0.63 \\
\hline 
\end{tabular}
\end{table}

\subsubsection{Taurus Spectral Types}

Many of the most famous objects in Taurus have spectral types that date back to \citet{Cohen1979}, as listed in the compilations of \citet{HBC1988} and \citet{Kenyon1995}.  The \citet{Cohen1979} spectral coverage was optimal for early spectral types but insufficient for M stars.  Table~\ref{tab:newtauspt} lists the most significant changes in Taurus spectral types, relative to the compilation of spectral types by \citet{Luhman2010}. Our new spectral types are often 2--3 subclasses later than those from \citet{Cohen1979}, particularly when veiling affected the spectral typing at short wavelengths.  In cases of overlap with \citet{Dorazi2011}, our spectral types are consistent to within $0.5$ subclasses of the measured effective temperature.  

Several Taurus stars with spectral types earlier than K0 are challenging for spectral type measurements because accretion produces emission in the same lines (e.g., \ion{Ca}{2} infrared triplet, H Balmer lines) that are used for spectral typing.  For example, H$\alpha$ appears in emission from V892 Tau and RY Tau despite early spectral types.  While RY Tau has had numerous spectral types bewteen F7--G1 \citet{Mora2001,Calvet2004,Hernandez2004}, the \citet{Cohen1979} spectral type of K1 has been adopted in several recent compulations.  Our spectral type of G0 for RY Tau agrees with other recent spectral types.

\begin{table}
\caption{Discrepancies in Taurus Spectral Types}
\label{tab:newtauspt}
\begin{tabular}{lcc}
\hline
Star & This Work & Literature$^a$\\
\hline
CIDA 9  & M1.8 & K8 \\
DM Tau & M3.0 & M1\\
DS Tau & M0.4 & K5\\
FM Tau & M4.5 & M0\\
FN Tau & M3.5 & M5\\
FP Tau & M2.6 & M4\\
FS Tau & M2.4 & M0\\
GM Tau & M5.0 & M6.4\\
GO Tau & M2.3 & M0 \\
IRAS 04216+2603 & M2.8 & M0.5\\
IRAS 04187+1927 & M2.4 & M0\\
IS Tau & M2.0 & M0\\
LkCa 4 & M1.3 & K7\\
LkCa 3 & M2.4 & M1\\
RY Tau & G0 & K1$^b$ \\
LkHa 332 G1 & M2.5 & M1\\
LkHa 358 & M0.9 & K8\\
\hline
\multicolumn{3}{l}{$^a$Literature SpT as adopted by Luhman et al.~2010}\\
\multicolumn{3}{l}{$^b$Other recent works have measured early G.}\\
\end{tabular}
\end{table}

\subsubsection{TWA Association Spectral Types}

Our spectral types for the TWA are uniformly later than the spectral types obtained from Webb et al.~(1999, see Table~\ref{tab:twaspt}).  Our spectral types for TWA 8A, TWA 8B, TWA 9A, and TWA 9B are consistent with those obtained from high-resolution spectra by \citet{White2004}.  Our spectral types are also mostly consistent with the recent spectral types measured from X-Shooter spectra by \citet{Stelzer2013}, with the exception of TWA 14 (M1.9 here versus M0.5 in Stelzer et al.).
The outdated spectral types from \citet{Webb1999} have led to some confusion regarding membership. 
\citet{Weinberger2013} discuss that space motions may suggest that TWA 9A and 9B are not members of the TWA, which they support with ages of 63 and 150 Myr, respectively.  The later SpT measured here lead to younger ages estimates that are consistent with the $\sim 10$ Myr age of the TWA.  

In some cases, the age of a star as measured from an HR diagram may differ from the dynamical or global age of a cluster. For example, 
with their later spectral type, \citet{Vacca2011}  argue that the age of TW Hya is $\sim 3$ Myr.
Our age is now consistent with the global age of the TWA.  However, even if the estimated age of a single star were younger, the dynamical age and cluster age are both consistent with 7--10 Myr \citep[e.g.][]{Mamajek2005,Ducourant2014}.   Any deviations from this age for confirmed members are likely due to real scatter in observed photospheric luminosities and temperatures rather than the actual age of the star.  These uncertainties are discussed in more detail in \S 5.

\begin{table}
\caption{New TWA Spectral Types}
\label{tab:twaspt}
\begin{tabular}{ccc}
\hline
Star$^1$ & This Work & Webb  et al.~1999 \\
\hline
TWA 1 & M0.5 & K7 \\
TWA 2AB & M2.2 & M0.5(+M2)\\
TWA 3A & M4.1 & M3\\
TWA 3B & M4.0 & M3.5\\
TWA 4AabBab & K6  & K5\\
TWA 5AB & M2.7 &M1.5(+M8.5) \\
TWA 6 & M0.0 & K7 \\
TWA 7 & M3.2 &M1\\
TWA 8A &M2.9&M2 \\
TWA 8B &M5.2&M5\\
TWA 9A &K6 & K5 \\   
TWA 9B &M3.5 &M1\\
\hline
\multicolumn{3}{l}{$^1$Unresolved binaries listed as single combined SpT}\\
\multicolumn{3}{l}{TWA 2 and TWA 5 unresolved here and in Webb et al.}\\
\end{tabular}
\end{table}

\subsection{Comparison of Extinctions to Previous Measurements}

In this subsection, we compare our extinctions to literature extinctions.  Our uncertainties are repeatable to $\sim 0.1$ mag., when multiple spectra of the same star are analyzed assuming a constant spectral type.  Including uncertainty from spectral type and gravity, our extinctions should be reliable to $\sim 0.2-0.3$ mag.
 Literature uncertainties are commonly claimed to be $\sim 0.2-1.0$ mag., although statistical errors on the lower end of this range are based on photometric accuracy and are typically not realistic.  The primary sources of extinction errors are caused by uncertainty in spectral type, gravity mismatches between the target star and a template for M-dwarfs, and the estimates for the shape and strength of the accretion continuum.

\begin{figure*}[!th]
\epsscale{1.}
\plottwo{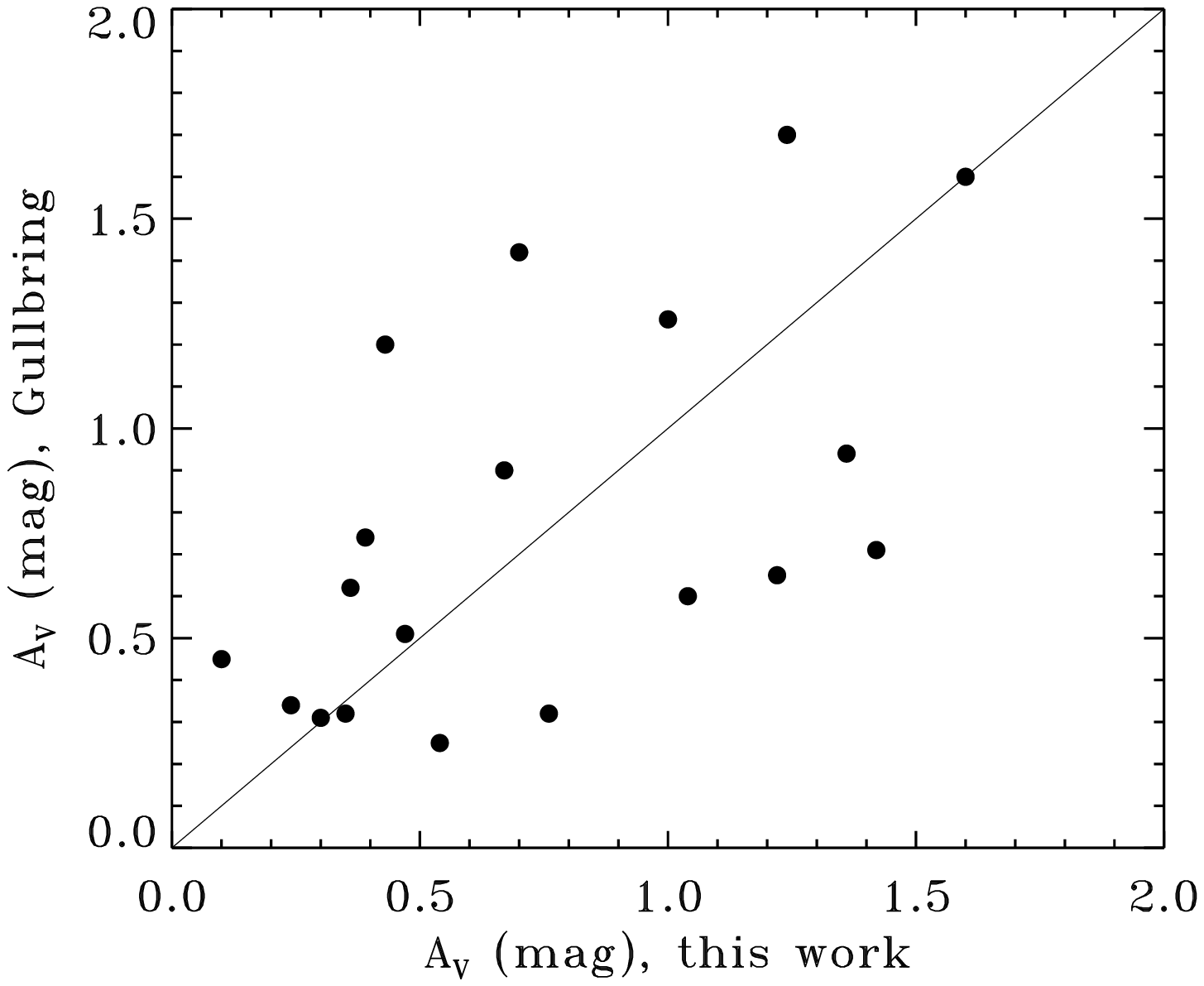}{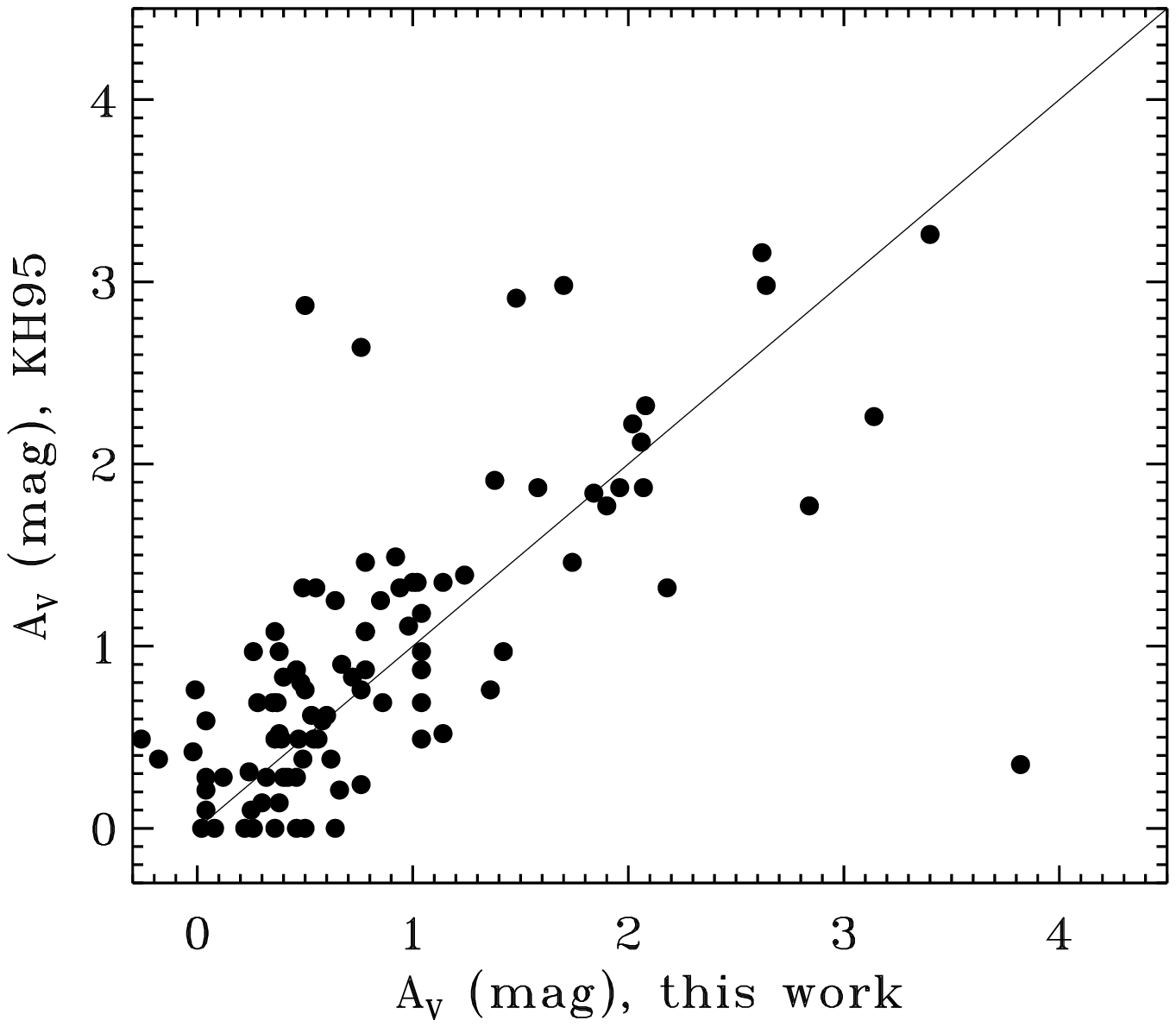}
\plottwo{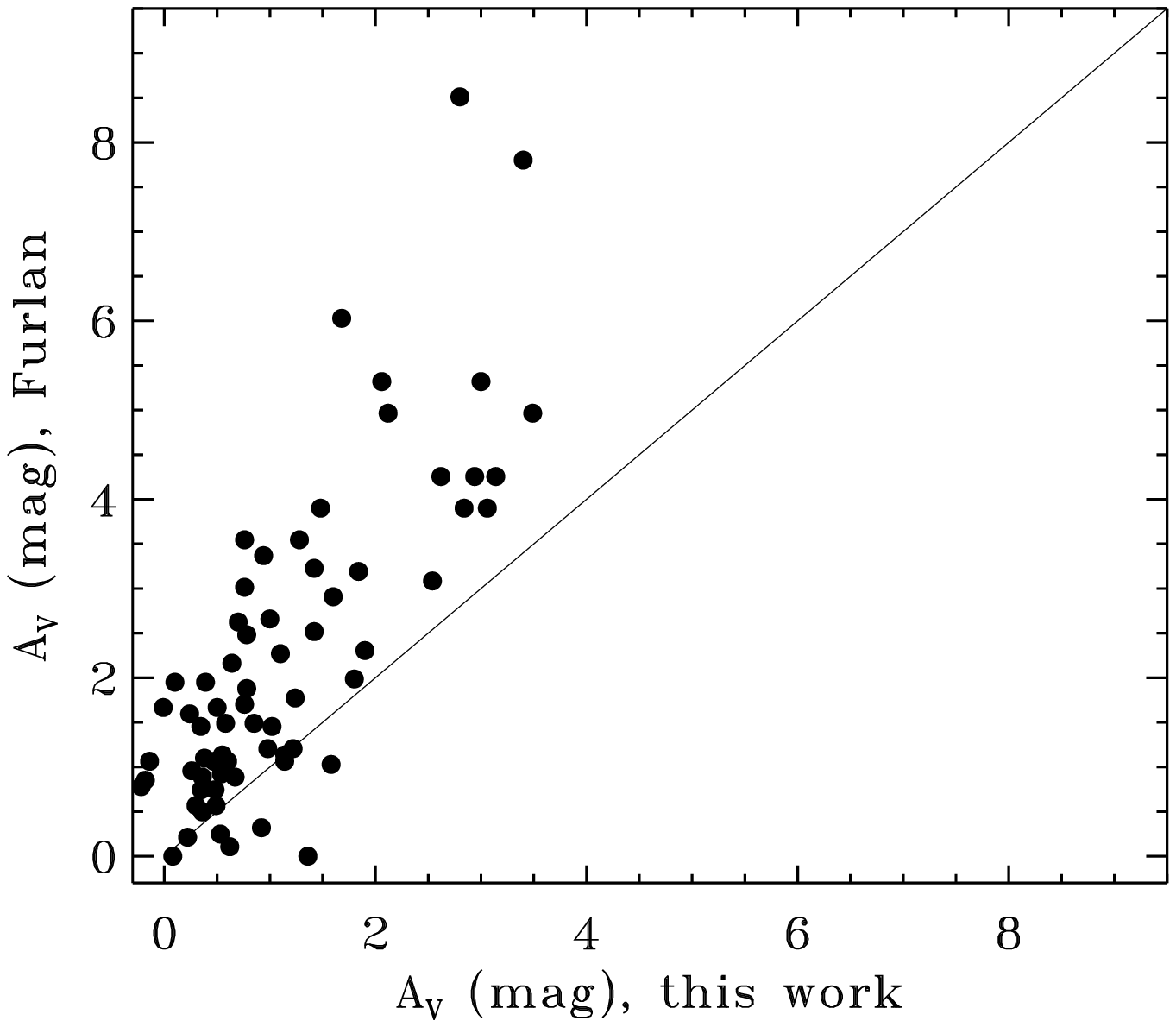}{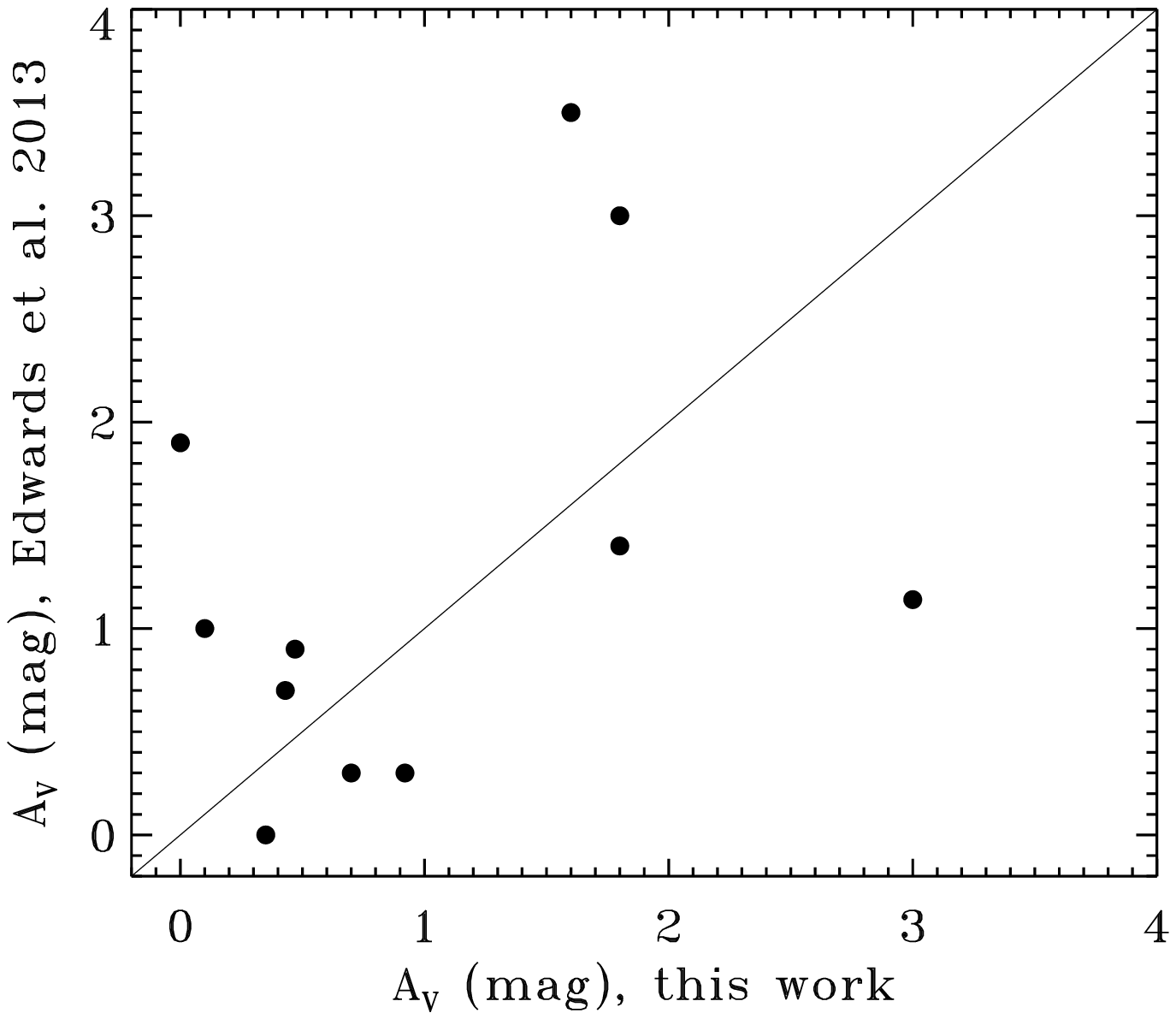}
\caption{A comparison of extinctions between this work and selected works in the literature.  Our extinctions are typically consistent with other optical extinction measurements (top left and right) but are systematically lower than near-IR extinction measurements based on photospheric colors (bottom left).  Comparisons with extinctions from H line flux ratios (bottom right) shows systematic problems for heavily veiled stars, but this new method may provide extinctions independent of photospheric colors.}
\label{fig:avcomp.ps}
\end{figure*}

Figure \ref{fig:avcomp.ps} shows the comparison of our extinctions with those from several different literature sources.  In general, our extinctions agree with literature estimates from optical extinction estimates, but discrepancies with near-IR extinction estimates are large and systematic. 

Gullbring et al.~(1998,2000) used optical photometry to measure extinctions.  The mean difference between our measurements and Gullbring et al. is 0.04 mag., with a standard deviation of 0.37 mag.  \citet{Kenyon1995} typically use V-R and V-I colors to measure extinctions for the bright stars that dominate overlap between our and their sample, with a difference of 0.1 mag. and a standard deviation of 0.7 mag. when compared to our results.  Given the uncertainties in their and our results, our extinctions are typically consistent with those of Gullbring and KH95.  The mean difference and scatter of our extinctions relative to Luhman extinctions are 0.10 and 0.93 mag., respectively.  

On the other hand, near-IR analyses yield significantly higher extinctions than those measured at optical wavelengths.  The large \citet{Furlan2011} survey of Spitzer IRS spectra included updated extinctions based on fitting photospheric and disk emission to JHK photometry and, in some cases, near-IR spectroscopy.  Their extinctions are typically 1.1 mag.~larger than the $A_V$ measured here, with 1.2 mag.~of scatter after accounting for that bulk shift.  
  Among the sources common to both studies, 5\% of extinctions are different by $>3$ mag., with large discrepancies especially common for strong accretors.   These differences are likely caused by the large near-IR excess associated with gas and dust in the inner disk, which can also be characterized by veiling \citep[e.g.][]{Meyer1997,Folha1999}.   Indeed, the J--H extinctions from the \citet{White2004} sample of evolutionarily young stars, which did not account for veiling, are an average of 3.6 mag.~larger than those measured here.  Large differences between optical and near-IR extinctions were also noted by \citet{McJunkin2013}.

\citet{Fischer2011} and \citet{McClure2013} developed a more robust and more painful approach to measure extinctions from flux calibrated near-IR photospheric spectra, after measuring and subtracting the veiling.  The veiling in the near-IR is caused by a combination of accretion and warm disk emission.  In both studies, the veiling is measured and subtracted off the observed spectrum.  Extinction is then measured by comparing the remaining photospheric spectrum to a standard star.  However, our extinctions are 1.3 mag lower than those of \citet{Fischer2011}, with a scatter of 1.1 mag., possibly because of differences in WTTSs templates.$^5$  \citet{McClure2013} obtain results closer to ours, with a bulk offset of $0.7$ mag and a scatter of 0.7 mag.  Our extinctions are actually larger than the near-IR extinctions of \citet{McClure2013} in 3 of the 9 stars in their sample.  These differences may be related to uncertainties in the near-IR colors of CTTSs and WTTSs, and the lower sensitivity of near-IR spectra to extinction.
\footnotetext[5]{This difference may be attributed to the lack of a sufficient grid of near-IR WTTS templates.  One of their two templates, V819 Tau was assigned $A_V=2.6$ mag (compared with 1.1 mag.~here) based on a comparison with a main sequence star.  Their other template, LkCa 14, was assigned an M0 spectral type (compared with K5 here).  \citet{Gullbring1998} also found anomalous near-IR colors for V819 Tau.}

Extinctions calculated from line flux ratios could in principle lead to more accurate measurements than photospheric-based extinctions, if the lines are optically thin or other easily modeled and have significantly different wavelengths.  \citet{Edwards2013} developed near-IR H Paschen and Brackett line fluxes as an extinction diagnostic.  The H line emission is usually dominated by the accretion flow and should usually have the same line of sight as the stellar photosphere.
Our extinctions are 0.27 mag. smaller than theirs with a standard deviation of 1.1 mag.  However, the agreement improves (0.6 mag.) when restricted to the 4 stars in both studies that do not have high veiling and powerful outflows.  The heavily veiled outflow sources have larger extinction uncertainties in this work and may have H line emission with significant outflow contributions.

While we consider optical extinctions more reliable than those in the near-IR, they are inappropriate to use when optical emission from a star is seen primarily scattered light.  This criteria applies especially to stars with disks viewed edge-on or stars with remnant envelopes.  Some systems like Sz 102 or HL Tau have very low $A_V$ measurements but are much fainter than would be expected for a Taurus TTS with their SpT.  In these cases, the extinction estimates likely require full SED modelling and in any case may not be relevant for interpreting the observed optical or near-IR emission from the star.

The extinction calculations presented here are more accurate than previous measurements for stars in our sample earlier than M5.  When veiling is negligible, photometry combined with a reliable spectral type and a template with similar gravity \citep{Pecaut2013} may yield a more reliable extinction than flux calibrated spectra.  Red or near-IR colors may be preferable to measure extinction to TTSs later than M5 because the optical emission is on the Wien tail of the blackbody distribution and changes quickly as a function of temperature.

\section{Observational Uncertainties in Stellar Properties and Cluster Luminosity Spreads}

Improvements in spectral types and extinctions lead to a more accurate placement on HR diagrams.  Whenever young stellar clusters have been placed on HR diagrams, a large luminosity spread is measured at a given spectral type
\citep[see reviews by][]{Hillenbrand2008,Preibisch2012}.  The observational contribution to luminosity spreads is typically estimated by creating a synthetic cluster of stars with temperatures and luminosities scattered by an amount consistent with the estimated uncertainties.  In many cases, the entire spread of luminosities may be explained by observational errors \citep{Hartmann1998,Slesnick2006,Preibisch2012}.  On the other hand,
\citet{Reggiani2011} found that the luminosity spread in {\it HST} optical photometry of the ONC could not be replicated with purely observational errors.
The observational uncertainties in stellar properties, and the uncertainties in the uncertainties, limit our ability to test pre-main sequence evolutionary tracks, the effect of accretion histories, and the timescale over which star formation occurs within a cluster.

In this section, we describe how the observational uncertainties in spectral type, extinction, and veiling measured in this paper relate to luminosity spreads.   Listed uncertainties refer to $\sim 1\sigma$ error bars, although these measurements are not always rigorous.  This description does not include some of the most important uncertainties: multiplicity, partial disk obscuration of the star, cluster membership (see \S 6.2 for a discussion), and stellar spots.  In \S 5.5, we present results of improved stellar parameters on the HR diagrams for the TWA and MBM 12.

\subsection{Direct Luminosity Uncertainties from Distance, Flux Calibration, and Extinction}
The approximate uncertainty in distance is $\sim 10\%$ to any given star in the Taurus Molecular Cloud, which leads directly to a 20\% uncertainty in luminosity.  The depth of the Taurus cloud in our line of sight is likely large compared with the median distance, so the percentage uncertainty in distance is large.  Many of the TWA objects have parallax distances with $<5$\%  uncertainties.  On the other hand, large systematic uncertainties plague the absolute, but not the relative, photometric distance to the MBM12 Association.

The absolute flux calibration, here $\sim 10$\%, leads directly to the same 10\% uncertainty in luminosity.  The relative flux calibration also leads to an uncertainty in the extinction, in this work about 0.1 mag.~in $A_V$.

Typical extinction uncertainties in $A_V$ are $\sim 0.2$ mag. (or $\sim0.4$ mag.~when veiling is significant), which here leads to a 13\% (28\%) luminosity uncertainty from the 7510 \AA\ photospheric  flux.

\begin{figure}[!t]
\epsscale{1.}
\plotone{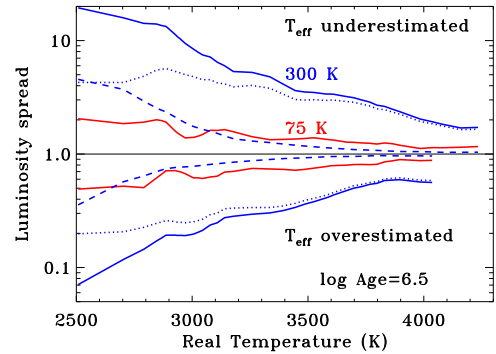}
\caption{The effect of spectral type and temperature uncertainty on luminosity spread.  Stars that have effective temperatures overestimated (or underestimated) will appear underluminous (or overluminous) in an HR diagram because (a) the expected luminosity at that higher temperature is higher than the expected luminosity at the real temperature, and (b) the bolometric correction is smaller at higher temperatures.  The combined curves (solid lines) show the ratio of the measured luminosity and the expected luminosity for a 3.2 Myr old star with temperature overestimated by 75 K or $\sim 0.5$ subclasses (red) and 300 K or $\sim 2$ subclasses (blue), based on the \citet{Baraffe2003} pre-main sequence tracks.  The individual components (a, dotted lines) and (b, dashed lines) are shown for the 300 K error.}
\label{fig:ages1.ps}
\end{figure}

\subsection{Methodological Uncertainties}

These uncertainties are introduced in our approach to fitting spectral type, accretion continuum flux, and extinction simultaneously in each spectrum from a grid of standard WTTSs.  The variables are correlated, so changing the spectral type or accretion continuum flux also lead to changes in the extinction.

\subsubsection{Veiling and Stellar Luminosity}

Veiling of the photospheric emission by the accretion continuum and any disk emission increases the observed flux.  If the accretion continuum flux is not subtracted, then the measured flux will overestimate the photospheric flux.  In this work, the photospheric luminosity is always corrected for veiling and as a result are usually lower than previous estimates.  
Uncertainties of $\sim$ 20\% in the strength of the accretion continuum typically lead to $\sim 5$\% uncertainties in the final luminosity, with larger uncertainties for higher veiling.  This error can be assessed for each target by comparing the veiling to the 7510 \AA\ photospheric flux (see Appendix C and  Table~\ref{tab:props.tab}).  
Failure to subtract the accretion continuum off from the measured flux will lead to systematically overestimating the stellar luminosity in a way that correlates with veiling and accretion.

\subsubsection{Spectral Type and Luminosity}
We assess internal SpT uncertainties of 0.2 subclasses for M-dwarfs, 0.5 subclasses between K8--M0.5, and 1 subclass for stars between G0 and K8.  The uncertainty for M-dwarfs relative to literature estimates is $\sim 0.5$ subclasses.  The spectral types are repeatable to those levels of precision for stars with multiple spectra.  These spectral types are optimized with a quantified inclusion of both the accretion continuum flux and reddening, and should have smaller uncertainties than spectral types obtained by eyeball comparisons of spectra to a grid of standard spectra.  

Veiling affects spectral type measurements.
Our largest change in spectral type is $4.4$ subclasses (for FM Tau), though errors larger than 2 subclasses would be surprising for the type of red spectra that have been obtained in most studies over the past decade.  However, differences of 2 subclasses for veiled spectra can be subtle, as demonstrated for UScoCTIO 33 and DM Tau above.

For an error in spectral type, the bolometric correction and therefore the estimated luminosity of the star does not change significantly, unless the photospheric flux is measured in the Wien tail for the relevant temperature.
However the stellar luminosity at a given age is sensitive to the effective temperature.  The comparison between the expected and estimated luminosity thereby introduces a luminosity spread.  Figure~\ref{fig:ages1.ps} shows that 3500 K stars with temperatures overestimated by 300 K and 75 K ($\sim 1.5$ and 0.2 SpT, respectively) would appear underluminous by a factor of $\sim 2.8$ and $\sim 1.3$, respectively.  This uncertainty may lead to veiled stars, such as UScoCTIO 33, systematically appearing underluminous and old if veiling is not accounted for in the SpT.  Some of this underluminosity may be balanced if veiling is also mistakenly unaccounted when converting a band flux to the luminosity.

\begin{figure}[!t]
\plotone{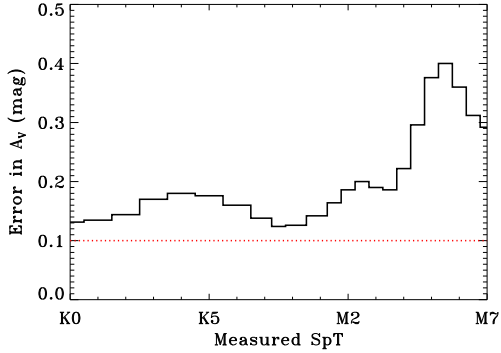}
\caption{Approximate uncertainty in $A_V$ introduced by spectral type uncertainty (solid line) and errors in the relative flux calibration (dotted line).  The spectral type uncertainty used here is 1 subclass at spectral types earlier than K7, 0.5 subclasses between K7--M1, and 0.3 subclasses for stars later than M1.  An incorrect spectral type will lead to an incorrect extinction because the comparison template has a different photospheric temperature and spectral shape.}
\label{fig:exterrors}
\end{figure}

\subsubsection{Spectral Type and Extinction}
 An error in the spectral type introduces a mismatch between the template and object spectra, thereby causing an error in the extinction.  Fig.~\ref{fig:exterrors} shows the uncertainty in $A_V$ introduced by the average spectral typing error.
In optical spectra or photometry, the uncertainty in spectral type dominates extinction uncertainties beyond~$\sim$M5 because the spectral slope changes sharply in the Wien tail.  In terms of the luminosity spread, some of the uncertainty introduced by a spectral type error may be partially balanced by the extinction error, which pushes the expected luminosity in the other direction \citep[e.g.][]{daRio2010}.

\subsubsection{M star gravity mismatches and extinction}
\begin{figure}
\epsscale{1.}
\plotone{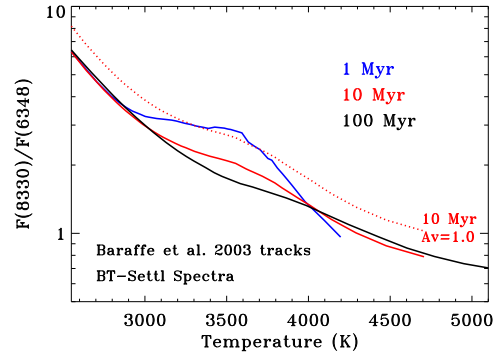}
\caption{The color dependence, here as $F_{red}=F(8330)/F(6348)$, for stars with gravity for 1, 10, and 100 Myr old stars.  The different colors change the SpT-effective temperature relation \citep[e.g.][]{Forestini1994} and extinction estimates -- even when comparing 1 Myr and 10 Myr old stars.  The gravities are obtained from the PMS tracks of \citet{Baraffe1998} and the colors are calculated from the BT-Settl synthetic spectra \citep{Allard2012}.}
\label{fig:gravity.ps}
\end{figure}

The gravity of the spectral template can also introduce significant errors, even when non-accreting T Tauri stars are used as photospheric templates.  Low mass stars with ages of 1--10 Myr have $\log g\sim 3.4-4.2$, depending on the age and mass (see Fig.~\ref{fig:grav1}).  The gravity difference between $\log g\sim 3.4-4.2$ (1--10 Myr for 0.7 $M_\odot$ star) leads to maximum errors of $A_V$=1.1 in $F_{\rm red}=F(8330)/F(6448)$ for M stars (Figure~\ref{fig:gravity.ps}).

\subsection{Assumed Standard Relations}

The uncertainties in extinction law and the shape of the accretion continuum are errors that apply systemically to stellar temperature and luminosity measurements.  We briefly desribe the effects of errors in these assumptions.

\subsubsection{Extinction Law}  The extinction law is assumed to be that of median interstellar grains, with a total-to-selective extinction of $R_V=3.1$.  Most spectra in our sample can be well fit with $R_V=3.1$ and have $A_V<3$, where the mean interstellar extinction law should apply.  
 The differences in the relative flux attenuation between extinction laws is particularly significant at $<5000$ \AA.
Grain growth in high extinction regions makes the extinction curve much more gray, with $R_V$ as high as 5.5 \citep{Indebetouw2005}.  The few stars in our sample that are heavily extincted ($A_V>5$) are only well fit with $R_V\geq4$.

In our optical spectra, for a star with a measured $A_V=1.0$, applying extinction laws with $R_V=5$ from \citet{Cardelli1989} or $R_V=5.1$ from \citet{Weingartner2001} would lead to $A_V=1.2$ and $1.15$, respectively.  The difference in luminosity is minimal for low extinctions.  However, an extinction $A_V=10$ and $R_V=5$ will be measured (in red-optical spectra or colors) as $A_V=8.3$ if $R_V=3.1$ is assumed, yielding a factor of 5.6 difference in luminosity if assessed at 7510 \AA.

\begin{table}
\caption{SpT, $A_V$ and the accretion continuum slope}
\label{tab:veilprob}
\begin{tabular}{ccccccc}
\hline
Star$^a$ &  \multicolumn{2}{c}{Red Slope} & \multicolumn{2}{c}{Flat Slope}$^b$ & \multicolumn{2}{c}{Blue Slope}\\
 & SpT & $A_V$ & SpT & $A_V$ & SpT & $A_V$\\
\hline 
UScoCTIO 33 & M4.6 &0.06 & M4.4 & 0.38 & M4.3 & 0.52\\
DF Tau & M3.4 &  0.20 & M2.7 & 0.18 & M2.5 & 0.18\\
DM Tau & M3.4 & 0.12 & M3.0 & 0.08 & M2.8 & 0.08\\
DP Tau &  M1.7 & 0.60 & M1.0 & 0.68 & M0.3 & 0.90\\
DR Tau & (K6) & 0.56 & (K6) & 0.50 & (K6) & 0.46 \\
GM Aur & K8.5 & 0.14 & K6.5 & 0.36 & K6.5 & 0.40\\
TW Hya & M1.1 & 0.0 & M0.9 & 0.08 & M0.7 & 0.12\\
ZZ Tau$^d$ & M4.4 & 0.54 & M4.4 & 0.56 & M4.3 & 0.58\\
\hline
\multicolumn{7}{l}{$F(2\lambda)=2F(\lambda)$ and $0.5 F(\lambda)$ for red and blue slopes.}\\
\multicolumn{7}{l}{$^a$All observations except UScoCTIO 33 from 29 Dec.~2008.}\\
\multicolumn{7}{l}{$^b$Results may differ slightly from best fits using all dates.}\\
\multicolumn{7}{l}{$^c$High veiling, so K6 SpT assumed for DR Tau}\\
\multicolumn{7}{l}{$^d$Example of little change because of weak veiling}\\
\end{tabular}
\end{table}

\subsubsection{Uncertain shape of accretion continuum}
Our analysis relies on an assumption that the accretion continuum flux is constant in erg cm$^{-2}$ s$^{-1}$ \AA$^{-1}$ versus wavelength.  While this assumption is reasonable, it may not apply to some sources (see Fig.~\ref{fig:veil1.ps}).  A negative slope (stronger emission at 4000 \AA\ than 8000 \AA) would lead to the inference of an earlier spectral types because the veiling would be weaker, so the photospheric TiO features would not be as deep.  For sources with moderate or strong veiling, the extinction would be underestimated in our paper.  

DM Tau is used here as an example of the effect of the shape of the veiling continuum for a moderately veiled star.   If the accretion continuum is two times weaker at 8000 \AA\ than at 4000 \AA, then the best fit model  is M2.7 with $A_V=-0.02$ mag.  If instead the accretion continuum is two times stronger at 8000 \AA\ than at 4000 \AA, then the best-fit model is M3.6 and $A_V=0.2$ mag.  The different spectral types are caused by different TiO absorption depths.  The extinction does not change significantly because the color change in the accretion continuum is offset by a color change in spectral type.  

Table~\ref{tab:veilprob} describes a similar analysis for a few stars with a range of veilings.  The synthetic spectra with an accretion continuum that get brighter to longer wavelengths are typically bad fits to the observed spectra.
For heavily veiled stars, the change in extinction could be as large as $A_V=0.9$ mag.  Spectral types of heavily veiled stars around K7 are especially dependent on the shape of the veiling continuum.
 The real uncertainty in SpT and $A_V$ are likely smaller than the differences described in this paragraph because the accretion continuum is likely much closer to a flat spectrum at optical wavelengths.

\subsubsection{Conversion from Spectral Type to Temperature}

The temperature scale for pre-main sequence likely has an uncertainty of $\sim 50$ K for early M-dwarfs and $\sim 100$ K for late M-dwarfs based on the comparisons between different temperature scales described in \S 3.2.1.  
In addition to these uncertainties, the models themselves have some uncertainty.  Any error in the temperature scale will apply systematically throughout the entire sample and does not introduce a luminosity spread for stars with similar spectral types.
The conversion from spectral type to temperature applied here is measured from our spectral type scale.

\subsubsection{Bolometric Corrections}

Our bolometric corrections were calculated from the BT-Settl model spectra \citep{Allard2012}.  The mismatch between models and real spectra may introduce small systematic errors into our luminosity calculations.  As with the conversion from spectral type to temperature, this error should not introduce a significant luminosity spread for stars with similar spectral types.

\subsection{Stellar Properties of Heavily Veiled Stars}

Some stars are so heavily veiled that the veil is similar to a burqa, almostly complete hiding the photosphere.  These stars pose particularly difficult problems for HR diagrams.   In this section we highlight two problems that may preferentially affect measuring photospheric emission.

\begin{figure*}
\epsscale{1.}
\plottwo{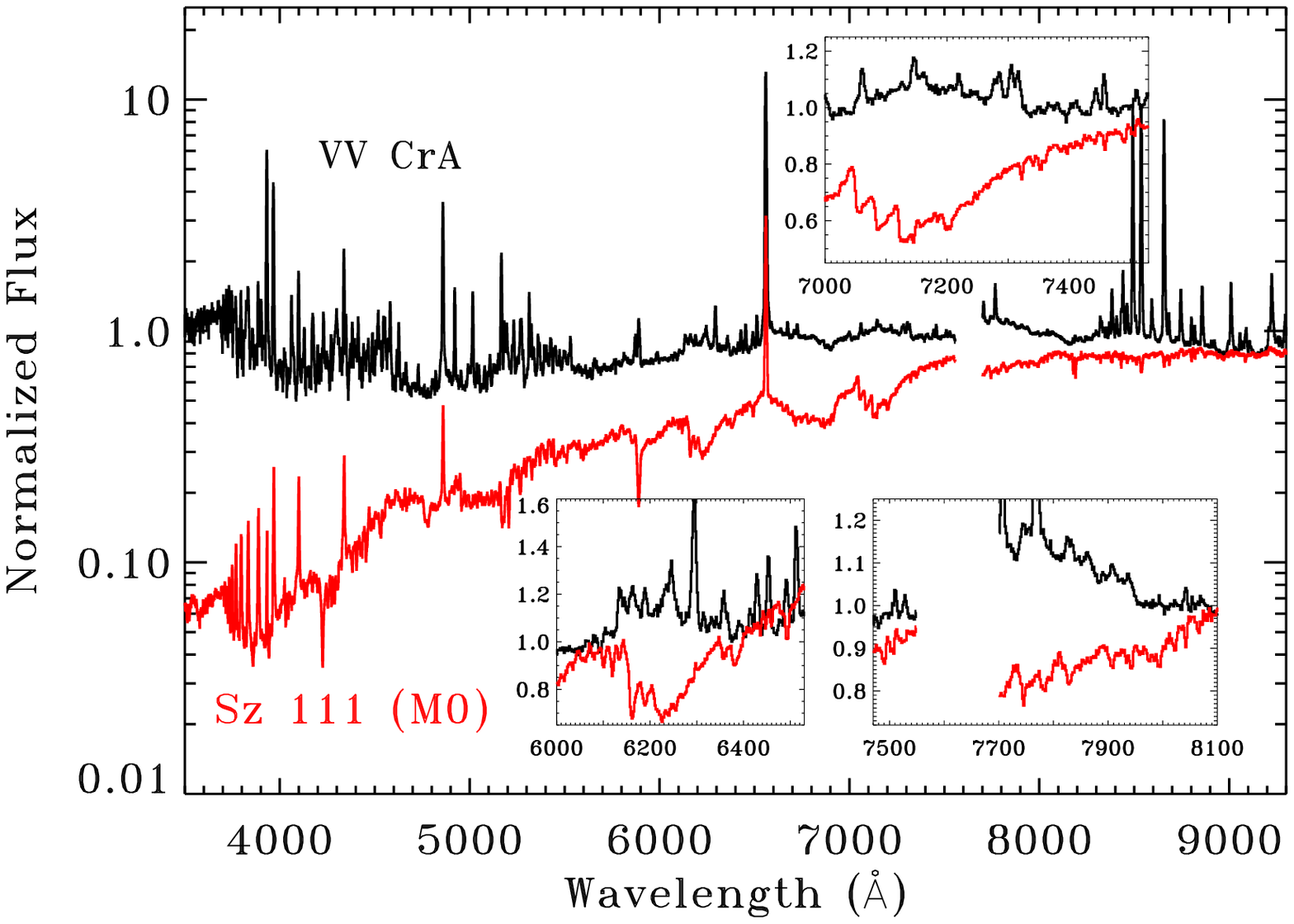}{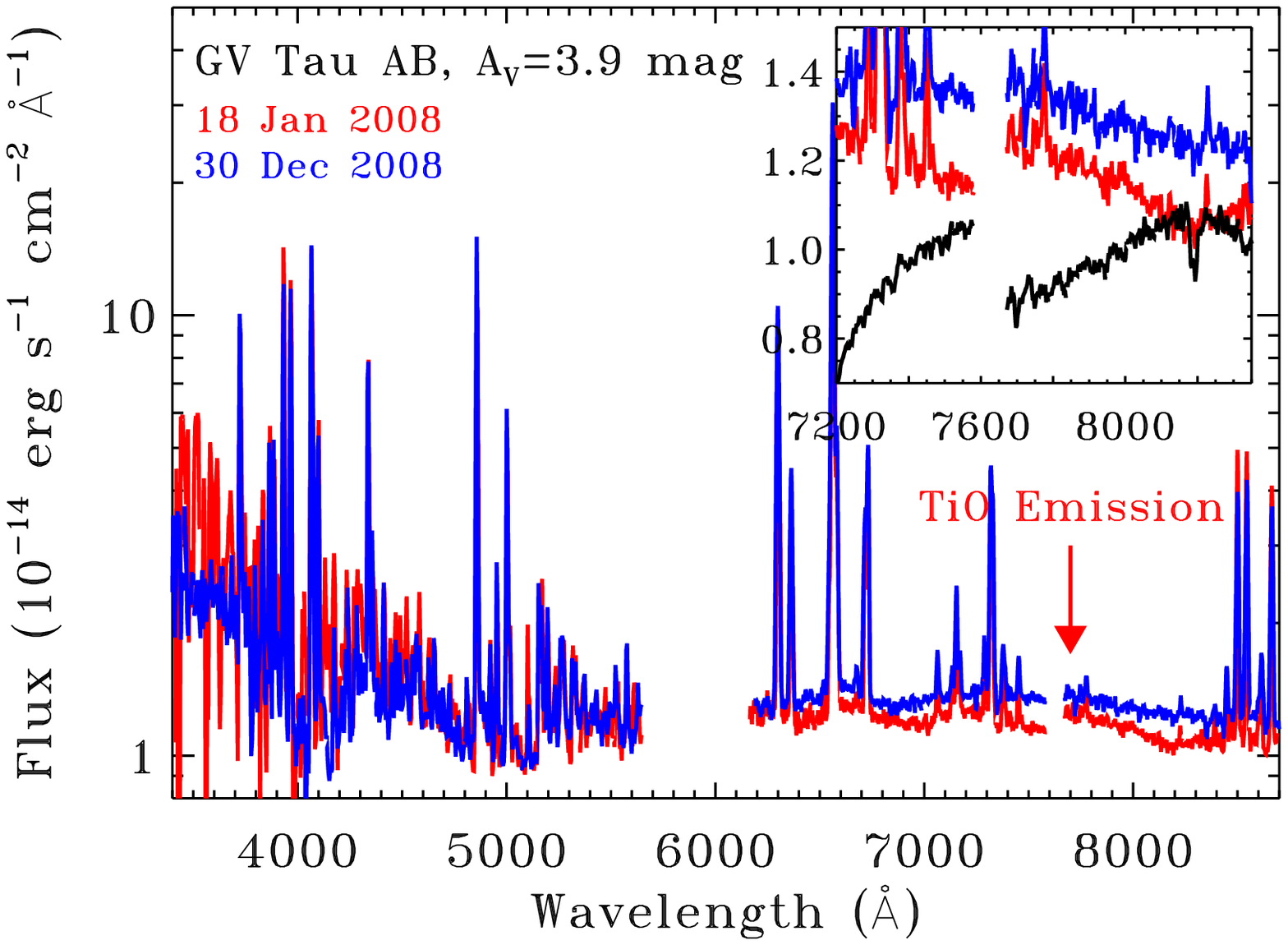}
\caption{TiO emission from extinction corrected spectra of VV CrA (left) and GV Tau AB (right).  TiO emission is detected at the same location as TiO absorption in the M0 star Sz 111 (left, red line) and M1 star TWA 13A (black line in inset panel).  For GV Tau, only the Jan.~2008 observation clearly shows TiO in emission.  The blue emission lines are also stronger in the Jan.~2008 observation than in Dec.~2008.}
\label{fig:vvcra.ps}
\end{figure*}

\subsubsection{Photospheric Emission of Heavily Veiled Stars}
The age of a pre-main sequence star is calculated from the contraction timescale and the effective temperature.  In most comparisons between data and model spectra, the photospheric luminosity is used as a proxy for the radius.  This surface area does not include the fraction of the star covered by the spot.  In the shock models of Gullbring et al. (1998), corrections are less than $\sim 20$\% and are not significant.  However, the shock models of \citet{Ingleby2013} include components at lower density than Gullbring et al.~in order to explain veiling at red wavelengths, in excess of previous models.  In three cases RW Aur A, DR Tau, and CV Cha, the accretion shock covers $20-40$\% of the stellar surface.  In extreme cases, especially for outbursts or class I objects, some estimate of this covering fraction would need to be combined with the photospheric surface area to calculate a stellar radius.  This uncertainty is ignored in this work.

\subsubsection{TiO in Emission and Spectral Types}
\citet{Hillenbrand2012} found TiO in emission from two Class I sources and one CTTS undergoing an outburst \citep[see also][]{Covey2011}.  In our sample, VV CrA and the Jan.~2008 GV Tau$^6$ spectrum show TiO in emission (Fig.~\ref{fig:vvcra.ps}).  Emission lines blanket the optical spectra of VV CrA, GV Tau, and three objects described by \citet{Hillenbrand2012}.  All objects with TiO emission have evidence from their spectral energy distributions that an envelope is present.
\footnotetext[6]{
The Dec.~2008 observation of GV Tau does not show obvious TiO emission, although this emission may be masked by additional red continuum emission.  The variability may be real or attributable to different slit positions and seeing.  GV Tau is the one source in our sample that is clearly extended in emission lines beyond what would be expected for a $1\farcs2$ binary, even in poor seeing.}

The TiO emission must be related in some way to strong accretion.  The warm TiO gas is likely located in a warm disk surface, which may be viscously heated by the accretion flow.  TiO emission has only been detected in clear cases.  Presumably other CTTSs have weak TiO emission that would require a dedicated search to detect.  The possible complications of TiO emission filling in absorption bands has not been considered here or in other work, but would severely complicate spectral typing.  Most likely, these complications arise for only class I stars.

\subsection{Luminosity Spreads of Loose Associations}

Young stars start to grow out of adolescence when the velocity dispersion of their parent cluster leads them to venture far from their birthplace.  At this stage, they are in loose associations with their sublings and are typically free of extinction.
In this subsection, we apply the previous description of luminosity spreads to stars in two nearby associations, the TW Hya Association and the MBM 12 Association.  Both associations are relatively small, with 10--30 known members.  The luminosity spread of Taurus is not discussed here because our sample is incomplete and biased and because Taurus includes many subclusters with large age differences.

\subsubsection{Luminosity Spread of the TW Hya Association}

The nearby TW Hya Association (TWA) is a loose association of $\sim 30$ stars with an age of $\sim 10$ Myr \citep{Webb1999}.  The members are especially meaningful for age estimates because of proximity, prevalence of parallax measurements, negligible extinction, and near-complete accounting of binarity.  

Since the association is not near any molecular cloud, extinction is assumed to be 0 for most members.  Two stars, TWA 30A and 30B, have disks that are nearly edge-on and may attenuate photospheric emission (Looper et al.~2010ab) and are therefore ignored in this analysis.  Although \citet{Ducourant2014} calculate and apply extinction corrections to several TWA members, the color corrections may be introduced by small errors in spectral type and at present are not accurate enough to use anything other than $A_V=0$ for members of this association.

\begin{figure*}
\epsscale{1.}
\plottwo{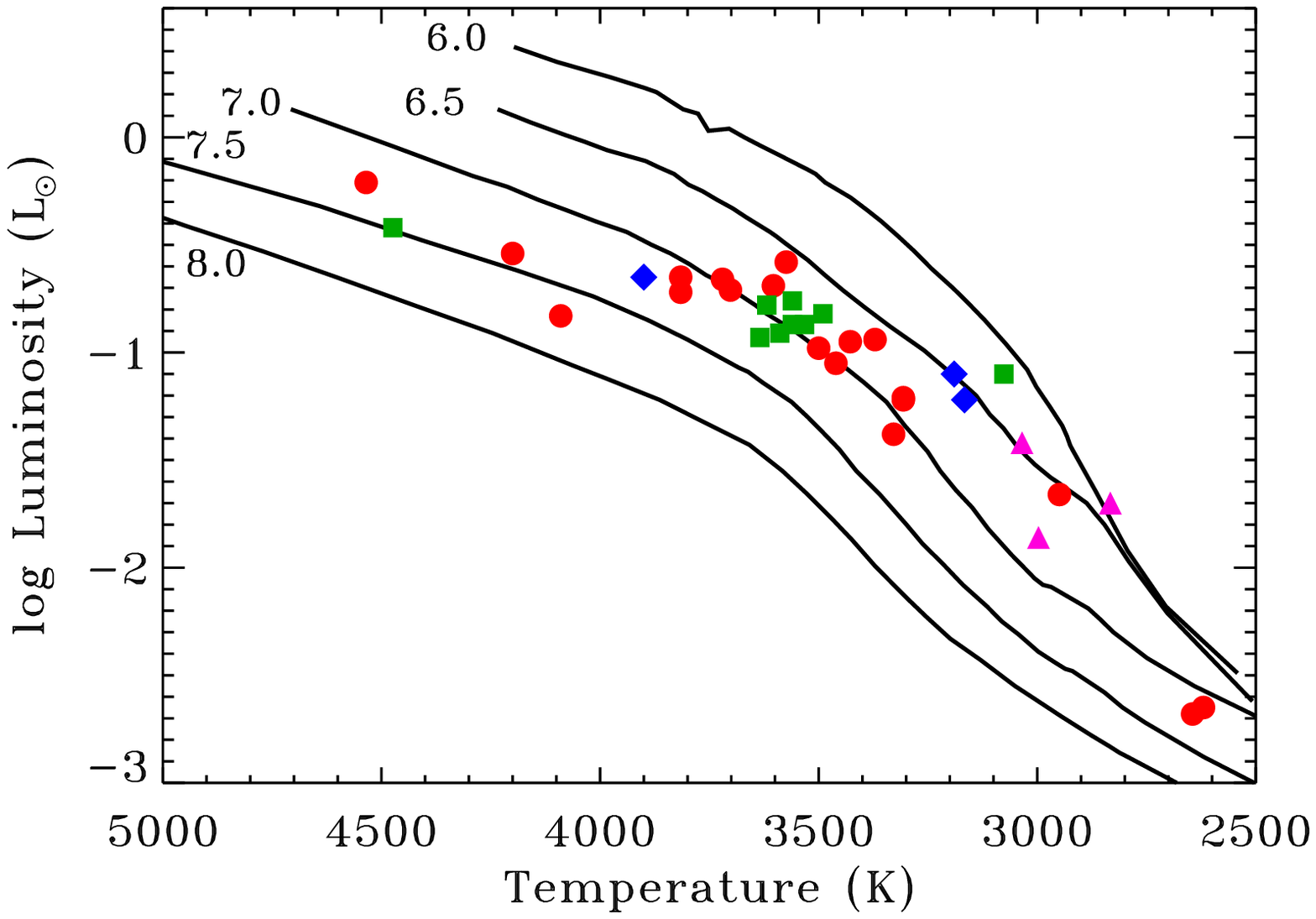}{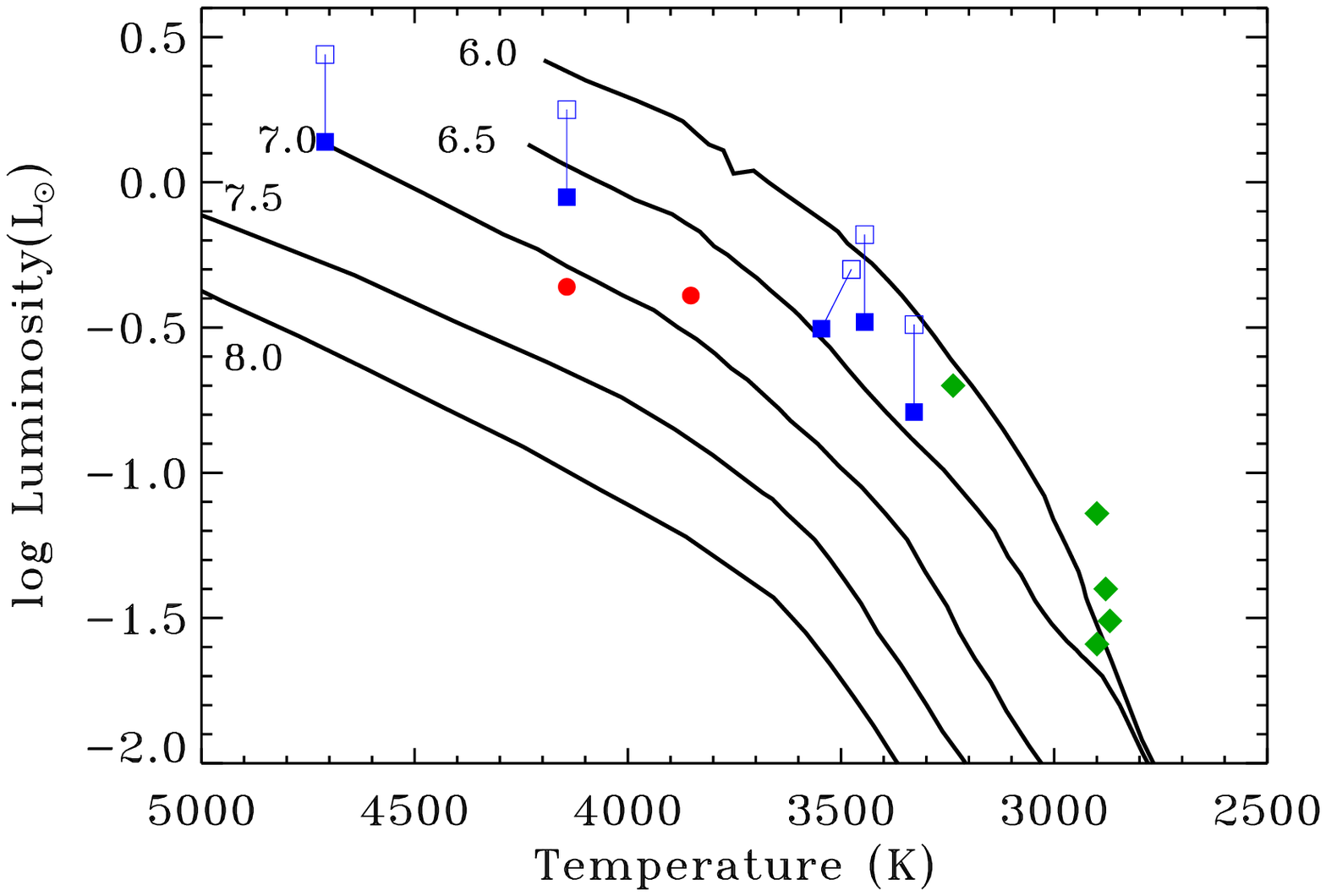}
\caption{HR diagram of known members of the TW Hya Association later than K0 (left) and in the MBM 12 Association (right).  For the TWA, red circles are stars with spectral types measured here and parallax distance measurements, blue diamonds are stars with spectral types measured here and dynamical distance measurements, green squares are literature spectral types and parallax distance measurements, and purple trianges are literature spectral types with dynamical distance measurements. For the MBM 12 Association, the members shown here are single stars (red circles), estimates of the primary properties for binaries (solid squares, empty squares mark the initial position), and stars without known binary properties (green diamonds).   Isochrones in both plots are shown at 0.5 dex intervals (as labeled in $\log$ Age in yr) from the \citet{Baraffe2003} pre-main sequence tracks. }
\label{fig:hrtwa}
\end{figure*}

Figure~\ref{fig:hrtwa} shows the HR diagram and luminosity spread of the TWA.  Table ~\ref{tab:twaages} lists the stellar parameters of TWA members, including an age from the \citet{Baraffe2003} pre-main sequence evolutionary tracks and a ratio $L/L10$ of the measured luminosity to $L10$, the 10 Myr luminosity for the relevant temperature.  Most of the luminosities are calculated from our optical spectra.  The luminosities of several close binaries are obtained from {\it HST} narrow-band imaging at 1.64 $\mu$m \citet{Weintraub2000}, while others are obtained from the J-band magnitude measured in 2MASS.  All total fluxes are calculated using bolometric corrections calculated from the BT-Settl spectra.  The binary systems TWA 3Aab, TWA 5Aab, TWA 16AB, TWA 23 AB, and TWA 32AB are roughly equal mass \citep{Muzerolle2000,Zuckerman2001,Shkolnik2011,Weinberger2013}, so the luminosities used for this analysis are divided by 2.  For HD 98800 Aa, Ba, and Bb, we used the temperatures calculated by \citet{Laskar2009}.  The luminosities of HD 98800 Bab are calculated from the absolute K-band magnitudes of \citet{Boden2005}, adjusted for distance.  The luminosity of HD 98800 Aa is calculated from the K-band flux in \citet{Prato2001}.

The combined TWA 2AB spectrum is classified here as an M2.2 object.  The color difference suggests a 1.5--2 subclass difference between the pair.  To determine the spectral type of the primary, we added scaled template spectra together and subsequently calculated a new spectral type and extinction.
As the secondary-to-primary mass ratio decreases, the secondary reduces the inferred temperature of the total system, thereby also decreasing the 
expected luminosity.  With complete optical wavelength coverage, the resulting spectral type is typically 0.5 subclasses later than the primary SpT and the extinction is within 0.1 mag.  TWA 2A is assigned a spectral type of M1.7 and used in the analysis below. TWA 2B is tentatively assigned a spectral type of M3.5 and is not used in the analysis.

Between K5--M3.5, the average scatter in luminosity is 0.39 dex relative to a 9 Myr isochrone.  This scatter is dominated by TWA 9A and 9B  and TWA 14 (see Table ~\ref{tab:twaages}).  \citet{Weinberger2013} suggest that TWA 9A and 9B are too old to be TWA cluster members.   \citet{Ducourant2014} also note that the kinematic distance for TWA 9A is highly discrepant with the parallax distance.  However, \citet{Malo2013} assign a dynamical membership probability of 99\%.  H$\alpha$ emission, Li absorption strength, and gravity indicators show that both stars are young.
Our ages of TWA 9A and 9B are younger than the \citet{Weinberger2013} estimate because of later spectral types measured here, though the pair is still underluminous.  
Between M4--M7, most stars become significantly overluminous for a 10 Myr age, according to these isochrones.  Some of this overluminosity could be reduced if the objects are hotter than the SpT-effective temperature conversion (\S 3.2.1).  The brown dwarfs at $\sim$M8 are near the predicted HR diagram location for the 10 Myr isochrone.

A more empirical approach suggests that the slope of luminosity versus temperature is much flatter between K0--M5 than the isochrone.  A best fit line to the K0--M5 objects yields a luminosity spread of only 0.13 dex and is able to recover the objects down to 3000 K (compared with 3300 K for the 10 Myr isochrone).  Only one object, TWA 9B, is severely underluminous relative to the line.  Such a fit would suggest that the contraction timescale for very low mass stars is longer than predicted, relative to the contraction timescales of solar mass stars and brown dwarfs.  Empirical isochrones crossing theoretical isochrones in this manner is not uncommon \citep[e.g.][]{Hillenbrand2008}.

Of the resolved stars in multiple systems, the four closest in spectral type are co-eval to within a luminosity of 20\%.  The exceptions, such as TWA 8A/8B and TWA 5Aab/5B, have large differences in spectral type, often with one in the problematic M4--M6 spectral type range.  The discrepancy points to an error in either effective temperature measurements or in pre-main sequence tracks rather than a real luminosity spread.

\subsubsection{Luminosity Spread of the MBM 12 Association}

The MBM 12 cloud was initially found by \citet{Magnani1985}.   A census of MBM 12 revealed a total of 12 stellar systems \citep{Luhmanmbm12}.  The approximate distance of 275 pc was calculated by comparing the luminosities to other nearby star forming regions.  The age ($\sim 1-5$) Myr and the distance are degenerate parameters.  Seven of the 12 objects retain a disk \citep{Meeus2009}, which is consistent with a $<5$ Myr age.

A binary census of the eight brightest targets revealed 7 multiple star systems \citep{Chauvin2002,Brandeker2003}.  Based on near-IR photometry of these multiple systems, four (MBM12 1, 3, 5, and 10) are roughly equal mass stars, so their measured fluxes here are divided by 2 to estimate the luminosity of a single star.  MBM12 12 (S018) is a triple system where the primary is a single star while the secondary is a resolved binary.  Based on the near-IR colors,  we divide the optical flux by a factor of 1.6 and shift the primary spectral type to M2.1 (from M2.6).  Two other binaries, MBM12 4 (LkH$\alpha$ 264) and MBM12 2 (LkH$\alpha$ 262) are widely separated are not affected by possible companions.

The resulting HR diagram (Fig.~\ref{fig:hrtwa}) is roughly consistent with a $\sim 3$ Myr age, if the 275 pc distance is accurate.  As with the TW Hya Association, stars hotter than 4000 K are fainter than expected from the 3 Myr isochrone.  Brown dwarfs cooler than 3000 K are brighter than the 3 Myr isochrone and may be affected by unaccounted binarity.  For stars warmer than 3000 K, the luminosity spread about a best fit line is 26\%.  The slope of the line is $5.8\times10^{-4}$ $\log L$ K$^{-1}$, almost exactly the same as the slope of $6.6\times10^{-4}$ $\log L$ K$^{-1}$  for the TWA.   The most underluminous star, MBM 2 (LkH$\alpha$ 262), is an accretor with an anomalously large extinction ($A_V=1.75$) relative to other stars in the association.  Some obscuration by the central star or a gray extinction law could lead to this underluminosity.

\begin{table}
\caption{Stellar Ages in the TWA$^a$}
\label{tab:twaages}
\begin{tabular}{lccccc}
Star & d (pc) & SpT & $\log L/L_\odot$ & $\log$ Age & $L/L10$\\
\hline
\multicolumn{6}{c}{SpT and Flux from this work}\\
 \hline
TW Hya & 54 & M0.5 & -0.72 & 7.21 & 0.70\\  
TWA 2A & 42 & $\sim$M1.7$^b$ & -0.69$^c$ & 6.79 & 1.39\\  
TWA 2B & 42 & $\sim$M3.5$^b$ &  -1.22$^c$ & 6.86 & 1.31\\  
TWA 3Aab & (35) & M4.1 & -1.22 & 6.57 & 2.83\\  
TWA 3B &  (35)  & M4.0 & -1.10 & 6.49 & 3.34\\  
TWA 5Aab & 49 & M2.7 & -1.05 & 6.80 & 1.36\\  
TWA 6 & (67) & M0 & -0.65 & 7.25 & 0.67\\  
TWA 7 & 34 & M3.2 &-0.94 & 6.64 & 1.80\\  
TWA 8A & 43 & M2.9 &-0.95 & 6.80 & 1.36\\ 
TWA 8B & 39 & M5.2 & -1.66 & 6.55 & 2.76\\  
TWA 9A & 47 & K6   & -0.83 & 7.71 & 0.30\\ 
TWA 9B & 52 & M3.4 & -1.38 & 7.11 & 0.81\\  
TWA 13A & 59 & M1.0 &  -0.66 & 6.98 & 1.03\\  
TWA 13B & 56 & M1.1 & -0.71  & 7.02 & 0.97 \\  
HR 4796A & 73 & A0 & 1.20 & -- & --\\
TWA 14 & 96 & M1.9 &  -0.58 & 6.61 & 1.96 \\  
TWA 23 AB & 49 & M3.5 & -1.21 &  6.84 & 1.37\\  
TWA 25 & 54 & M0.5 & -0.65 & 7.12 & 0.83\\  
TWA 27 & 52 & M8.25 & -2.68 & 7.18 & 0.73\\  
TWA 28 & 55 & M8.5 &-2.65 & 7.09 & 0.84\\  
\hline
\multicolumn{6}{c}{SpT and Flux from literature}\\
\hline
TWA 5B & (49) & M9 & -2.83$^b$ & 7.15 & 0.76 \\  
TWA 10 & 62 & M2 & -0.87 & 6.97 & 1.06\\ 
TWA 11B & (67) & M2.5 & -0.82$^a$ & 6.75 & 1.51\\  
TWA 11C & 69 & M4.5 & -1.10$^a$ & 6.28 & 5.99\\   
TWA 12   & 64 & M1.6 & -0.78$^a$ & 6.96 & 1.07 \\  
TWA 15A  &110$^c$ & M1.5 & -0.93$^a$ & 7.21 & 0.71\\  
TWA 15B  & 117$^c$ & M2.2 & -0.87$^a$ & 6.90 & 1.18\\ 
TWA 16AB & 78 & M1.8 & -0.91$^a$ & 7.09 & 0.87 \\  
TWA 20 & 77 & M2.0 & -0.76 & 6.82$^a$ & 1.35\\ 
TWA 21 & 51 & K3.5 & -0.42 & 7.48$^a$ & 0.42\\  
TWA 26 & 38 & M9.0 & -2.85 & 7.22 & 0.74\\  
TWA 29 & 79  & M9.5 &  -2.95 & 7.27 & 0.70\\  
TWA 30A &  (56) & M5 & \multicolumn{2}{c}{Edge-on disk} &\\
TWA 30B  & (56) &M4 & \multicolumn{2}{c}{Edge-on disk} & \\
TWA 32 AB & (77) & M6.3 &  -1.70 & $<6$ & 3.88\\  
TWA 33 & (52.6) & M4.7 & -1.42 & 6.48 & 3.56\\  
TWA 34 & (50) & M4.9 & -1.86 & 6.77 & 1.59\\  
\hline
HD 98800 Aa & 45 & 4535 & -0.21 & 7.28 & 0.60\\
HD 98800 Ba & 45 & 4200 & -0.54 & 7.40 & 0.51 \\
HD 98800 Bb & 45 & 3500 & -0.98 & 6.91 & 1.16 \\
\hline
\multicolumn{6}{l}{Restricted to TWA members with distances }\\
\multicolumn{6}{l}{Parallaxes from Weinberger et al.~(2013), }\\
\multicolumn{6}{l}{~~~~~Malo et al.~(2013), Teixeira et al.~(2008),}\\
\multicolumn{6}{l}{~~~~~Gizis et al.~(2007), Biller \& Close (2007),}\\
\multicolumn{6}{l}{~~~~~van Leeuwen~(2007), and Ducourant et al.~(2014)}\\
\multicolumn{6}{l}{Dynamical distances in () \citep{Mamajek2005}}\\
\multicolumn{6}{l}{Literature SpT from \citet{Konopacky2007}}\\
\multicolumn{6}{l}{~~~~~\citet{Webb1999}, \citet{Kastner2008}}\\
\multicolumn{6}{l}{~~~~~\citet{Shkolnik2011}, \citet{Gizis2002}}\\
\multicolumn{6}{l}{~~~~~\citet{Zuckerman2001}, \citet{Zuckerman2004}}\\
\multicolumn{6}{l}{~~~~~\citet{Looper2010a,Looper2010b},\citet{Bonnefoy2009}}\\
\multicolumn{6}{l}{~~~~~\citet{Schneider2012},\citet{Allers2013}}\\
\multicolumn{6}{l}{$^a$$L$ from J-band magnitude \citep{Skrutskie2006}}\\
\multicolumn{6}{l}{$^b$$L$ from H-band magnitude \citep{Weintraub2000}}\\
\multicolumn{6}{l}{$^c$Large distance may indicate non-membership}\\
\end{tabular}
\end{table}

\section{DISCUSSION}

\subsection{The limited affect of accretion histories on pre-main sequence evolution}

The pre-main sequence models of \citet{Baraffe2010} demonstrate that stars contract faster if they form primarily by large accretion events with rates of $>10^{-4}$ $M_\odot$ yr$^{-1}$, as opposed to steady accretion throughout the protostellar lifetime.  If the large accretion events are episodic and randomly distributed in strength, then this evolution predicts a significant luminosity spread in observed HR diagrams \citep{Baraffe2012}.  The evolutionary effects will remain large at 10 Myr for $0.1$ M$_\odot$ stars, with luminosity differences up to a factor of 25, but may be minimal for solar mass stars, depending on the seed mass of the star and size of episodic accretion bursts.

The stars in the TWA are roughly coeval, as are those in MBM 12.  Stars in multiple star systems in the TWA also tend to be co-eval with each other.  Similarly, in Taurus 2/3 of binaries have ages consistent to 0.16 dex, with many of the outliers attributable to veiling or other observational uncertainties (Kraus \& Hillenbrand~2009; see also Hartigan et al.~1994, 2003; White 2001). 

If episodic accretion dominates stellar growth to an extent that the evolutionary tracks are severely altered, these results  would require that the effects of episodic accretion are similar for the majority of stars.  In this case, the ages of pre-main sequence stars would be significantly and uniformly overestimated in all regions.  However, the $\sim 7-10$ Myr age of the TWA obtained from comparison to PMS tracks is similar to the dynamical expansion age of $7.5\pm0.7$ Myr \citep{Ducourant2014}.

These comparisons suggest that large episodic accretion outbursts do not significantly alter pre-main sequence evolution over general populations of stars with masses 0.3--0.7 $M_\odot$.  Any affects on entire populations are minimal by ages of 5--10 Myr.  However, such events may alter the evolutionary course of a minority of stars, and could in principle explain the underluminosity of a star like TWA 9B.  

\subsection{Minimizing observational errors in HR diagrams}

The uncertainties in effective temperature and luminosity explains at least some of the luminosity spreads measured within young clusters.  In addition to the uncertainties described in \S 5, membership, binarity, and disk obscuration can severely affect measured luminosity spreads in HR diagrams.

The choice of star forming region and observed wavelength largely determines which uncertainties are minimized and which are problematic.  At present, 
the spectroscopic and direct imaging binary census in Taurus is relatively complete, at least for the well known solar mass members \citep{White2001,Kraus2009,Nguyen2012}.  In ideal cases where the two stars are diskless and have the same extinction, a binary accounting to $0.2$ times the mass of the primary star yields a $\sim 0.1$ dex error in age.
 Binarity is a much more severe problem for more distant regions, and likely requires the use of population synthesis models to interpret luminosity spreads.  Use of near-IR colors to calculate mitigates the effect of extinction uncertainties, however systematic extinction offsets may be prevalent.
The relative distance uncertainty in nearby regions is currently much larger than that for more distant regions.

The near-IR has some advantages over optical spectra. Veiling corrections are less important for red or near-IR flux measurements of late M-dwarfs. The emission produced by accretion and warm dust reaches a minimum between 1--1.5 $\mu$m, while the photospheric flux from late M-dwarfs peak at those same wavelengths.  For these late M-dwarfs, red or near-IR colors should be used for extinction estimates because they are relatively constant with spectral type \citep[e.g.][]{Leggett1992} and usually include negligible contribution from veiling.  However, extinction estimates are less sensitive and less reliable when measured from near-IR observations.

Probing ages of young ($<5$ Myr) clusters requires an assessment of the effects of disk parameters on the measured stellar properties.
 Partial disk obscuration of the star significantly decreases measured luminosities and is particularly difficult to account for.  Within our biased sample, $\sim 10$\% of accreting objects are significantly underluminous relative to the expected luminosity of a pre-main sequence star.  Some of these faint objects have disks viewed edge-on, which blocks the light from the star.  In cases such as AA Tau \citep{Bouvier2013} and perhaps CW Tau, a disk warp periodically or stochastically blocks the light from the central star \citep[see also][]{Alencar2010,Findeisen2013}.  The measured luminosities of these objects are not realistic and should be discarded from population studies of ages obtained from HR diagrams.  
The difficulty is in knowing which stars to discard.  Easy cases will appear below the zero age main sequence on an HR diagram, but many cases will not be so obvious.    The inclusion of this uncertainty will require measurements of the frequency and scale of such events from monitoring observations, such as those done by COROT \citep{Alencar2010}.  When variability information is available, the most straightforward technique is to simply calculate the average stellar brightness.  However, in the case of disk obscuration the maximum photospheric luminosity is likely more appropriate.

\section{PROSPECTS FOR FUTURE IMPROVEMENTS AND CONCLUSIONS}

This paper provides a consistent and robust set of spectral types and extinctions for 283 young stars, including many of the most well studied.  The primary advances in this paper are the implementation of simultaneous measurements of the extinction, accretion continuum flux, and spectral type for accreting stars and a sufficient sample size to obtain a robust set of extinction-corrected spectral templates.  The effects of veiling on spectral type and extinction are reduced when analyzing spectra with coverage from 4000--9000 \AA.  A similar approach was recently used by \citet{Manara2013b} to investigate two stars with previously reported ages of 30 Myr.  Their accurate spectral type and luminosity yielded an age of 2--3 Myr, consistent with the age of the parent cluster.

An updated grid of photospheric M star templates will eventually be needed to account for the evolution of colors with pre-main sequence contraction.  Unfortunately, this problem is challenging to solve because the T Tauri stars with known $A_V=0$ mag.~are those in the 7--10 Myr old TWA and the $\eta$ Cham association.  No T Tauri star in a young ($<3$ Myr) region can be assumed to have $A_V=0$ (or any other $A_V$) based only on its colors, independent of a model template.  The high binary fraction of young WTTSs \citep{Kraus2012} also affects their use as photospheric templates.   Although we minimize the effects of gravity dependence by using T Tauri stars as templates, the gravity dependence between 1--10 Myr may still be significant and is not accounted for.  Photometric samples of non-accretors are likely reliable, but degeneracy between spectral type and accretion continuum flux can lead to spectral type uncertainties of at least two subclasses.

The approach to measuring spectral types and extinction in this paper can reach a luminosity accuracy of $\sim 0.1-0.2$ dex for most classical T Tauri stars, and should serve as a particularly useful guide in the analysis of broadband spectra obtained by {\it VLT}/X-Shooter (e.g., Manara et al. 2013ab) and for analysis of GAIA observations.  
The grid of spectral types should be improved and based on more direct measurements of effective temperature by comparing high resolution spectra to models.  The spectral type-effective temperature conversions are also uncertain at present because  model atmopsheres fail to reproduce some large spectral features for spectral types later than M4.
The relationship between spectral type and extinction needs particular improvement between K5--M0.5, where the accuracy of our grid relative to other publications is especially uncertain.  Our results rely upon the assumption that the accretion continuum flux is flat.  However, the strength of the broadband accretion continuum should be measured with simultaneous broadband spectra.  Finally, extinction measurements should include the effect of gravity on photospheric emission, following the gravity-dependent colors obtained by \citet{Covey2007} and \citet{Pecaut2013}.  Ideally, some optically thin line ratios could be found and used to measure extinction, independent of gravity.

\section{Acknowledgements}

We appreciate valuable discussions with Suzan Edwards, Adam Kraus, Sylvie Cabrit, Kevin Covey, Davide Fedele, and Michael Rugel, and also thank Kraus for help with a Taurus membership database.  GJH appreciates financial support for this project provided by the Youth Qianren Program of the National Science Foundation of China and the Observatoire de Paris for hosting him as a visiting astronomer.

\section{Appendix A}

Our observations include two possible members of Taurus, a background supergiant that was a candidate member of Lupus, and a reflection nebulosity.  These tangential results are discussed below.

\subsection{GK Tau B}

GK Tau B is a visual companion located $2\farcs4$ from GK Tau \citep{Hartigan1994}.  Based on optical photometry, \citet{White2001} suggested that the star may be a visual binary that is not associated with the Taurus star forming region.  

Our 20.~Jan. 2008 observations of GK Tau A and GK Tau B were obtained with slits placed perpendicular to the position angle of the two stars.  Some bleeding from GK Tau A likely affects the GK Tau B spectrum.  The MgH band, \ion{Ca}{2} IRT absorption, and H$\alpha$ emission are all detected from GK Tau B at a level that is inconsistent with possible bleeding from GK Tau A.  GK Tau B is consistent with a $\sim$ K3 SpT and $A_V=2.1$.  The H-$\alpha$ emission indicates that GK Tau B is accreting, is a likely member of the Taurus Molecular Cloud, and is likely associated with GK Tau A.

\subsection{2MASS J04162709+2807313}
We decided to observe 2MASS J04162709+2807313 on a whim, since it close (14$^{\prime\prime}$) from LkCa 4, had not been previously discussed in the literature, and was reasonably noticed to be bright during our acquisition of LkCa 4.  LkCa 4 S has a similar spectral type as LkCa 4 AB but is a factor of 3.5 fainter.  Weak H$\alpha$ emission suggests chromospheric activity, an indicator of youth.  The K I 7700 \AA\ and Na I 8200 \AA\ doublets are weak and indicate a low gravity, consistent with the pre-main sequence.  If this star is a wide binary companion to LkCa 4, half of the luminosity difference is accounted for by the binarity of LkCa 4.

The optical brightness of LkCa 4 S during our observation is surprising.  The star is 6 magnitudes fainter than LkCa 4 in 2MASS JHK and in USNO-B, with 1 magnitude of variability.  The JHK color difference is consistent with $A_V=1.9$ mag.  LkCa 4 S is also not listed in the WISE all sky catalogue, which rules out variability and extinction from an edge-on disk.  Perhaps our observation occurred when the star peaked out of what is normally an opaque ISM.

\subsection{2MASS J16003440-4225386}
2MASS J16003440-4225386 was listed as a candidate member of Lupus based on an IR excess and colors that are consistent with a late M star \citep{Chapman2007}.  The star has broad TiO bands but a huge absorption band around 8200 \AA, characteristic of a late M star with a very low surface gravity.  We classify the star as an $\sim$M9 I supergiant and a likely Cepheid variable.

\subsection{Reflection Nebulosity of Sz 68 A}

Sz 68 is a triple system, with the second and third components located $0\farcs126$ and  $2\farcs808$ from Sz 68 A \citep{Correia2006}.  The star drives a bright jet, known as HH 186 \citep{Heyer1989}.

\begin{figure}
\epsscale{1.}
\plotone{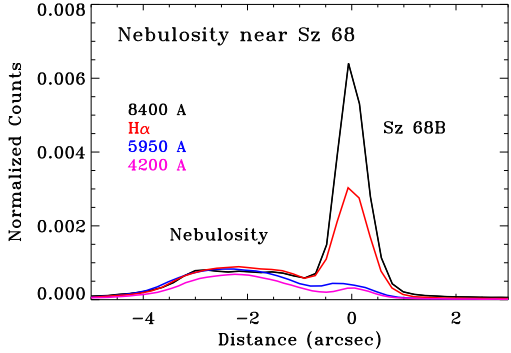}
\caption{The shape of emission from Sz 68 B and the Sz 68 A reflection nebulosity in the dispersion direction at four different wavelengths.  Within the slit, nebulosity extends over $\sim 4^{\prime\prime}$ and reflects 1.5\% of the total light from Sz 68A at all optical wavelengths.}
\label{fig:sz68neb.ps}
\end{figure}

A bright reflection nebula was located between Sz 68 AB and Sz 68 C at the time of our observations.  The spectrum of the reflection nebula is exactly the same as the Sz 68 AB spectrum.  Figure~\ref{fig:sz68neb.ps} shows the spatial profile of emission in the slit of our Sz 68 C observation.  The nebulosity within our slit accounts for 1.5\% of the total emission from Sz 68 AB.  The total emission from the nebulosity is likely much higher.  The fraction of light from Sz 68 AB scattered into our line of sight by the reflection nebulosity does not depend on wavelength.  Extended dust emission was independently found in {\it Herschel}/PACS observations of the Sz 68 system \citep{Cieza2013}.

The Sz 68 C spectrum was extracted from the image by fitting a 2nd order polynomial to the edge of the nebulosity, subtracting off the model flux, and then extracting the leftover stellar flux.

\section{Appendix B:  Comparisons between synthetic and observed spectra}

\begin{table*}
\caption{Temperature Measurements for Spectral Type Grid}
\label{tab:sptteff1.tab}
\begin{tabular}{llcccccccc}
Star & Date & SpT &  Blue+Red & Red only & Blue+red, no TiO & TiO Only & Adopted & Fac\\
\hline
HBC 407 & 28 Dec. & K0 & 5110 & 4910 & -- & 5100 & 5110 & 0.99\\
HBC 372 & 28 Dec.  & K2 & 4710 & 4570 & -- & 4850 & 4710 & 0.99\\
LkCa 14    & 28 Dec. & K5 &4200 & 4240 & 4220 & 4540 &4220 &  1.02\\
MBM12 1 & 19 Jan. & K5.5 & 4160 & 4240 & 4190 & 4420 & 4190& 0.99\\
TWA 9A    & 20 Jan. & K6.5 & 4140 & 4160 & 4160 & 4300 &4160 & 1.03 \\
V826 Tau &30 Dec. & K7 & 4020 & 4130 & 4020 & 4300 & 4020 & 1.05\\
V830 Tau & 30 Dec & K7.5 & 3940 & 4050 & 3930 & 4200 & 3930 & 1.03 \\
TWA 6      & 29 Dec.& M0 & 3930 & 3990 & 3910 & 4200 & 3910 & 1.05 \\
TWA 25   &  28 Dec. & M0.5 & 3840 & 3850 & 3850 & 3680 & 3770 & 1.05\\
TWA 13A & 18 Jan.& M1.0 & 3750 & 3810 & 3740 & 3620 & 3690 & 1.05\\
LkCa 4   & 30 Dec. & M1.5 & 3730 & 3800 & 3720 & 3620 & 3670 & 1.05\\
LkCa 5     & 28 Dec. & M2.2 & 3550 & 3660 & 3530 & 3500 & 3520 & 1.05 \\
LkCa 3     & 28 Dec. & M2.4 & 3530 & 3650 & 3510 & 3500 & 3510 & 1.06\\
TWA 8A    & 18 Jan. & M3.0 & 3380 & 3530 & 3370 & 3410 & 3390 & 1.04\\
TWA 9B    & 20 Jan. & M3.4 & 3340 & 3440 & 3360 & 3330 & 3340 & 1.05\\
2M J1207-3247 & 28 Dec. & M3.5 & 3320 & 3430 & 3320 & 3380 & 3350 & 1.10\\
Hen 3-600 B & 28 May & M4.1 & 3140 & 3320 & 3140 & 3100 & 3120 & 1.05\\
XEST 16-045 & 28 Dec. & M4.4 & 3070 & 3140 & 3100 & 3100 & 3100 & 1.06\\
J2 157          & 28 Dec. & M4.7 & 3020 & 3080 & 3060 & 3040 & 3050 & 1.03\\
TWA 8B        &30 Dec. & M5.2 & 2840 & 3000 & 2840 & 2990 & 2910 & 1.05\\
MBM12 7       & 30 Dec. & M5.6 & 2830 & 2990 & 2830 &2960 & 2890 & 1.05 \\
V410 X-ray 3 &30 Dec. & M6.5 & 2770 & 2800 & 2780 & 2880 & 2830 & 1.09\\
Oph 1622 A &28 May & M7.25 & 2680 & 2710 & 2680 & 2820 & 2750 & 1.08\\
\hline
\end{tabular}
\end{table*}

\begin{figure*}
\plotone{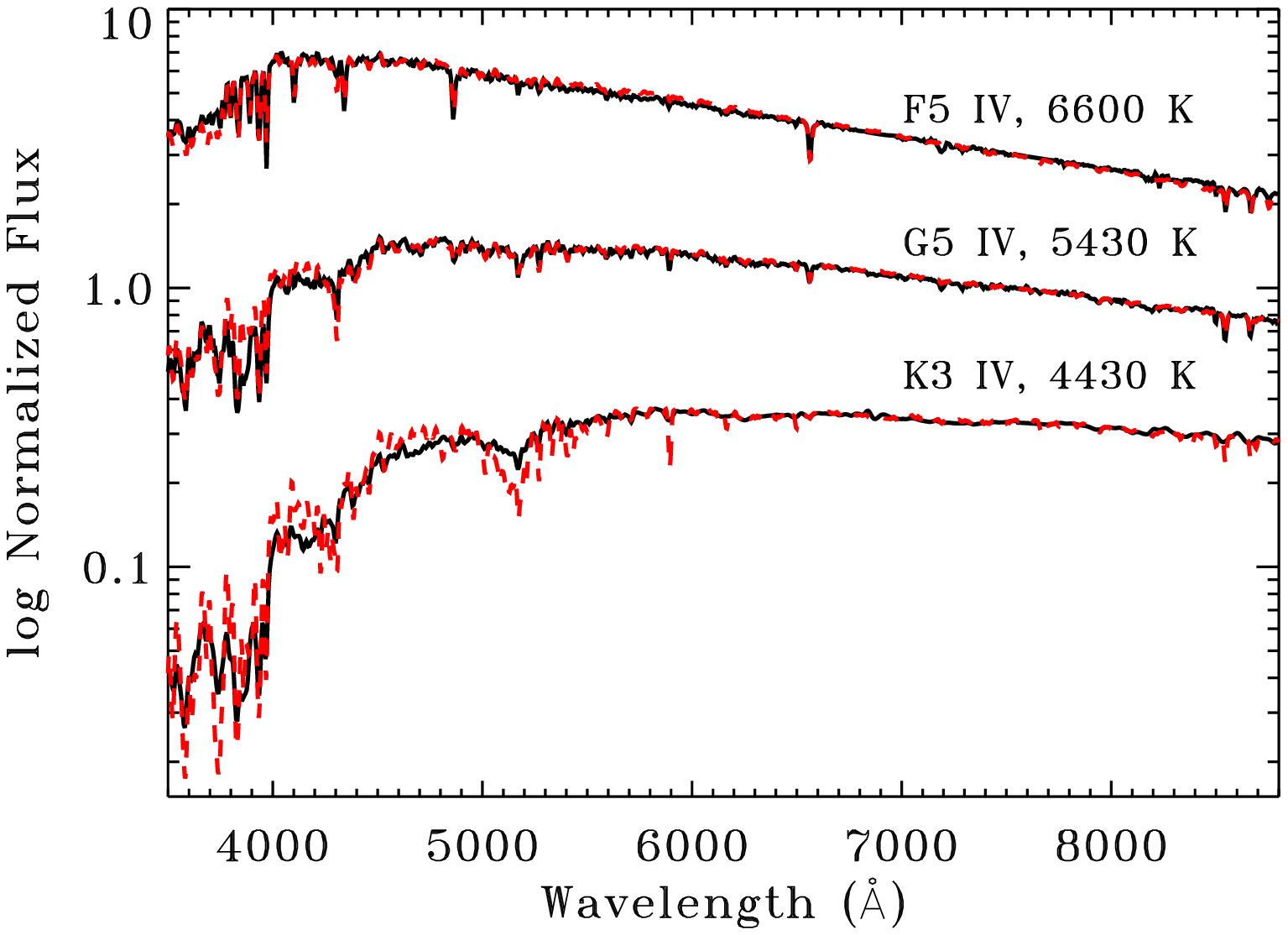}
\epsscale{1.1}
\plottwo{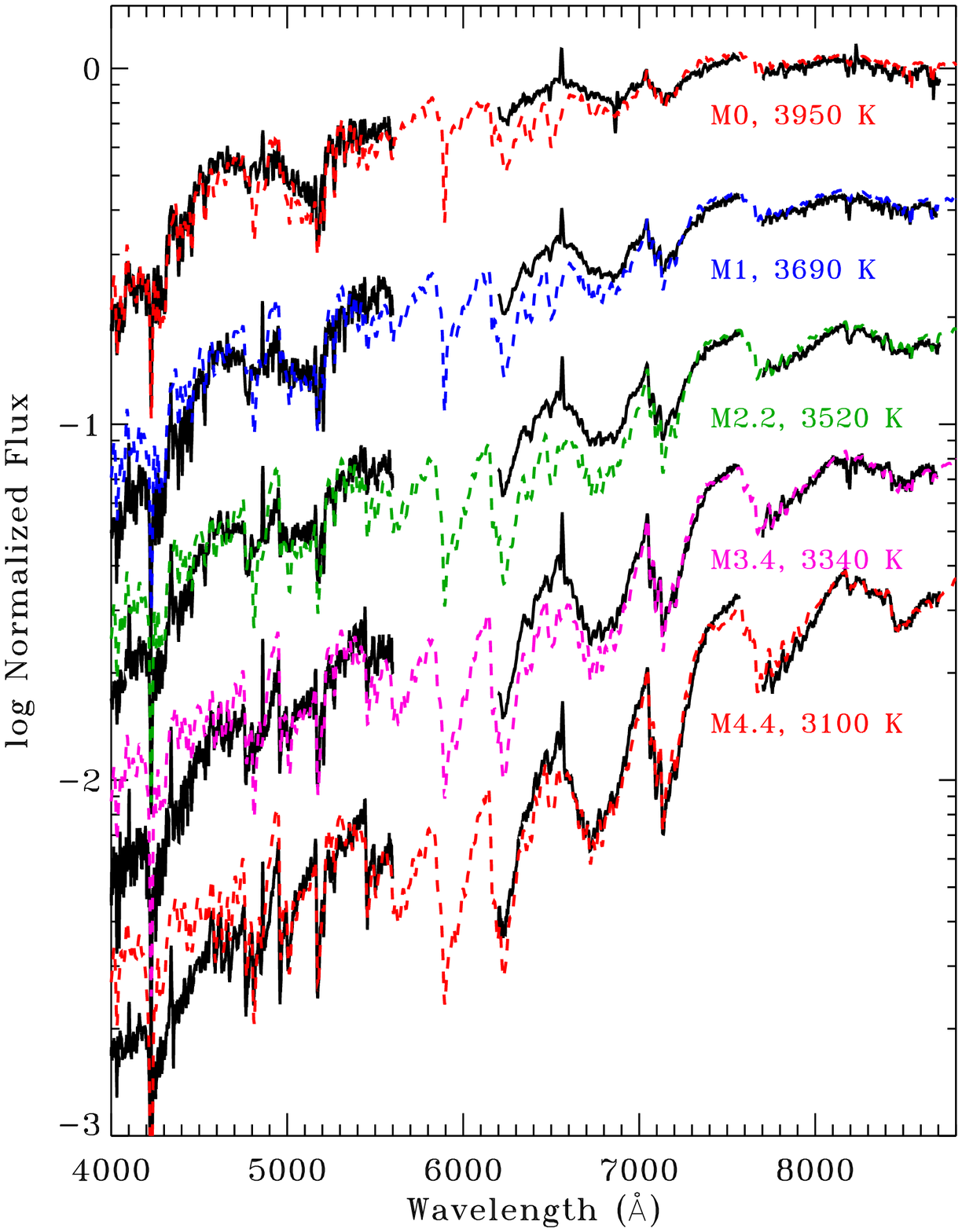}{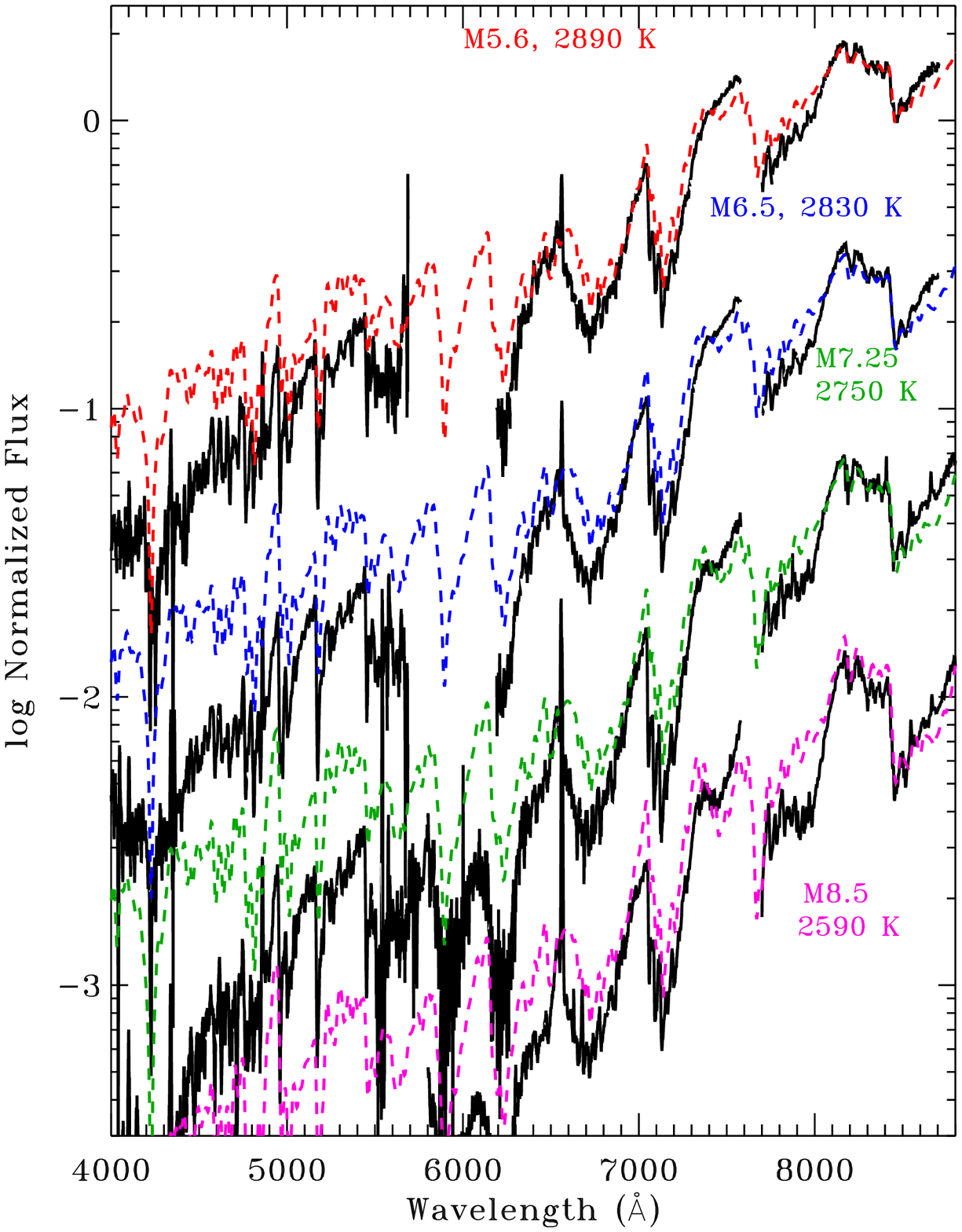}
\caption{Top:  Comparisons between luminosity class IV stars in the Pickles library and the best fit BT-Settl synthetic spectra, normalized at 7300 \AA.  
Bottom:  Comparisons between K5.5--M4 stars (left), M5-M8.5 stars (right, M8.5 is red only) from our grid of spectral templates (Table~\ref{tab:gridstars}), and the best fit BT-Settl synthetic spectra.  Spectra are normalized at 7350 \AA.  The model spectra are then scaled by the best-fit parameter from Table~\ref{tab:sptteff1.tab}.  At spectral types later than M4, the blue spectra are much stronger in the BT-Settl models than in the observed spectra.}
\label{fig:tempcomp2}
\end{figure*}

This appendix and Figure \ref{fig:tempcomp2} describe in detail our comparisons between observed spectra and the BT-Settl models with $\log g=4.0$ of \citet{Allard2012}.  The BT-Settl models in \citet{Allard2012} are calculated at 100 K intervals.  Intermediate temperatures are calculated by linearly interpolating between tempreatures at 10 K intervals.   Our spectral type grid is listed in Table~\ref{tab:gridstars} and is supplemented with an M9.5 spectrum of KPNO 4 (Luhman, private communication).  The observed M8.5 and M9.5 spectra only cover red wavelengths (5700--9000 \AA).

For stars earlier than M0, synthetic spectra at some temperature can be found that reproduces the spectral shape and most features of the observed spectra.  Some small differences occur at locations of strong lines, particularly the MgH/Mg b complex at 5200 \AA.  

For stars later than M0, the synthetic spectra are less good
at reproducing the observed young star spectra.  Specifically,
with normalization at 7300 A,
the synthetic spectra lack opacity and are much stronger than the observed spectra shortward
of 5000 \AA.
This discrepancy increases towards cooler stars.  The synthetic spectrum is also slightly fainter than the observed spectrum at 5500--6500 \AA.  Later than $\sim M4$, the three strong TiO bands at 7140, 7600, and 8500 \AA\ in the synthetic spectra no longer match the observed band depth.  The discrepancy at blue wavelengths also becomes larger.  The VO absorption band at 7500 \AA\ is also much stronger in the synthetic spectra than in the observations, so the 7400--7600 \AA\ region is avoided in our temperature calculations.

 The temperatures are calculated by considering both
the overall spectral energy distribution and the absorption
line and band strengths.   Specifically, we find a
synthetic spectrum and a normalization that best
fits the observed spectra considering several ranges 
in wavelengths within the spectral range
of the data:   (a) for blue+red, fits to the full 4300--8700 \AA\ spectrum, (b) for red only, fits to the 6300--8700 \AA\ spectrum, (c) for the blue+red spectrum, where the fits exclude the spectral locations of deep TiO bands, and (d) for the depth of TiO bands, with each band normalized to a nearby wavelength region so that the TiO-only fit is independent of the broadband colors.  Results from these fits are presented in Table~\ref{tab:sptteff1.tab}.  The adopted temperatures for (1) spectral types earlier than K5 are obtained from (a), the full blue+red fit; (2) for spectral types K5--M0, from (c), the blue+red spectrum that excludes TiO bands, and (3) for spectral types later than M0, the average temperature from (c), the blue+red fits excluding TiO and (d), the TiO-only fits.  For M-dwarf models, the absolute scaling is based on the fit from (c), the blue+red without TiO, calculated for the adopted temperature.
If the molecular data is insufficient to reproduce the depth of the strong molecular bands, then the fits that focus on continuum regions and  exclude molecular bands may be more accurate.  For stars earlier than M0, the fit to the full spectrum is used to convert SpT to temperature.  For stars later than M0, the spectral locations outside of TiO bands and the depth of the TiO bands are averaged to provide our conversion to temperature.  The scaling parameter between the observed and model spectra is also listed in Table~\ref{tab:sptteff1.tab} and is based on the 7350--7400 \AA\ spectral range.

The fits to the TiO spectrum and to the spectrum outside of the TiO bands differ by as much as 200 K, which suggests a $\sim 100$ K uncertainty in our conversion and that the model atmospheres do not yet reproduce spectra of pre-main sequence stars that are cool enough for TiO and other molecules to provide significant opacity in the atmosphere.  Increasing $\log g$ from 4.0 to 4.5 leads to a decrease of $\sim 50$ K in temperature.
The discrepancies between the synthetic and observed spectra are significant for objects later than M1, increase substantially at spectral types later than M4, and are especially large at M9.5.  The uncertainty in the SpT-temperature conversion also leads to uncertainty in the bolometric corrections and in the luminosity from evolutionary models that use these atmospheres.

\clearpage

\section{Appendix C:  Table of Stellar Properties}

Table~\ref{tab:props.tab} lists the an abbreviated name, number of spectra $N$ on different nights, distance, spectral type, extinction, flux at 7510 \AA corrected for extinction, veiling $r$ at 7510 \AA, and the $\log$ of the photospheric luminosity for all stars in our sample.
An electronic-only table lists for each spectrum the full target name and position, the observation date, and mass and age estimates.
Negative extinctions and the calculated extinctions to the TW Hya Association are listed in parenthesis and are treated as $A_V=0$ when calculating luminosities.
Several stars only have red spectra because of either detector failure or the star was too faint, and are noted with ``r'' after the spectral type.  These red-only spectra have larger uncertainties in the spectral type and extinction measurements.
For unresolved multiple stars systems (e.g., TWA 4 AabBab, GG Tau Aab, GG Tau Bab, LkCa 3AabBab, etc.), the spectral type and stellar properties are global measurements of the entire system.

The mass and age estimates are obtained by comparing the photosphere temperature and luminosity to the \citet{Baraffe2003} pre-main sequence tracks for masses $<0.2$ $M_\odot$, \citet{Tognelli2011} tracks for masses $>0.4$ M$_\odot$, and interpolated between those models for 0.2--0.4 $M_\odot$.   Multiplicity is not accounted for in these mass and age estimates.  Severely underluminous stars located below the main sequence have no age listed and a mass (listed in parenthesis) assessed by assuming a 3 Myr.

For the 62 stars observed on multiple nights, in Table~\ref{tab:props.tab} the veiling and flux at 7510 \AA\ are averages, and the luminosity is calculated from the average flux.  These values are each listed independently for each night in the electronic Table.   The SpT and, when possible, the extinction are the average of those measured for each spectrum.  In several cases, the listed extinctions are different, indicating that no one extinction could accurately explain all spectra from the object.  The observed flux at 7510 \AA\ and consequent photospheric luminosity show variability, some of which is attributed to uncertainties in the absolute flux calibration.

Stars with heavily veiling and no spectral type are listed as continuum (c) stars.  For these stars, extinction is calculated by assuming the continuum is flat.  The listed $F_{7510}$ corresponds to the extinction-corrected flux rather than the photospheric flux, and is listed in parenthesis.  In less extreme cases of heavily veiled stars, the spectral type may be estimated and is listed with a ``c'' following the spectral type.

Spectral types of M dwarfs are listed to 0.1 subclass, although our internal precision is $\sim 0.2-0.3$ subclasses.  Larger differences are likely when comparing spectral types to other studies.  Extinctions of stars later than K0 were measured against our spectral type grid and are listed to $0.05$ mag.~in $A_V$.  
Our extinctions are accurate to $\sim 0.2$ mag.~for stars with little or no veiling.   

The extinction measurements assume an extinction law based on a total-to-selective extinction of $R_V=3.1$ for most targets.  Targets with large extinctions ($A_V>5$) typically required higher $R_V$, indicative of larger grains.  V892 Tau could only be fit with an extinction law using $R_V>5$ and is assumed to be $R_V=5.5$.  DoAr 21 and SR 21 (a star with a transition disk) required fits with extinction laws using $R_V=4$.  For IRS 48, we assumed $R_V=5.5$ because of the high extinction.

When possible we rely on parallax distances: 120 pc for Ophiucus \citep{Loinard2008}, 131 pc for stars near the Lynds 1495 complex in Taurus \citep{Torres2012}, 147 pc for stars near T Tau \citep{Loinard2007}, 161 pc for the stars near the HP Tau complex in Taurus \citep{Torres2009}, 140 pc for all other Taurus objects, and 416 pc for Orion \citep{Menten2007,Kim2008}.  Distances for TWA members are listed in Table~\ref{tab:twaages}.
We also use 150 pc for Lupus 1 and 200 pc for Lupus 3 \citep{Comeron2008}, 130 pc for CrA \citep{Neuhauser2008}, 145 pc for Upper Sco OB Association \citep{Preibisch2008}, 275 pc for MBM 12  \citep{Luhmanmbm12}, and 200 pc for AT Pyx in the Gum Nebula \citep{Kim2005}.

The distance to RR Tau is not well constrained and is left blank here.  The commonly cited distance of 800 pc has often been credited to several more recent publications but was actually calculated by \citet{Herbig1960}.  The distance to RR Tau was assumed to equal to the distance to the A6 star BD+26 887, located $3^\prime$ away.  The distance to BD+26 887 was then calculated by comparing its magnitude to that of a main sequence B8 star, based on the spectral type at the time.  \citet{Hernandez2004} later adjusted the distance of BD+26 887, but not RR Tau, to 2 kpc based on rough proximity to molecular clouds with distances inferred by \citep{Kawamura1998}.  If we assume that AB Aur and RR Tau have the same luminosity, RR Tau would be located at $\sim 670$ pc.  On the other hand, \citet{Slesnick2006} identified pre-main sequence stars 3 degrees to the south of RR Tau (RR Tau was not covered in their survey) that have brightnesses consistent with the $\sim 140$ pc distance to Taurus.

\clearpage
\pagebreak

\begin{table}[!h]
\caption{Stellar Properties}
\label{tab:props.tab}
{\footnotesize 
\begin{tabular}{lccccccc}
\hline
Object  & N & d & SpT & $A_V$ & $\log F$  & $r$ & $\log L$\\
   & \# & pc &    & mag & $^a$  &$^b$ & $L_\odot$\\
\hline
MBM12 1								 &    1 &   275 & K5.5 &       0.10 &     -13.18 &       0.00 &     0.25 \\
MBM12 2								 &    1 &   275 & M0.3 &       1.65 &     -13.83 &       0.18 &    -0.39 \\
MBM12 3								 &    1 &   275 & M2.8 &       0.55 &     -13.66 &       0.02 &    -0.18 \\
MBM12 7								 &    1 &   275 & M5.6 &  (-0.10) &     -14.79 &       0.00 &    -1.14 \\
LkHa 264							 &    1 &   275 & K5.5 &  (-0.25) &     -13.80 &       0.57 &    -0.36 \\
MBM12 8								 &    1 &   275 & M5.9 &       0.30 &     -15.19 &       0.00 &    -1.51 \\
\end{tabular}}
\end{table}

\begin{table}[!h]
{\footnotesize \begin{tabular}{lccccccc}
\hline
Object  & N & d & SpT & $A_V$ & $\log F$  & $r$ & $\log L$\\
   & \# & pc &    & mag & $^a$  &$^b$ & $L_\odot$\\
\hline
2M 0420+2804			 &    1 &   131 & M3.5 &       0.25 &     -13.66 &       0.00 &    -0.80 \\
XEST 16-045				 &    1 &   131 & M4.5 &  (-0.05) &     -13.80 &       0.00 &    -0.87 \\
J2 157									 &    1 &   131 & M4.6 &       0.35 &     -14.13 &       0.00 &    -1.19 \\
IRAS 04187+1927 &    1 &   131 & M2.4 &       3.10 &     -13.38 &       0.08 &    -0.56 \\
DE Tau									 &    4 &   131 & M2.3 &       0.35 &     -13.09 &       0.05 &    -0.28 \\
RY Tau									 &    1 &   131 & G0 &       1.85 &     -11.74 &       0.00 &     1.03 \\
\end{tabular}}
\end{table}

\begin{table}[!h]
{\footnotesize \begin{tabular}{lccccccc}
\hline
Object  & N & d & SpT & $A_V$ & $\log F$  & $r$ & $\log L$\\
   & \# & pc &    & mag & $^a$  &$^b$ & $L_\odot$\\
\hline 
FY Tau									 &    2 &   140 & M0.1 &       3.05 &     -13.24 &       0.16 &    -0.39 \\
FZ Tau									 &    3 &   140 & M0.5c &       3.50 &     -13.33 &       0.30 &    -0.48 \\
UZ Tau B							 &    1 &   140 & M3.1 &       0.70 &     -13.10 &       0.03 &    -0.21 \\
UZ Tau A							 &    1 &   140 & M1.9 &       0.90 &     -13.26 &       0.14 &    -0.40 \\
HBC 403								 &    1 &   140 & K6.0 &       0.85 &     -13.20 &       0.00 &    -0.36 \\
JH 112B								 &    4 &   140 & K5.5 &       3.10 &     -13.27 &       0.00 &    -0.42 \\
\end{tabular}}
\end{table}

\begin{table}[!h]
{\footnotesize \begin{tabular}{lccccccc}
\hline
Object  & N & d & SpT & $A_V$ & $\log F$  & $r$ & $\log L$\\
   & \# & pc &    & mag & $^a$  &$^b$ & $L_\odot$\\
\hline 
DR Tau									 &    4 &   140 & K6c &       0.45 &     -13.33 &       0.51 &    -0.49 \\
DS Tau									 &    4 &   140 & M0.4 &       0.25 &     -13.57 &       0.36 &    -0.72 \\
UY Aur									 &    1 &   140 & K7.0 &       1.00 &     -12.91 &       0.07 &    -0.07 \\
ST 34										 &    1 &   140 & M3.4 &       0.50 &     -13.70 &       0.14 &    -0.79 \\
GM Aur									 &    5 &   140 & K6.0 &       0.30 &     -13.16 &       0.18 &    -0.31 \\
LkCa 19								 &    1 &   140 & K2.0 &       0.50 &     -12.69 &       0.00 &     0.13 \\
\end{tabular}}
\end{table}

\begin{table}[!h]
{\footnotesize \begin{tabular}{lccccccc}
\hline
Object  & N & d & SpT & $A_V$ & $\log F$  & $r$ & $\log L$\\
   & \# & pc &    & mag & $^a$  &$^b$ & $L_\odot$\\
\hline 
Sz 65										 &    1 &   150 & K6.0 &       0.80 &     -12.95 &       0.00 &    -0.05 \\
Sz 66										 &    1 &   150 & M4.3 &       0.50 &     -13.77 &       0.04 &    -0.73 \\
Sz 68 A								 &    1 &   150 & K2.0 &       1.00 &     -12.14 &       0.00 &     0.74 \\
Sz 68 B								 &    1 &   150 & M5.9 &  (-0.10) &     -14.19 &       0.00 &    -1.04 \\
GW Lup									 &    1 &   150 & M2.3 &       0.55 &     -13.57 &       0.08 &    -0.63 \\
HM Lup									 &    1 &   150 & M2.9 &       0.60 &     -13.75 &       0.10 &    -0.80 \\
\end{tabular}}
\end{table}

\clearpage
\pagebreak

\clearpage
\pagebreak

\begin{table}[!h]
\caption{Electronic Table:  Stellar Properties}
\label{tab:props_electronic.tab}
\begin{tabular}{lccccccccccc}
\hline
Name & 2MASS Name & Date & d & SpT & $A_V$ & $\log F$ & $r_{7510}$ & $\log L$ & R & M & $\log$ Age\\
\hline
MBM12 1 & 2MASS J02552579+2004516 & 19 Jan. & 275 & K5.5 & 0.10 & -13.18 & 0 & 0.25 & 2.57 & 0.75 & 6 \\
MBM12 2 & 2MASS J02560799+2003242 & 19 Jan. & 275 & M0.3 & 1.65 & -13.83 & 0.18 & -0.39 & 1.43 & 0.64 & 6.5 \\
MBM12 3 & HBC 9 & 19 Jan. & 275 & M2.8 & 0.55 & -13.66 & 0.02 & -0.18 & 2.28 & 0.32 & 5.8 \\
MBM12 7 & RX J0256.3+2005 & 30 Dec. & 275 & M5.6 & (-0.10) & -14.79 & 0 & -1.14 & 1.06 & 0.07 & 6 \\
LkHa 264 & 2MASS J02563755+2005371 & 19 Jan. & 275 & K5.5 & (-0.25) & -13.8 & 0.57 & -0.36 & 1.26 & 0.94 & 6.9 \\
MBM12 8 & 2MASS J02574903+2036076 & 28 Dec. & 275 & M5.9 & 0.30 & -15.19 & 0 & -1.51 & 0.71 & -0.02 & 6.8 \\
MBM12 5 & 2MASS J02581122+2030037 & 28 Dec. & 275 & K2.0 & 0.90 & -12.96 & 0 & 0.44 & 2.51 & 1.49 & 6.2 \\
\end{tabular}
\end{table}

\begin{table}[!h]
\begin{tabular}{lccccccccccc}
\hline
Name & 2MASS & Date & d & SpT & $A_V$ & $\log F$ & $r_{7510}$ & $\log L$ & R & M & $\log$ Age\\
\hline
LkCa 5 & 2MASS J04173893+2833005 & 28 Dec. & 131 & M2.2 & 0.05 & -13.46 & 0 & -0.64 & 1.28 & 0.42 & 6.5 \\
V410 X-ray 1 & 2MASS J04174965+2829362 & 28 Dec. & 131 & M3.7 & 1.70 & -14.42 & 0.04 & -1.55 & 0.53 & 0.2 & 7.1 \\
V410 X-ray 3 & 2MASS J04180796+2826036 & 30 Dec. & 131 & M6.5 & 0.20 & -14.54 & 0 & -1.45 & 0.79 & -0.05 & 6.8 \\
V409 Tau & 2MASS J04181078+2519574 & 20 Jan. & 131 & M0.6 & 1.00 & -12.94 & 0 & -0.15 & 1.96 & 0.52 & 6.2 \\
V409 Tau & 2MASS J04181078+2519574 & 29 Dec. & 131 & M0.6 & 1.00 & -13.01 & 0 & -0.22 & 1.8 & 0.54 & 6.2 \\
V409 Tau & 2MASS J04181078+2519574 & 30 Dec. & 131 & M0.6 & 1.00 & -12.98 & 0 & -0.19 & 1.87 & 0.53 & 6.2 \\
HBC 372 & 2MASS J04182147+1658470 & 28 Dec. & 147 & K2.0 & 0.65 & -13.55 & 0 & -0.69 & 0.68 & -- & -- \\
\end{tabular}
\end{table}

\begin{table}[!h]
\begin{tabular}{lccccccccccc}
\hline
Name & 2MASS & Date & d & SpT & $A_V$ & $\log F$ & $r_{7510}$ & $\log L$ & R & M & $\log$ Age\\
\hline
XEST 15-034 & 2MASS J04293623+2634238 & 28 Dec. & 140 & M4.1 & 0.20 & -14.11 & 0 & -1.15 & 0.89 & 0.19 & 6.5 \\
DH Tau & 2MASS J04294155+2632582 & 29 Dec. & 140 & M2.3 & 0.65 & -13.53 & 0.4 & -0.66 & 1.26 & 0.41 & 6.5 \\
DI Tau & 2MASS J04294247+2632493 & 20 Jan. & 140 & M0.7 & 0.70 & -13.02 & 0 & -0.17 & 1.92 & 0.52 & 6.2 \\
DI Tau & 2MASS J04294247+2632493 & 29 Dec. & 140 & M0.7 & 0.70 & -13 & 0 & -0.14 & 1.98 & 0.51 & 6.2 \\
IQ Tau & 2MASS J04295156+2606448 & 29 Dec. & 140 & M1.1 & 0.85 & -13.92 & 0.6 & -1.06 & 0.72 & 0.61 & 7.4 \\
IQ Tau & 2MASS J04295156+2606448 & 30 Dec. & 140 & M1.1 & 0.85 & -13.25 & 0.11 & -0.39 & 1.55 & 0.51 & 6.4 \\
CFHT 20 & 2MASS J04295950+2433078 & 30 Dec. & 140 & M5.2 & 2.30 & -14.27 & 0 & -1.23 & 0.93 & 0.08 & 5.1 \\
\end{tabular}
\end{table}

\begin{table}[!h]
\begin{tabular}{lccccccccccc}
\hline
Name & 2MASS & Date & d & SpT & $A_V$ & $\log F$ & $r_{7510}$ & $\log L$ & R & M & $\log$ Age\\
\hline
GH Tau & 2MASS J04330622+2409339 & 18 Jan. & 140 & M2.3 & 0.40 & -13.07 & 0 & -0.19 & 2.16 & 0.36 & 6.1 \\
V807 Tau & 2MASS J04330664+2409549 & 18 Jan. & 140 & K7.5 & 0.50 & -12.58 & 0.05 & 0.26 & 2.83 & 0.6 & 6 \\
V830 Tau & 2MASS J04331003+2433433 & 30 Dec. & 140 & K7.5 & 0.45 & -13.09 & 0 & -0.24 & 1.58 & 0.73 & 6.5 \\
GI Tau & 2MASS J04333405+2421170 & 20 Jan. & 140 & M0.4 & 2.55 & -13.09 & 0.03 & -0.24 & 1.72 & 0.58 & 6.3 \\
GI Tau & 2MASS J04333405+2421170 & 29 Dec. & 140 & M0.4 & 1.05 & -13.05 & 0.05 & -0.2 & 1.81 & 0.57 & 6.3 \\
GI Tau & 2MASS J04333405+2421170 & 30 Dec. & 140 & M0.4 & 2.55 & -13.16 & 0.08 & -0.31 & 1.59 & 0.6 & 6.4 \\
GK Tau AB & 2MASS J04333456+2421058 & 28 Dec. & 140 & K6.5 & 1.35 & -13.03 & 0.11 & -0.18 & 1.63 & 0.79 & 6.4 \\
\end{tabular}
\end{table}

\begin{table}[!h]
\begin{tabular}{lccccccccccc}
\hline
Name & 2MASS & Date & d & SpT & $A_V$ & $\log F$ & $r_{7510}$ & $\log L$ & R & M & $\log$ Age\\
\hline
LkCa 15 & 2MASS J04391779+2221034 & 18 Jan. & 161 & K5.5 & 0.30 & -12.95 & 0.31 & 0.02 & 1.97 & 0.82 & 6.2 \\
LkCa 15 & 2MASS J04391779+2221034 & 28 Dec. & 161 & K5.5 & 0.30 & -13.14 & 0 & -0.17 & 1.58 & 0.89 & 6.5 \\
LkCa 15 & 2MASS J04391779+2221034 & 29 Dec. & 161 & K5.5 & 0.30 & -13.09 & 0.09 & -0.12 & 1.67 & 0.87 & 6.5 \\
LkCa 15 & 2MASS J04391779+2221034 & 30 Dec. & 161 & K5.5 & 0.30 & -13.1 & 0.06 & -0.13 & 1.66 & 0.87 & 6.5 \\
GN Tau & 2MASS J04392090+2545021 & 18 Jan. & 140 & M2.5 & 3.05 & -13.28 & 0.36 & -0.4 & 1.73 & 0.37 & 6.3 \\
GN Tau & 2MASS J04392090+2545021 & 21 Jan. & 140 & M2.5 & 3.05 & -13.38 & 0.12 & -0.5 & 1.54 & 0.38 & 6.3 \\
ITG 15 & 2MASS J04394488+2601527 & 18 Jan. & 140 & M5.0 & 2.65 & -13.64 & 0 & -0.61 & 1.86 & 0.11 & 5 \\
\end{tabular}
\end{table}

\begin{table}[!h]
\begin{tabular}{lccccccccccc}
\hline
Name & 2MASS & Date & d & SpT & $A_V$ & $\log F$ & $r_{7510}$ & $\log L$ & R & M & $\log$ Age\\
\hline
TWA 6 & 2MASS J10182870-3150029 & 19 Jan. & 67 & M0.0 & 0.05 & -12.9 & 0 & -0.69 & 0.99 & 0.76 & 7.1 \\
TWA 6 & 2MASS J10182870-3150029 & 29 Dec. & 67 & M0.0 & 0.05 & -12.83 & 0 & -0.62 & 1.07 & 0.75 & 7 \\
TWA 7 & 2MASS J10423011-3340162 & 18 Jan. & 34 & M3.2 & -0.10 & -12.63 & 0 & -0.96 & 0.97 & 0.32 & 6.7 \\
TWA 7 & 2MASS J10423011-3340162 & 29 Dec. & 34 & M3.2 & -0.10 & -12.59 & 0 & -0.92 & 1.02 & 0.32 & 6.7 \\
TWA 1 & 2MASS J11015191-3442170 & 18 Jan. & 54 & M0.5 & 0.00 & -12.78 & 0.15 & -0.76 & 0.96 & 0.69 & 7.1 \\
TWA 1 & 2MASS J11015191-3442170 & 19 Jan. & 54 & M0.5 & 0.00 & -12.79 & 0.17 & -0.77 & 0.94 & 0.69 & 7.1 \\
TWA 1 & 2MASS J11015191-3442170 & 20 Jan. & 54 & M0.5 & 0.00 & -12.77 & 0.23 & -0.74 & 0.97 & 0.69 & 7.1 \\
\end{tabular}
\end{table}

\begin{table}[!h]
\begin{tabular}{lccccccccccc}
\hline
Name & 2MASS & Date & d & SpT & $A_V$ & $\log F$ & $r_{7510}$ & $\log L$ & R & M & $\log$ Age\\
\hline
Sz 65 & 2MASS J15392776-3446171 & 28 May & 150 & K6.0 & 0.80 & -12.95 & 0 & -0.05 & 1.86 & 0.79 & 6.3 \\
Sz 66 & 2MASS J15392828-3446180 & 28 May & 150 & M4.3 & 0.50 & -13.77 & 0.04 & -0.73 & 1.46 & 0.13 & 5.2 \\
Sz 68 A & 2MASS J15451286-3417305 & 28 May & 150 & K2.0 & 1.00 & -12.14 & 0 & 0.74 & 3.5 & 1.57 & 5.9 \\
Sz 68 B & Sz 68 B & 28 May & 150 & M5.9 & -0.10 & -14.19 & 0 & -1.04 & 1.23 & 0.06 & 5 \\
GW Lup & 2MASS J15464473-3430354 & 28 May & 150 & M2.3 & 0.55 & -13.57 & 0.08 & -0.63 & 1.3 & 0.41 & 6.5 \\
HM Lup & 2MASS J15475062-3528353 & 28 May & 150 & M2.9 & 0.60 & -13.75 & 0.1 & -0.8 & 1.13 & 0.36 & 6.6 \\
Sz 73 & 2MASS J15475693-3514346 & 28 May & 150 & K8.5 & 2.75 & -13.67 & 0.36 & -0.77 & 0.89 & 0.76 & 7.3 \\
\end{tabular}
\end{table}

\end{document}